# THE DESIGN OF OPTICAL STOCHASTIC COOLING FOR IOTA

V. Lebedev[1], J. Jarvis, H. Piekarz, A. Romanov, J. Ruan, Fermilab, Batavia, IL 60510, U.S.A.
M. Andorf, Cornell University, NY 14850, USA

Abstract[2]: The paper presents a journal version of the Design Report on the Optical Stochastic Cooling experiment to be carried out at IOTA ring in Fermilab later this year. It discusses the theory which experiment is based on, beam parameters, major requirements to the storage ring systems and technical details of the experiment implementation.

---

[1] val@fnal.gov
[2] Work supported by (1) Fermi Research Alliance, LLC under Contract No. De-AC02-07CH11359 with the United States Department of Energy, and (2) U.S. National Science Foundation under Award No. PHY-1549132, the Center for Bright Beams.

# Contents

# 1. Introduction

The Optical Stochastic Cooling (OSC) [1] is based on the same principles as the well-tested microwave stochastic cooling [2,3,4], but it uses much smaller optical wavelengths resulting in a possibility of cooling much denser beam. In the case of stochastic cooling operating at the optimal gain, the maximum emittance cooling rate of a bunched beam can be estimated as [5]:

$$\frac{1}{\tau} \approx \frac{W \sigma_s}{NC} ,  \qquad (1)$$

where $W$ is the bandwidth of the system, $N$ is the number of particles in the bunch, $\sigma_s$ is the rms bunch length, and $C$ is the machine circumference. A transition from the microwave stochastic cooling, which uses typical wavelengths of about 5 cm and relative bandwidth of 50%, to the OSC with a wavelength of 1 - 2 μm and relative bandwidth of 5-20% allows one to increase the cooling rates by more than 3 orders of magnitude. That creates ways to obtain cooling rates required by future hadron colliders.

Due to much smaller wavelength, a usage of usual pickups and kickers in the OSC is impossible. Instead, undulators are used for both the pickup and the kicker. In the OSC a particle emits e.-m. radiation in the first (pickup) undulator. Then, the radiation is amplified in an optical amplifier (OA) and is focused into the second (kicker) undulator where it produces a longitudinal kick to the same particle which generated the radiation. A magnetic chicane between the undulators appropriately delays each particle in the bunch with respect to its radiation, such that the longitudinal kick is corrective.

The IOTA ring is a multipurpose storage ring of 40 m circumference [6]. It is designed to operate both with protons and electrons. The OSC experiment takes place in one of eight straight sections of the ring. The length of the straight is about 6 m. A conceptual layout of IOTA OSC straight is shown in Figure 1. The magnetic chicane also provides space for the OA and optical telescope focusing radiation to the kicker undulator.

The chicane bends particles in the horizontal plane. The beam optics is arranged so that horizontal and longitudinal motions are coupled in the chicane resulting in particle cooling in both degrees of freedom. Cooling in the vertical plane is produced through coupling of betatron motions in the horizontal and vertical planes outside the cooling straight.

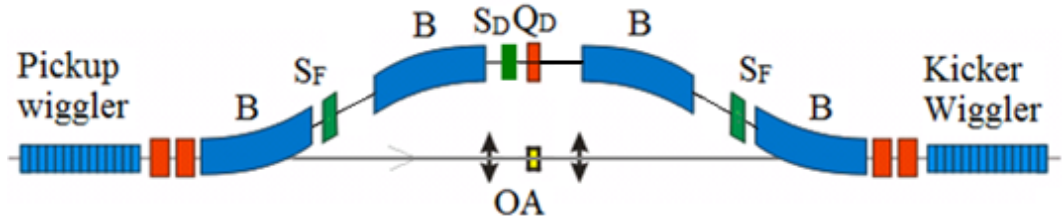

Figure 1: OSC schematic: B – chicane dipoles, $S_F$ and $S_D$ – path length correction sextupoles, $Q_D$ – $x$-$s$ coupling quadrupole, OA – optical amplifier.

Although the OSC was proposed as a cooling method to be used for hadrons or nuclei the OSC

experiments in IOTA use electrons. The usage of 100-MeV ($\gamma = 200$) electrons instead of protons greatly reduces the cost of the experiment but does not limit its generality and applicability to future hadron colliders and storage rings. The IOTA research scope should allow us to test most experimental techniques required for the OSC application in hadron (proton) colliders. Although the emittance and momentum spread in the IOTA electron beam are significantly smaller than in a typical 100 MeV electron storage ring their values are close to a typical emittance and momentum spread of high energy hadron colliders. The demonstration of OSC in IOTA does not require the same large gain of OA as for proton collider. That greatly simplifies the OA design.

The OSC studies are split into two stages. At the first stage, there is no an optical amplifier. In this case just refocusing the e.-m. radiation of the pickup undulator into the kicker undulator amplifies cooling due to synchrotron radiation (SR) by well above an order of magnitude. We call such arrangement the passive OSC. A usage of optical amplifier planned for the second stage will demonstrate cooling with an amplifier. We call this arrangement the active OSC.

A delay of radiation required for the passive OSC is significantly smaller than for the active OSC. That allows us to use shorter wavelengths for the passive OSC resulting in faster cooling. For the active OSC an additional delay in the OA yields a larger delay of radiation and, as will be seen below, requires longer wavelengths. The basic wavelength is chosen to be 0.95 μm for the passive OSC and 2.2 μm for the active OSC. The choice of 2.2 μm wavelength was also determined by availability of wideband optical amplification at this wavelength.

Sections 2 – 8 discuss both the passive and active OSC. Section 10 discusses the optical amplifier for the active OSC. Other sections discuss only passive OSC which starts first and therefore it is in significantly more advanced state. Its mechanical design is mostly complete, and installation work initiated.

## 2. Basics of Beam Optics for the OSC

In this section, we shortly review the theoretical basis required to understand arrangements of the OSC straight and its optics. We refer a reader to Refs. [7,8] for details.

The relative longitudinal momentum change, which a particle receives in the kicker undulator from its amplified radiation, can be approximated as:

$$\frac{\delta p}{p} = -\kappa \sin(k_0 s) \ , \tag{2}$$

where $k_0 = 2\pi/\lambda_0$ is the radiation wave number at the kicker undulator, $s$ is the particle longitudinal displacement relative to the reference particle on the way from pickup to kicker, and the value of the parameter $\kappa$ will be determined later in Section 6. We assume here that the reference particle obtains zero kick. The particle longitudinal displacement depends on its momentum and betatron amplitude. In the linear approximation one can write:

$$s = M_{51}x + M_{52}\theta_x + \left(M_{56} - \frac{L_{pk}}{\gamma^2}\right)\frac{\Delta p}{p} \ , \tag{3}$$

where $M_{5n}$ are the elements of 6x6 transfer matrix from pickup to kicker, $L_{pk}$ is the pickup-to-kicker distance, $\gamma$ is the relativistic factor, $x$, $\theta_x$ and $\Delta p/p$ are the particle coordinate, angle and relative momentum deviation in the pickup. In the absence of betatron oscillations one obtains

$$s = S_{pk}\frac{\Delta p}{p}, \quad S_{pk} = M_{51}D + M_{52}D' + M_{56} - \frac{L_{pk}}{\gamma^2} \ , \tag{4}$$

where $D$ and $D'$ are the dispersion and its derivative in the pickup undulator. Consequently, in the linear approximation the emittance cooling rate[3] for the longitudinal degree of freedom is:

$$\lambda_s = f_0 \kappa k_0 S_{pk} \ . \tag{5}$$

where $f_0$ is the revolution frequency. In the absence of coupling between the horizontal and longitudinal degrees of freedom ($D = D' = 0$) the transverse cooling rate is equal to zero. Consequently, the sum of cooling rates is:

$$\lambda_s + \lambda_x = f_0 \kappa k_0 \left(M_{56} - \frac{L_{pk}}{\gamma^2}\right) \ . \tag{6}$$

The theorem about sum of friction decrements [9] states that the sum of cooling rates is not changed if cooling rates are redistributed between degrees of freedom. A direct proof of this statement in the application to the OSC can also be found in Ref. [7]. Considering that the sum of cooling rates is not changed one obtains the horizontal emittance cooling rate:

---

[3] In this document we are using the cooling rates for emittances and squared momentum deviation. The amplitude cooling rates are twice smaller than the emittance cooling rates.

$$\lambda_x = -f_0 k_0 \kappa \left( DM_{51} + D'M_{5,2} \right) = f_0 k_0 \kappa \left( M_{56} - S_{pk} - \frac{L_{pk}}{\gamma^2} \right). \tag{7}$$

For IOTA the term $L_{pk}/\gamma^2$ is small and we neglect it in further consideration although it is kept in the calculations of real cooling parameters. Eqs. (5) and (7) yield the ratio of cooling rates:

$$\frac{\lambda_x}{\lambda_s} = \frac{M_{56}}{S_{pk}} - 1. \tag{8}$$

For a given $M_{56}$ the ratio is uniquely determined by the dispersion and its derivative at the pickup undulator.

To find cooling rates for large amplitude oscillations we rewrite Eq. (2) in the following form:

$$\frac{\delta p}{p} = \kappa \sin\left( a_x \sin(\psi_x + \psi_c) + a_p \sin(\psi_p) \right), \tag{9}$$

where $a_x$, $a_p$, $\psi_x$ and $\psi_p$ are the dimensionless amplitudes (expressed in the units of radiation phase) and phases of pickup-to-kicker path lengthening due to betatron and synchrotron motions, respectively, and $\psi_c$ is the phase shift between a momentum kick and particle betatron motion. The dimensionless amplitude of synchrotron motion directly follows from Eqs. (2) and (4):

$$a_p = k_0 \left( M_{51} D + M_{52} D' + M_{56} \right) \left( \frac{\Delta p}{p} \right)_m, \tag{10}$$

where $(\Delta p/p)_m$ is the amplitude of momentum oscillations. To find the dimensionless amplitude due to betatron motion we express particle coordinates through its Courant-Snyder invariant[4], $\tilde{\varepsilon}$, and the betatron motion phase, $\psi$:

$$\begin{aligned} x &= \sqrt{\tilde{\varepsilon}\beta} \cos\psi, \\ \theta_x &= -\sqrt{\tilde{\varepsilon}/\beta} \left( \sin\psi + \alpha \cos\psi \right), \end{aligned} \tag{11}$$

where $\beta$ and $\alpha$ are the beta- and alpha-functions in the pickup undulator. Substituting these expressions to the equitation describing the longitudinal displacement due to betatron motion, $M_{51}x + M_{52}\theta_x$, and performing simple transformations one obtains the dimensionless amplitude due to betatron motion:

$$a_x = k_0 \sqrt{\tilde{\varepsilon} \left( \beta M_{51}^2 - 2\alpha M_{51} M_{52} + \left(1+\alpha^2\right) M_{52}^2 / \beta \right)}. \tag{12}$$

Averaging momentum kicks over betatron and synchrotron oscillations one obtains the fudge factors for the transverse and longitudinal cooling rates:

---

[4] The Courant-Snyder invariant is defined as following: $\tilde{\varepsilon} = x^2 \left(1+\alpha^2\right)/\beta + 2\alpha x \theta + \beta \theta^2$.

$$\begin{bmatrix} F_x(a_x,a_p) \\ F_s(a_x,a_p) \end{bmatrix} \equiv \begin{bmatrix} \lambda_x(a_x,a_p)/\lambda_x \\ \lambda_s(a_x,a_p)/\lambda_s \end{bmatrix} = \begin{bmatrix} 2/a_x \cos\psi_c \\ 2/a_p \end{bmatrix} \oint \sin\left(a_x \sin(\psi_x+\psi_c)+a_p \sin\psi_p\right) \begin{bmatrix} \sin\psi_x \\ \sin\psi_p \end{bmatrix} \frac{d\psi_x}{2\pi} \frac{d\psi_p}{2\pi},$$
(13)

Computation of the integrals yields:

$$\begin{bmatrix} F_x(a_x,a_p) \\ F_s(a_x,a_p) \end{bmatrix} = 2 \begin{bmatrix} J_0(a_p)J_1(a_x)/a_x \\ J_0(a_x)J_1(a_p)/a_p \end{bmatrix}.$$
(14)

Here $J_0(x)$ and $J_1(x)$ are the Bessel functions. Note that the integration yielded that $\psi_c$ does not affect the horizontal cooling rate. As one can see from Eq. (14) the cooling rates oscillate with growth of amplitudes. For a given degree of freedom the cooling rate changes its sign the first time at its own amplitude equal to $\mu_{11} \approx 3.832$ and for the other plane at the amplitude of $\mu_{01} \approx 2.405$. Here $\mu_{01}$ and $\mu_{11}$ are the first roots of $J_0(x)$ and $J_1(x)$, respectively. Requiring both cooling rates be positive one obtains the stability condition,

$$a_{x,p} \leq \mu_{01} \approx 2.405 .$$
(15)

That yields the stability boundaries for the emittance and the momentum spread:

$$\varepsilon_{max} = \frac{\mu_{01}^2}{k_0^2 \left(\beta M_{51}^2 - 2\alpha M_{51}M_{52} + (1+\alpha^2)M_{52}^2/\beta\right)},$$
(16)

$$\left(\frac{\Delta p}{p}\right)_{max} = \frac{\mu_{01}}{k_0 S_{pk}}.$$
(17)

We will call these values by the horizontal cooling acceptance and the momentum acceptance of cooling, respectively.

We need to note that the actual cooling acceptances are larger than it is specified by Eqs. (15) – (17). Even if cooling in one plane is absent the cooling in the other plane may bring a particle to a point where cooling is present in both planes. Figure 2 shows particle trajectories in the plane of particle dimensionless amplitudes in the course of cooling for different ratios of cooling rates in the absence of diffusion. The particle trajectories were computed with help of Eqs. (13) and (14), and their shape is characterized by a single parameter – the ratio of cooling rates. As one can see in the case of equal cooling rates all particles inside a circle of radius $\sqrt{2}\mu_{01} \approx 3.4$ are cooled. For non-equal rates shapes of cooling trajectories are distorted but the overall cooling acceptances are weakly affected. In the further analysis to determine the stability boundary we will use a conservative estimate of Eqs. (15) - (17).

We introduce the cooling ranges as ratios of cooling area acceptances $(\Delta p/p)_{max}$ and $\varepsilon_{max}$ to the rms values of momentum spread, $\sigma_p$, and horizontal emittance, $\varepsilon_0$, which are set by the synchrotron radiation in the absence of OSC and x-y coupling. That yields:

$$n_{\sigma s} = (\Delta p/p)_{max}/\sigma_p \ , \quad n_{\sigma x} = \sqrt{\varepsilon_{max}/\varepsilon_0} \ . \tag{18}$$

Note that the horizontal cooling range depends only on the dispersion and its derivative in the pickup undulator, while the transverse cooling range depends on the beta-function and its derivative in the pickup undulator but does not depend on the dispersion.

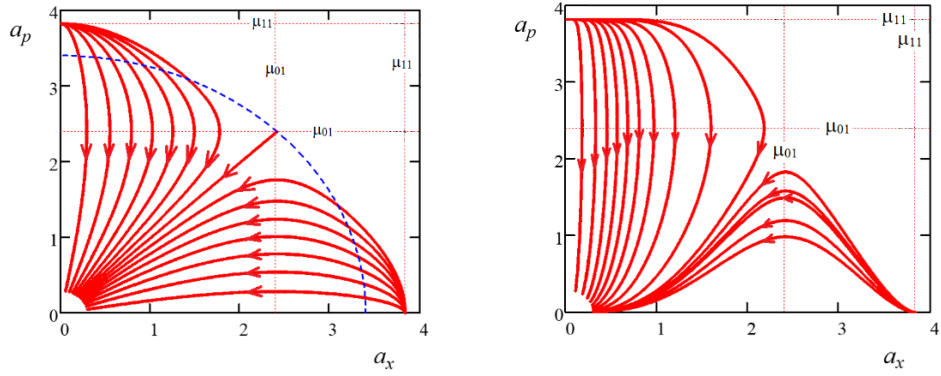

Figure 2: Amplitude trajectories in the course of OSC cooling; left – $\lambda_x/\lambda_s=1$; right – $\lambda_x/\lambda_s=0.3$. Blue dashed circle has radius of $\sqrt{2}\mu_{01}$.

## 3. Linear Beam Optics for OSC

The analysis of possible optics arrangements in the cooling area yielded that the layout presented in Figure 1 is not only the most straightforward but also represents a reliable and effective choice. The cooling chicane consists of four dipoles with parallel edges, which in the absence of other focusing elements does not produce focusing in the horizontal plane resulting in that $M_{56}=S_{pk}$. As one can see from Eq. (7) transverse cooling requires $M_{56}$ and $S_{pk}$ being different. It is achieved by placing a defocusing quad in the chicane center.

To make an estimate showing interdependency of cooling parameters we leave only leading terms in the thin lens approximation assuming also that the bends have zero length and do not produce horizontal focusing[5]. That yields:

$$M_{56} \approx 2\Delta s, \quad S_{pk} \approx 2\Delta s - \Phi D^* h, \tag{19}$$
$$\lambda_x / \lambda_s \approx \Phi D^* h / (2\Delta s - \Phi D^* h).$$

Here $\Delta s$ and $h$ are the path lengthening and the trajectory offset in the chicane, $\Phi = 1/F$ is the defocusing strength of the quad located in the chicane center, and $D^*$ is the dispersion in there. Similarly, using Eq. (17) one obtains estimates for the cooling ranges:

$$n_{\sigma s} \approx \frac{\mu_{01}}{(2\Delta s - \Phi D^* h) k_0 \sigma_p},$$
$$n_{\sigma x} \approx \frac{\mu_{01}}{2 k_0 h \Phi \sqrt{\varepsilon_0 \beta^*}}, \tag{20}$$

where $\beta^*$ is the beta-function in the chicane center. Above we assumed that the optics is symmetric relative to the chicane center, i.e. $dD/ds=0$ and $d\beta/ds=0$ in the chicane center. Such choice minimizes the maximal dispersion and beta-function in the cooling area.

As one can see from Eq. (19) the parameter $\Phi D^* h$ determines the ratio of cooling rates. Assuming equal cooling rates one obtains, $\Delta s = \Phi D^* h$, and, consequently, the cooling ranges are:

$$n_{\sigma s} \approx \frac{\mu_{01}}{k_0 \sigma_p \Delta s},$$
$$n_{\sigma x} \approx \frac{\mu_{01}}{2 k_0 \Delta s} \sqrt{\frac{D^{*2}}{\varepsilon \beta^*}} \equiv \frac{\mu_{01}}{2 k_0 \Delta s} \sqrt{\frac{A^*}{\varepsilon_0}}, \tag{21}$$

where $A^*=D^{*2}/\beta^*$ is the dispersion invariant[6] in the chicane center. Its value is conserved in a straight line where bending magnets are absent. As one can see from the above equations the cooling dynamics is determined by a handful of parameters: the initial rms momentum spread ($\sigma_p$) and emittance ($\varepsilon_0$), the wave number of optical radiation ($k_0$), the dispersion invariant ($A^*$), the trajectory offset in the chicane ($h$), and the path length delay ($\Delta s$). The value of $\Delta s$ is determined by signal delay in optical amplifier and focusing telescope. An amplifier increases the delay. Therefore, in the case of passive cooling we use a smaller delay and shorter wavelength than for the case of active cooling. After few iterations, compromise values of $\Delta s$=0.648 mm and $\Delta s$=2 mm were chosen. Table 1 presents basic beam and optics parameters computed for real optics of the IOTA ring. Computations with Eqs. (19) - (21) yield quite close results. As one can see a transition to a shorter wavelength resulted in a reduction of cooling aperture from 260 to 72 nm. Note that here we assume that the beam optics does not have coupling between the vertical and horizontal planes. Effects of *x-y* coupling are considered in a separate section of this paper.

---

[5] As will be seen below due to small bending angles of the OSC straight dipoles such approximation works quite well for the considered optics.
[6] The dispersion invariant is defined as follows: $A_x = D_x^2 (1+\alpha_x^2)/\beta_x + 2\alpha_x D_x' D_x + \beta_x D_x'^2$. The derivatives of beta-functions and dispersion are equal to zero in the chicane center. That yields: $A^* = D_x^2 / \beta_x$.

**Table 1: Electron beam and beam optics parameters for the IOTA OSC**

| | Passive OSC | Active OSC |
|---|---|---|
| Beam momentum | 100 MeV/c | |
| Rms momentum spread set by SR, $\sigma_p$ | 0.986·10$^{-4}$ | 1.06·10$^{-4}$ |
| Hor. rms emit. set by SR in the absence of OSC and *x-y* coupling, $\varepsilon_0$, nm | 0.857 | 2.62 |
| Delay in the cooling chicane, $\Delta s$, mm | 0.648 | 2.00 |
| Offset in the chicane, $h$, mm | 20.0 | 35.13 |
| Ratio of OSC rates in the absence of *x-y* coupling, $\lambda_x/\lambda_s$ | 1.06 | 1.16 |
| Basic radiation wavelength, $2\pi/k_0$, μm | 0.95 | 2.2 |
| Horizontal cooling acceptance, $\varepsilon_{max}$, nm | 72 | 264 |
| Longitudinal cooling acceptance, $(\Delta p/p)_{max}$, | 5.7·10$^{-4}$ | 4.7·10$^{-4}$ |
| Cooling ranges in the absence of OSC and *x-y* coupling, $n_{\sigma x}/n_{\sigma s}$ | 9.2/5.8 | 10 / 4.4 |
| Horizontal beta-function in the chicane center, $\beta^*$, m | 0.25 | 0.12 |
| Dispersion in the chicane center, $D^*$, m | 0.27 | 0.48 |
| Dispersion invariant in the chicane center, $A^*$, m | 0.29 | 1.92 |

Large value of the dispersion invariant required for the OSC leads to a collider type optics, *i.e.* optics with small value of the horizontal beta-function in the chicane center so that the large value of the invariant could be achieved with manageable value of dispersion. Figure 3 presents the beta-functions and dispersion in the cooling area for passive (0.95 μm) and active (2.2 μm) cooling. Figure 4 shows corresponding dispersion invariants for the IOTA half-ring. The equilibrium horizontal emittance is mainly excited by the SR, and its contribution to the diffusion is proportional to the average value of dispersion invariant in dipoles. Therefore, the strength of quadrupoles in the OSC chicane vicinity was adjusted so that to reduce the invariant as fast as possible outside of cooling area and, thus, to minimize the equilibrium horizontal emittance. That was especially important for the active cooling optics. In this case a reduction of $A_x$ from ~1.9 to

~0.3 m after the first IOTA dipole yielded sufficiently small equilibrium emittance resulting horizontal cooling range of 10σ. In the case of passive cooling the smaller value of dispersion invariant in the OSC straight, $A^*$, resulted in a significantly smaller horizontal equilibrium emittance set by SR. That mostly compensated a redaction of $n_{\sigma x}$ which happened due to redaction of horizontal cooling aperture.

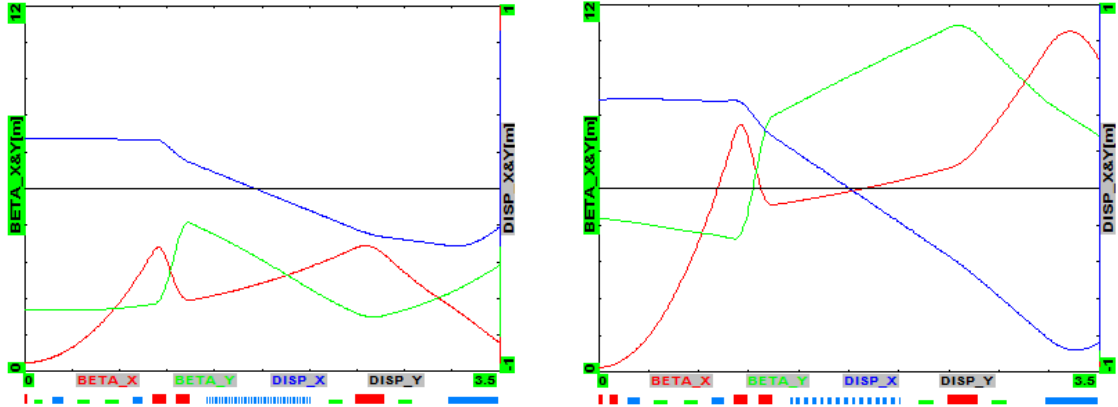

Figure 3: Beta-functions ($\beta_x$ – red, $\beta_y$ – green) and dispersion (blue) for half of cooling straight. Chicane center is located at the frame origin ($s = 0$); red squares at the bottom mark positions of quads, the blue squares – dipoles and undulator; left - optics for passive cooling and right – for active cooling.

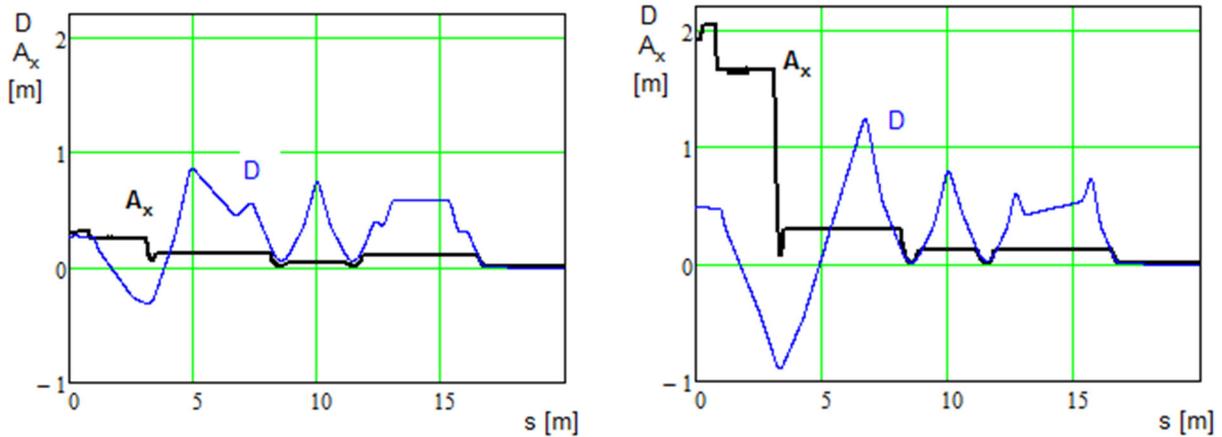

Figure 4: The dispersion (blue line) and the dispersion invariant (black line) for the IOTA OSC optics. Chicane center is located at $s = 0$; left - optics for passive cooling, and right – for active cooling.

Figure 5 presents dependences for $M_{56}$ and the partial slip factor $S = M_{51}D + M_{52}D' + M_{56}$ on the beam travel from pickup to kicker. One can see that $S$ has large variations of its value on the beam travel through the chicane. These variations are excited by large dispersion in the chicane. In the absence of focusing in the chicane center, $S$ and $M_{56}$ would be equal at the chicane end. Non-zero focusing makes them different. Large $S$ variations make resulting $S$ being quite sensitive to optics errors. The sample lengthening due to betatron motion is even more sensitive to optics errors. It is as shown in Figure 6. One can see that the final lengthening is about 100 times smaller than its peak value located between chicane dipoles.

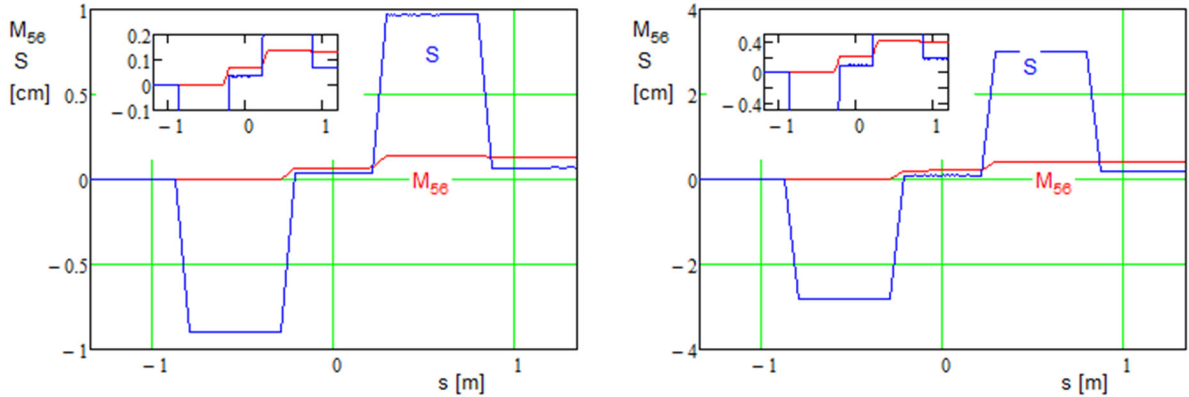

Figure 5: Dependence of $M_{56}$ and $S$ in the OSC straight; chicane center is located at $s = 0$; left - optics for passive cooling, and right – for active cooling; inserts show the same data with smaller vertical scale.

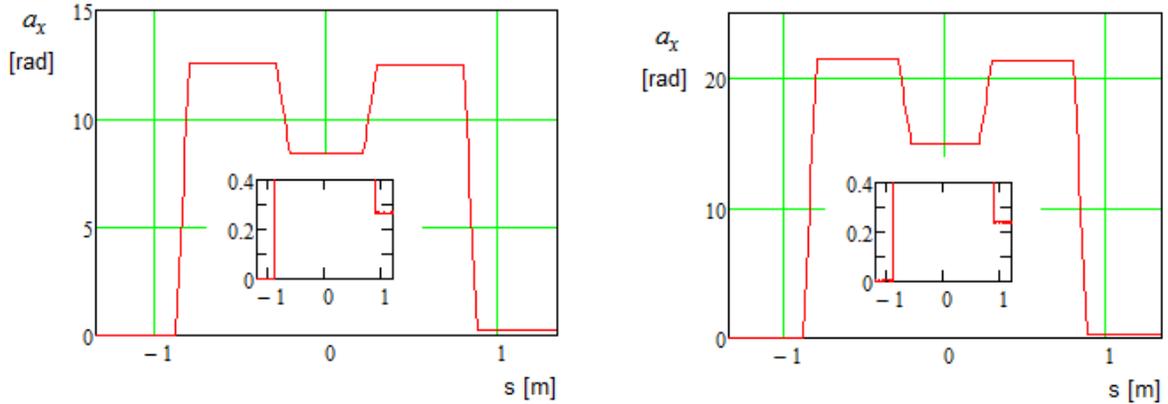

Figure 6: Dimensionless rms sample lengthening due to betatron motion, $a_p$, in the OSC straight; chicane center is located at $s = 0$; left - optics for passive cooling and right – for active cooling; inserts show the same data with smaller vertical scale.

For chosen parameters of the OSC chicane optics the OSC has moderate sensitivity to optics errors in the ring. Figure 7 and Figure 8 show a dependence of cooling acceptance and cooling rates ratio on the relevant optics parameters at the end of pickup undulator. We imply here that the focusing properties of all elements in the OSC straight are at the design values and that optics mismatches are introduced by focusing errors in the rest of the ring. As one can see $\Delta\beta/\beta < 10\%$ and $\Delta D < 10$ cm should be sufficient to achieve reliable OSC.

A variation of strength of the defocusing quad located in the center of OSC straight (marked $Q_D$ in Figure 1) allows one to adjust the ratio of cooling rates and cooling acceptances. Figure 9 shows how variations of this quad integral strength change the ratio of cooling rates and cooling ranges.

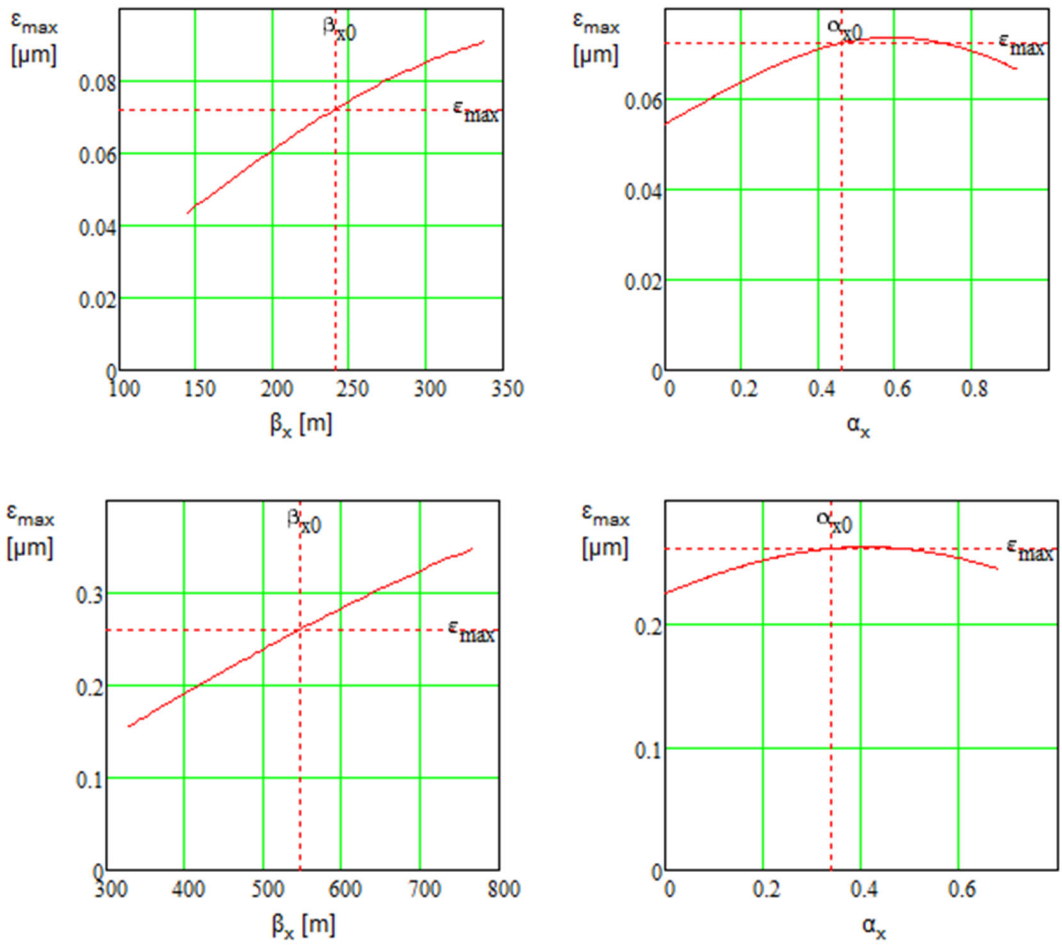

Figure 7: Dependence of cooling acceptance on the β- and α-functions at the end of pickup undulator; top - optics for passive cooling, and bottom – for active cooling. Vertical and horizontal lines show the design parameters.

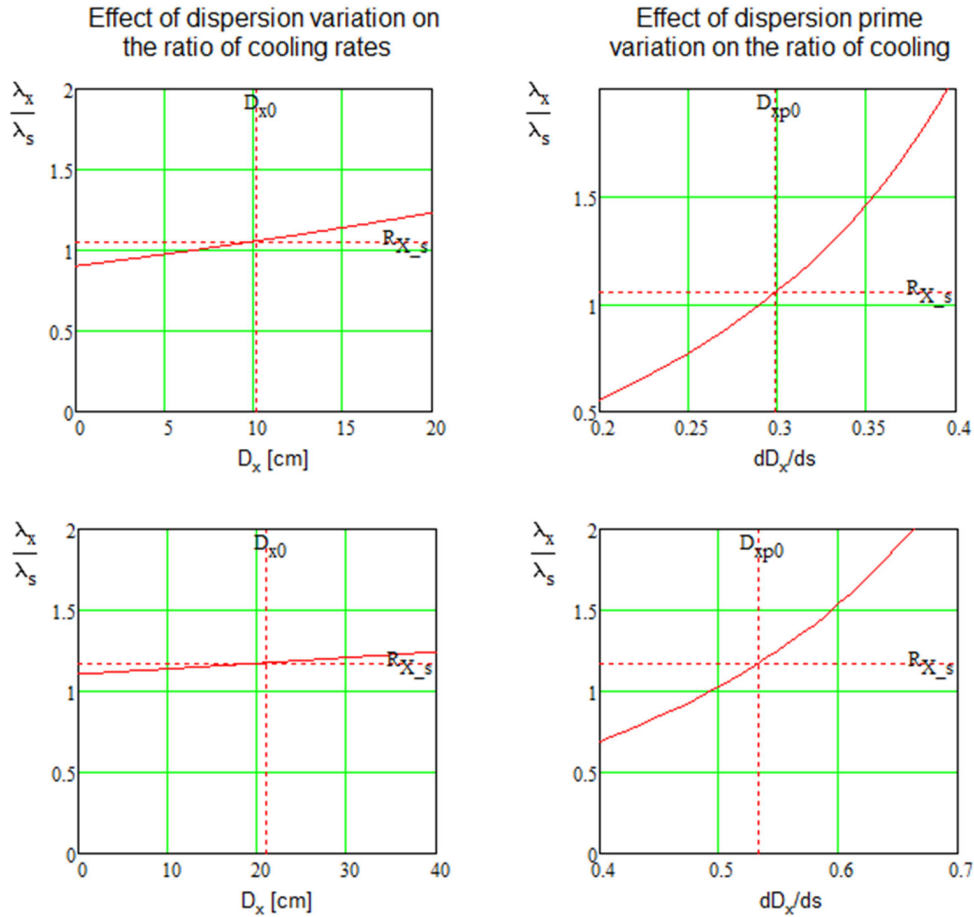

Figure 8: Dependence of cooling rates ratio on the dispersion and its derivative at the end of pickup undulator; top - optics for passive cooling and bottom – for active cooling. Vertical and horizontal lines show the design parameters.

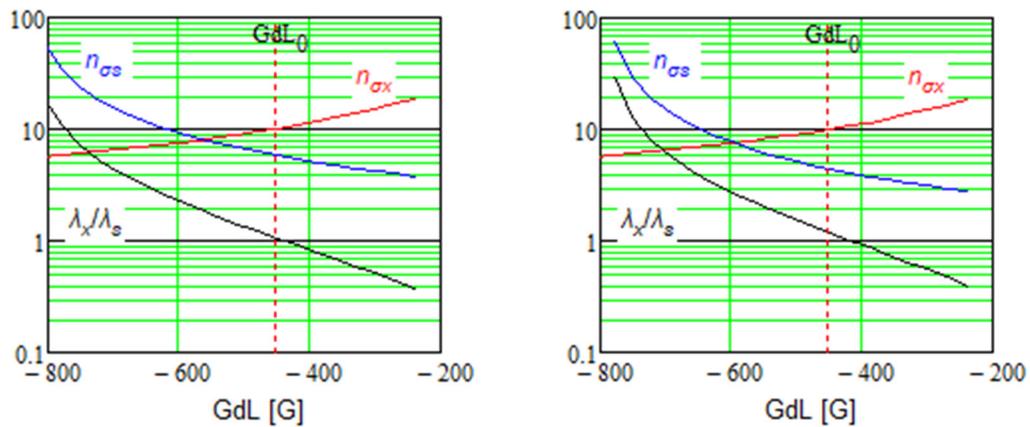

Figure 9: Dependences of horizontal (red) and longitudinal (blue) cooling ranges, and the ratio of cooling rates (black) on the integral strength of OSC quad: passive cooling – left, active cooling – right. Vertical dashed lines mark nominal values of the quad strength. The reference rms emittance and momentum spread are: 0.857 nm, $0.986 \cdot 10^{-4}$ for passive cooling, and 2.62 nm, $1.06 \cdot 10^{-4}$ for active cooling.

## 4. Non-linear Beam Optics for OSC

Another important limitation on the beam optics is associated with the higher order contributions to the path lengthening coming from the betatron and synchrotron motions. The major contribution comes from the transverse angles in particle motion. In a straight line the second order correction for the path lengthening is:

$$\Delta s_2 = \frac{1}{2}\int\left(\theta_x(s)^2 + \theta_y(s)^2\right)ds \ , \tag{22}$$

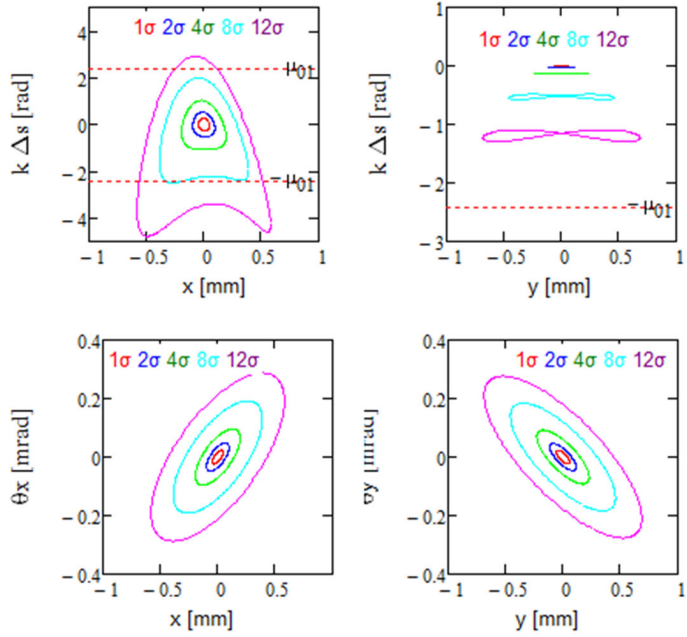

Figure 10: Dependence of path lengthening on the way from pickup to kicker (top) and the particle angle in the kicker (bottom) on the particle horizontal (left) or vertical (right) coordinate for different amplitudes of betatron motion in the absence of path length correction for the passive OSC; 1σ corresponds to the single particle emittance (Courant-Snyder invariant) of 0.857 nm. Horizontal dashed lines mark the boundary of cooling acceptance

where $\theta_x(s)$ and $\theta_y(s)$ are angles of a particle on the way from pickup to kicker. Path lengthening in the bends was computed assuming rectangular dipoles with hard edge and uniform magnetic field. Figure 10 shows a dependence of path lengthening on the way from pickup to kicker for different amplitudes of betatron motion in the absence of non-linear correction. Each closed curve represents the path lengthening for a given action (Courant-Snyder invariant) and betatron phase varying from 0 to 360 deg[7]. The value of invariant is determined by the betatron amplitude; and in finding initial particle locations in the phase space we assume linearity of the input phase space, *i.e.* particles are located at the ellipse determined by initial Twiss parameters and the action. Thus, for each location in horizontal axis we obtain two points at the vertical axis. Due to much smaller value of the horizontal beta-function than the vertical one (see Figure 3) and, consequently, much larger horizontal angles in the chicane center the effect of non-linearity in path lengthening is significantly larger in the horizontal plane than in the vertical one. Note that in the absence of non-linearities, path lengthening due to vertical angle would be equal to zero while path lengthening in the horizontal plane should describe an ellipse in the $k\Delta s$–$x$ plot as follows from Eq. (3) for $\Delta p/p = 0$ and $x$–$\theta_x$ coordinates located at the ellipse with constant value of Courant-Snyder invariant. However, as one can see in the top-left picture in Figure 10 and Figure 11 the phase trajectories resemble ellipse for small amplitudes only. This large non-linear contribution results in significant reduction of cooling acceptance in the horizontal plane and needs to be compensated. Note that in the absence of sextupole compensation the betatron motion is linear what can be seen in the bottom pictures of Figure 10 and Figure 11.

---

[7] Here we assume that the particles are located at input to the OSC region

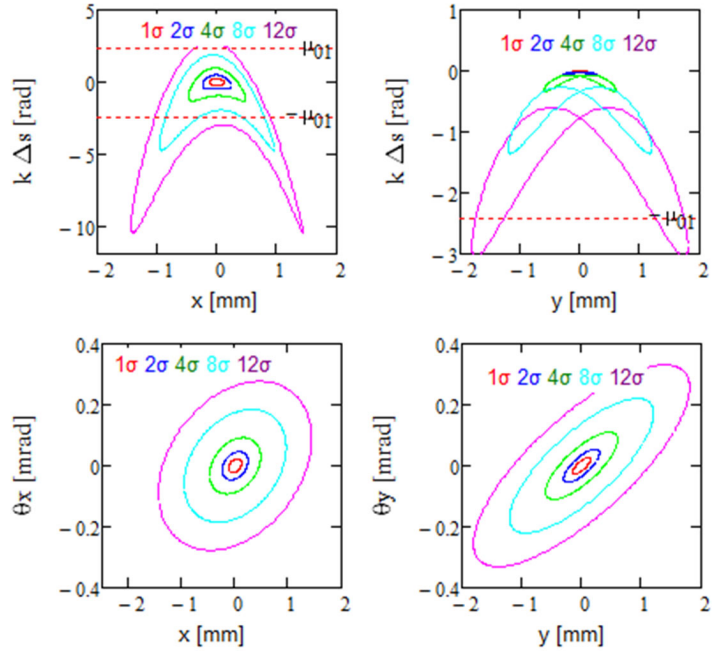

Figure 11: The same as Figure 10 but for the case of active OSC; 1σ corresponds to the single particle emittance (Courant-Snyder invariant) of 2.62 nm.

Effect of non-linearity is increased with decrease of the wavelength and β-function in the chicane center, and with an increase of the rms emittance. A combination of these three major contributions determines non-linearities of longitudinal displacements presented in Figure 10 and Figure 11.

Non-linear path lengthening is corrected by three sextupoles: two located between each pair of chicane dipoles and another one near the central quad as shown in Figure 1. To allocate space for other OSC systems only one of two possible sextupoles near the central quad is used. Large difference of horizontal and vertical beta-functions in the $S_F$ and $S_D$ locations enables a non-linear path lengthening compensation for both planes. Corresponding results are presented in Figure 12 and Figure 13 for the cases of passive and active OSC, respectively. The values of sextupole strength are quite large. Sextupoles significantly affects linearity of beam transport in the horizontal plane as one can see in these figures (left-bottom pictures). Consequently, these sextupoles reduce the dynamics aperture.

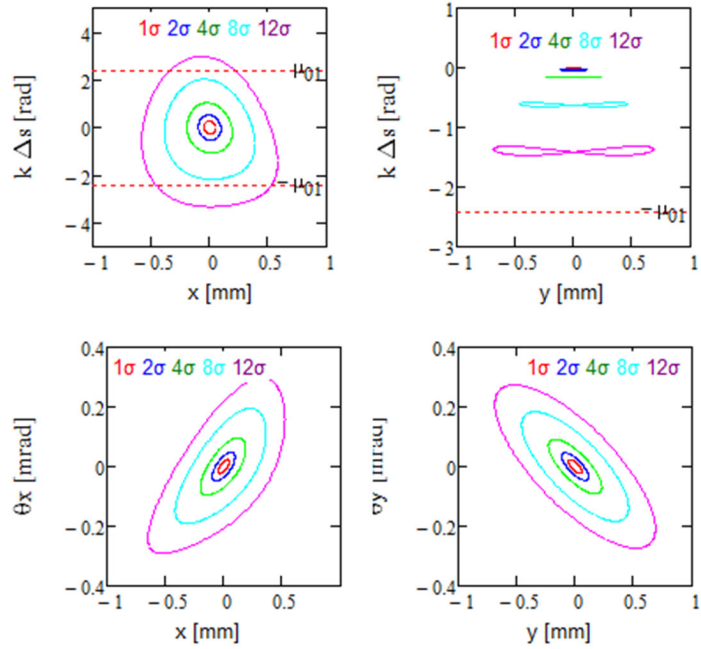

Figure 12: The same as Figure 10 but the non-linear path lengthening is corrected by sextupoles. The integral strength of sextupoles is: $S_FdL$ = 11 kG/cm (two sextupoles, L=10 cm), $S_DdL$ = −6 kG/cm (one sextupole, L=6 cm).

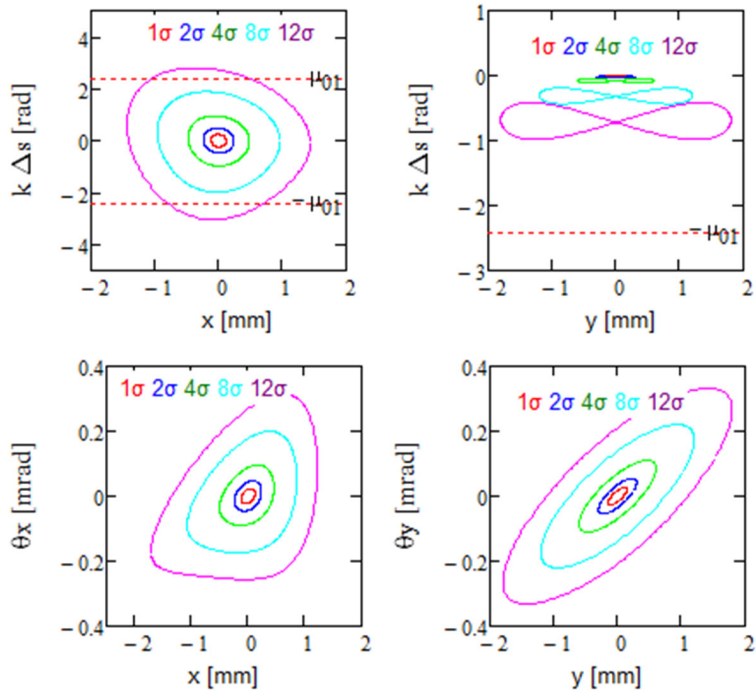

Figure 13: The same as Figure 11 but the non-linear path lengthening is corrected by sextupoles. The integral strength of sextupoles is: $S_FdL$ = 7.5kG/cm, $S_DdL$ = −5.8kG/cm (single sextupole).

To be effective for the non-linear path length correction the OSC straight sextupoles should be located between each pair of dipoles where the dispersion is large. That results in their large contributions to the ring chromaticity. If the chromaticity is corrected by the ring sextupoles at additionally may reduce the dynamic aperture (see details below in Section 5).

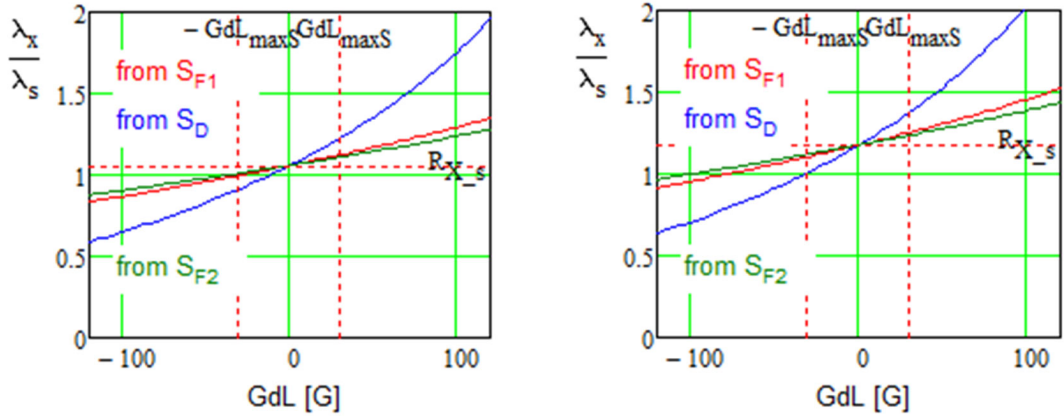

Figure 14: Dependence of the ratio of cooling rates on the quadrupole field integral at the location of corresponding sextupoles (see Figure 1): left – passive OSC, right – active OSC.

Another problem associated with the large values of the path length correction sextupoles is a feeddown of the sextupole field to the quadrupole field due to beam offsets in the sextupoles. It results in (1) a redistribution of cooling rates and (2) changes in the cooling acceptances. Figure 14 presents how the ratio of cooling rates is changing with appearance of quadrupole field at the location of each sextupole. One can see that it is desirable for the quadrupole integral strength do not exceed about 30 G. That corresponds to a 20-30 µm beam offsets in sextupoles and represents a very tight requirement which needs to be satisfied to achieve beam cooling in the entire cooling acceptance anticipated for the IOTA OSC.

## 5. IOTA Beam Optics

The basic layout of the IOTA ring is shown in Figure 15. Its design is dictated by enclosure space limitations and specific demands from various planned experiments. The ring geometry, positions of common elements, and powering of main quadrupoles and dipoles have mirror symmetry with respect to the vertical middle plane of the figure. The ring is composed of 8 bending magnets (four 30◦ and four 60◦). For the OSC experiment the ring has 41 quadrupoles grouped into 21 families. The correction system includes 16 skew-quadrupoles combined with horizontal and vertical frame-type correctors, 2 strong vertical correctors for bumping closed orbit during injection, 8 horizontal correctors in the main dipoles and 15 sextupoles. The diagnostic system has 20 electrostatic pickups, and 10 synchrotron light-based beam image acquisition stations installed at each of 8 dipoles and 2 OSC undulators. There are 6 straight sections in the ring. Two of them are taken by RF system and injection. Other four are planned to be used for the experiments: two for integrable optics, one for electron lens, and one for OSC.

Aperture limitations (half gap or radius) for the IOTA ring with exception of the OSC straight are determined by the following elements:
- 2.37 cm - vacuum chamber in straights made of stainless 2" OD steel tube,
- 3.1 x 2.0 cm – vertical kicker,
- 2.0 x 3.1 cm – horizontal kicker,
- 1.2 cm – octupoles for the integrable optics experiments.

For OSC optics the main aperture limitations are in the OSC straight. Compared to the acceptance determined by apertures pointed out above, they reduce acceptance by about 5 times resulting in the horizontal and vertical acceptances to be 10.5 and 6.6 μm, respectively.

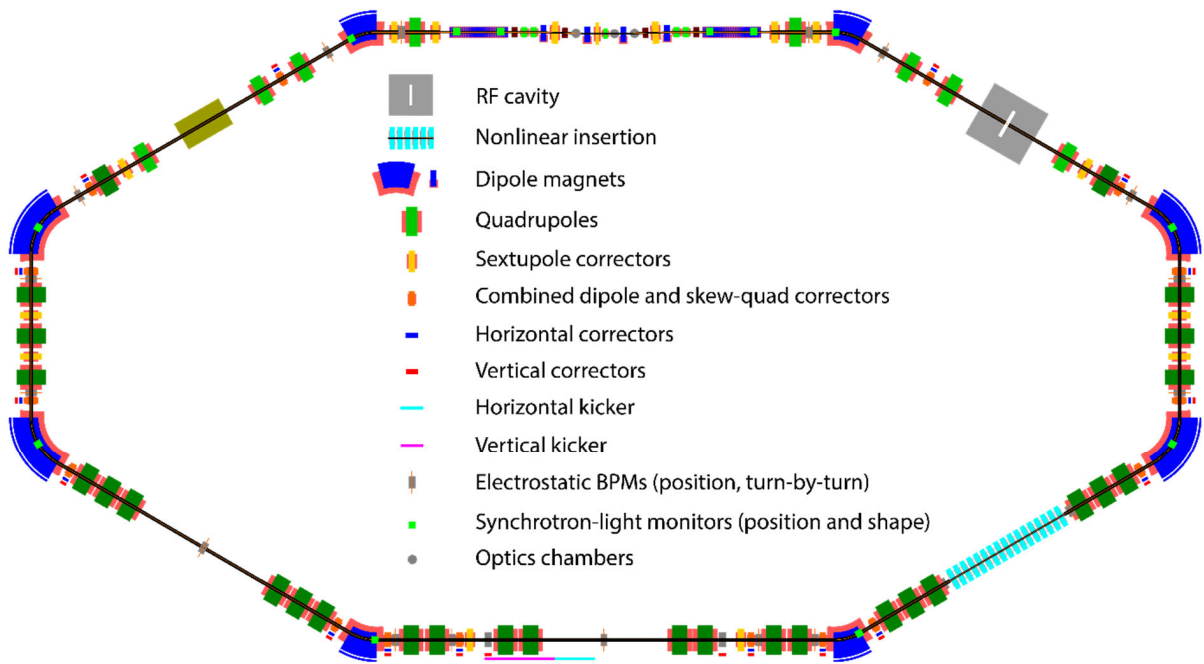

Figure 15: IOTA layout with basic set of magnets. OSC straight is at the top.

The optical stochastic cooling experiment uses a long straight section opposite to the injection straight (see Figure 15). All elements of this straight will be replaced with elements of the OSC insert presented in detail in Figure 16. To support the mirror symmetry of IOTA lattice, the magnets of OSC insert and their powering also have mirror symmetry with exception of sextupole sx1r. The dispersion and beta-functions at the middle of the OSC insertion determine the cooling parameters and therefore are fixed to the desired values (see Table 1). Using the beta-functions and dispersion in the OSC center as a starting point the optics of entire ring was built for the cases of passive and active OSC. Corresponding beta-functions and dispersions are presented in Figure 17. Table 2 presents the main ring parameters for the OSC optics.

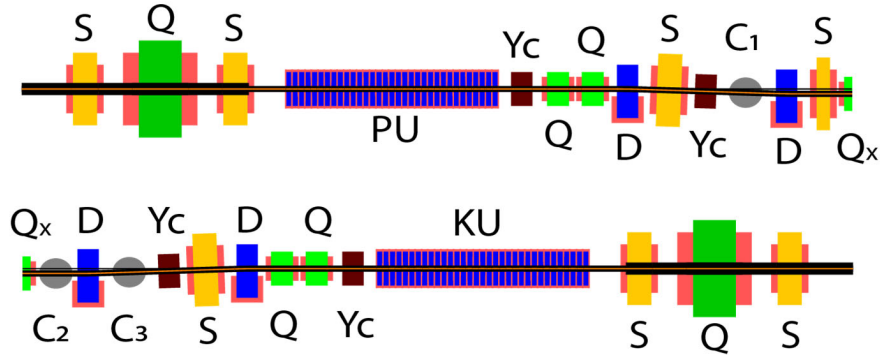

Figure 16: A schematic layout of the OSC straight: top – first half (upstream of the straight center), bottom – second half; qx1 is located in the OSC straight center and marked Qx; the sextupole sx1r is next to the left (upstream) from Qx and is marked by S.

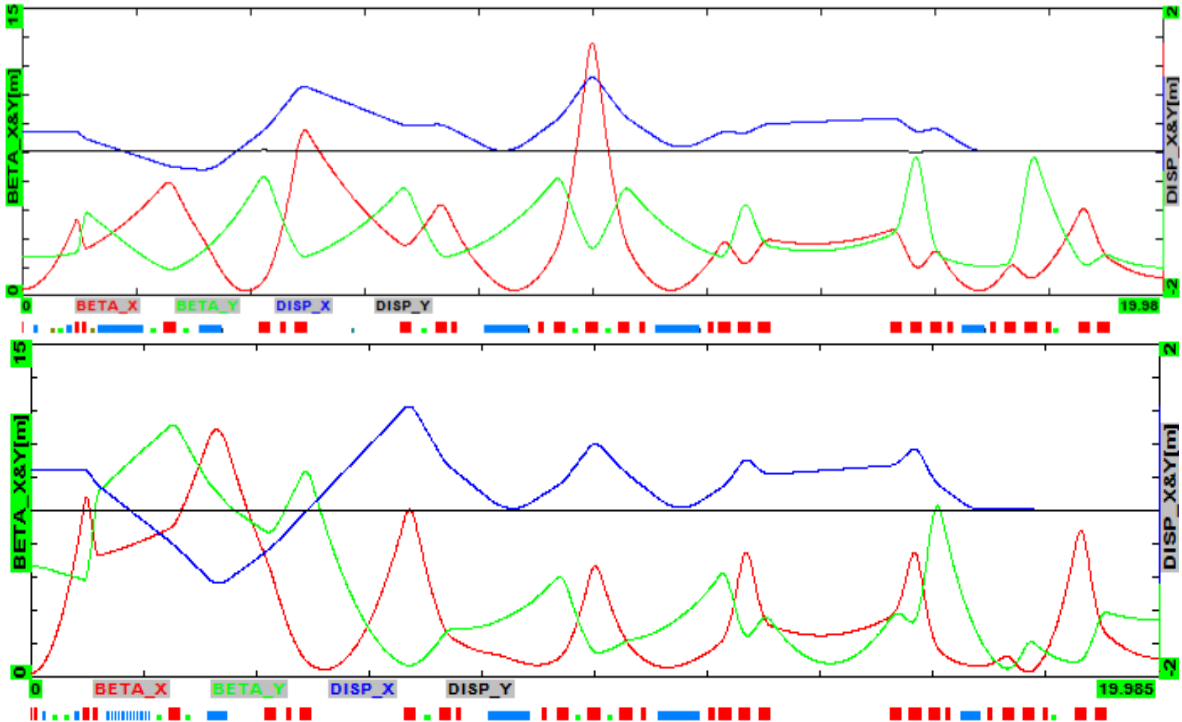

Figure 17: The beta-functions ($\beta_x$ – red, $\beta_y$ – green) and the dispersion (blue) for the passive (top) and active OSC optics for the half of IOTA ring starting from the center of the OSC straight.

There are considerable differences between passive and active OSC. In particular, these differences include: (1) different undulators due to different basic wavelengths, and (2) different vacuum chambers of central part of OSC straight due to different offsets in the chicane and replacement of optical lens by optical amplifier. Majority of other elements can be used in both experiments. To maximize the horizontal cooling range the optics optimization was aimed to minimize the equilibrium horizontal emittance. The horizontal emittance is proportional to the average value of dispersion invariant in the dipoles. The dispersion invariant is fixed in the OSC straight (see Section 2). Therefore, a minimization of the emittance implies a minimization of the dispersion invariant for the rest of the ring. As one can see in Figure 4 the dispersion invariant, which value is large in the OSC straight, is greatly decreased after the dipoles nearest to the OSC straight. It resulted in an acceptable value for the equilibrium horizontal emittance for both passive and active OSC.

**Table 2: Main IOTA ring parameters for OSC optics**

| | Passive OSC | Active OSC |
|---|---|---|
| Beam kinetic energy, MeV | 100 | |
| Circumference, m | 39.96 | |
| Energy drop per turn due to SR*, eV | 12.7 | |
| Betatron tunes, $\nu_x / \nu_y$ | 5.42/2.42 | 5.42 / 3.42 |
| Natural chromaticity**, $\xi_x / \xi_y$ | -10.2 / -8.1 | −10.5 / −10.2 |
| Contributions of OSC sextupoles to ring chromaticities, $\delta\xi_x / \delta\xi_y$ | 29.2 / -20.9 | 60.5 / −44.4 |
| Emittance cooling times in the absence of OSC, s ($\tau_x/\tau_y/\tau_s$) | 1.06 / 1.05 / 0.52 | 0.87 / 1.0 / 0.55 |
| Momentum compaction factor | 0.00171 | -0.0165 |
| Dynamic aperture, $\varepsilon_{xmax}/\varepsilon_{ymax}$ (nm) | 350/350 | 260 /260 |
| Geometric ring acceptances for OSC optics, hor. / vert. (μm) | 10.5/6.6 | - |
| RF voltage, V | 70 | 45 |
| RF frequency | 30.01 MHz | |
| Harmonic number | 4 | |
| Accelerating phase (to compensate SR loss), deg. | 10 | 16 |
| RF bucket height, $\Delta p/p$ | $7 \cdot 10^{-3}$ | $1.6 \cdot 10^{-3}$ |
| Synchrotron frequency | $2.72 \cdot 10^{-5}$/204 Hz | $6.7 \cdot 10^{-5}$/502 Hz |
| Rms bunch length in the absence of OSC, cm | 3.9 | 16.6 |
| Number of periods in an undulator | 16 | 7 |
| Peak magnetic field of undulator, kG | 2.22 | 1.005 |
| Undulator period, cm | 4.840 | 11.063 |
| Vertical betatron tune shifts introduced by undulators, $\delta\nu_y$ | 0.084 | 0.067 |

\* This number accounts for contribution coming from the OSC undulators.
\*\* We imply here that the natural chromaticity is determined by linear optics. The ring chromaticity measured in the IOTA Run II (spring of 2020) showed that there are sources of chromaticity which are presently unaccounted in optics. Most probable, these sources are non-linear fields of bending magnets. An estimate yields that their contribution to the chromaticities should not exceed 2 units for OSC optics.

As will be seen in Section 6 there is shallow maximum of the cooling rate dependence on the undulator parameter, *K*. The optimum value depends on the acceptance of light optics and is near K ≈ 1.5. The undulator parameter is directly related to the undulator magnetic field which (1) determines an addition to the equilibrium beam emittance set by SR and (2) and changes the vertical focusing in the ring. That represents additional limitations. The length of each undulator is limited by the available space to about 77 cm. For a chosen basic wavelength, the length of undulator period and, consequently, the number of periods in the given space depend on the magnetic field in the undulator. Figure 18 shows how the horizontal emittance increase and the vertical tune shift depend on the number of periods, where the undulator magnetic field was adjusted to obtain desired value of basic radiation wavelength. Basing on this data and the cooling rates dependence on the undulator parameter presented in Section 6 the number of periods was chosen to be 16 for the passive cooling and 7 for the active cooling. Note that the value of dispersion invariant in the OSC undulators is set by OSC requirements. That makes the horizontal emittance growth uniquely dependent on the number of periods. Note also that for the chosen number of periods the tune shift does not exceed 0.1. That makes optics correction relatively simple when the undulators are turned on and off.

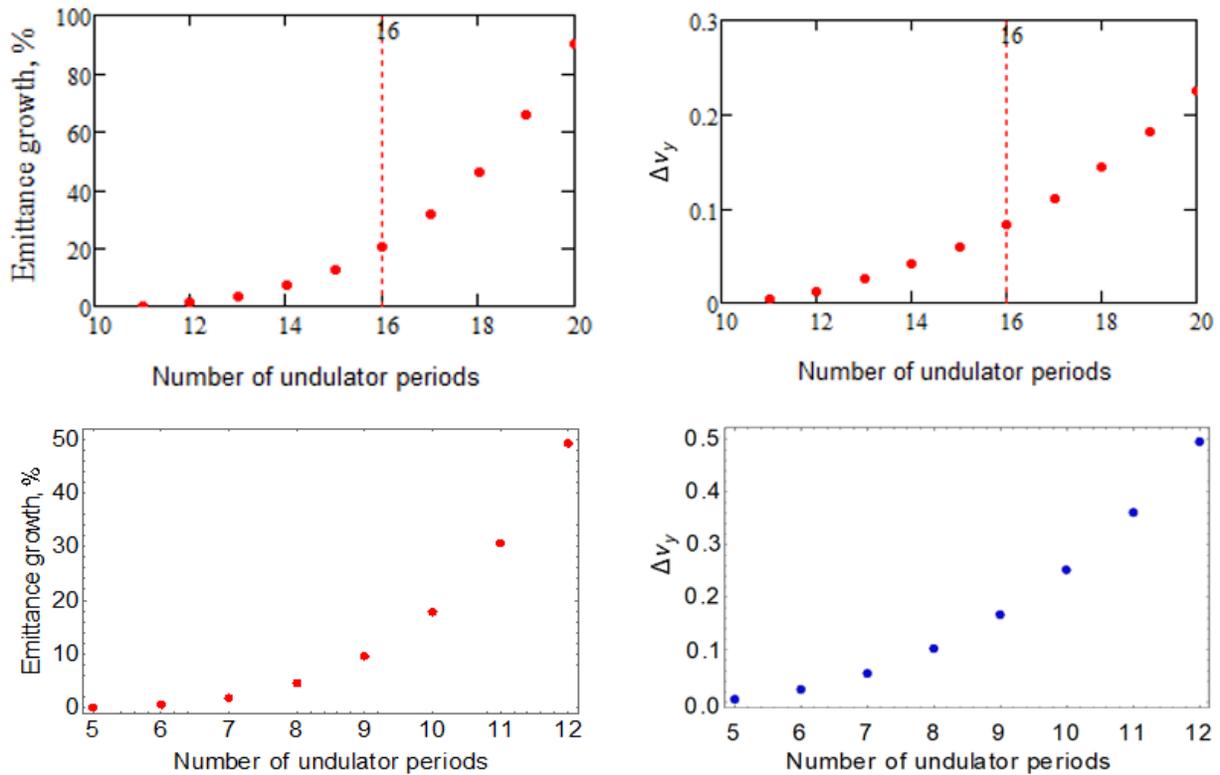

Figure 18: Dependences of emittance growth (left) and vertical tune variation (right) on the number of undulator periods for passive OSC (top) and active OSC (bottom); the total length of undulator is 77.5 cm.

Strengths of sextupoles in the OSC chicane are fixed at quite large values by a requirement to cancel the non-linearity of longitudinal motion introduced by transverse motion of electrons on their travel from pickup to kicker (see Section 4). For the active cooling these sextupoles introduce unacceptably large tune chromaticities, and therefore their effect on the chromaticity must be compensated by other sextupoles of the ring. Six families of sextupoles are available to control two chromaticities. That potentially could be used to minimize the driving terms of third order resonances. Unfortunately, the sextupole strengths are limited resulting in 3 families being set to its maximum values (see Table 3). An optimization of the other sextupoles resulted in acceptable dynamic aperture of about 10σ for both transverse degrees of freedom. The dynamic aperture study is based on the FMA analysis carried out with LifeTrack [10]. The results are presented in Figure 19. As one can see small betatron coupling does not affect the dynamic aperture leaving it at the same value of about 10σ or 150 nm in both transverse planes.

The dynamic aperture for the passive cooling is discussed in Section 5.3 below.

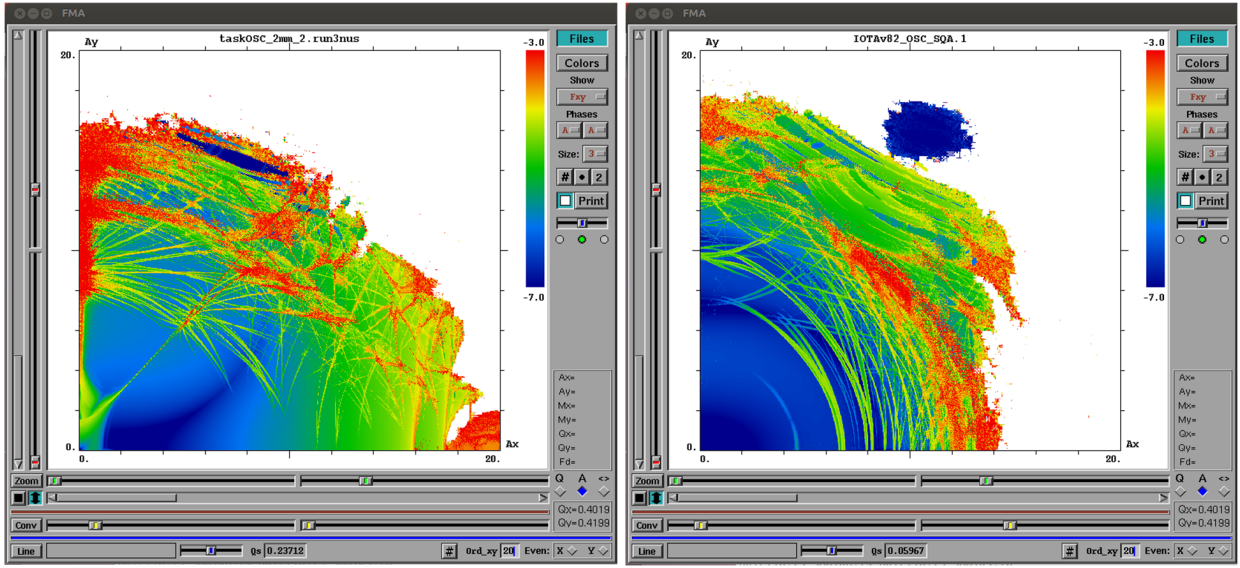

Figure 19: FMA analysis of dynamic aperture for ideal lattice (left) and lattice with small betatron coupling (tune split of 0.003) that fully excites coupling resonance (right); reference emittance is 1.5 nm; optics for active OSC. Blue color corresponds to the linear and stable motion which determines the longtime dynamic aperture. White color corresponds to the particle amplitudes where a particle did not survive to the end of tracking.

Table 3: Gradients of sextupoles for zero chromaticity ($S = d^2B_y/dx^2$ [kG/cm$^2$])

| ID | sa1* | sc1* | sc2* | sd1 | se1 | se2 | sx1 | sx2 |
|---|---|---|---|---|---|---|---|---|
| $S_{sext}$** | 0.335 | -0.335 | -0.335 | 0.118 | 0.331 | -0.319 | -1.05 | 1.79 |

\* Sextupole strength is limited.
\*\* The effective length of sextupoles is equal to 10 cm

## 5.1. Injection

The momentum of the electrons of about 100 MeV/c required for the OSC experiment is considerably smaller than the nominal momentum of 150 MeV/c initially planned for other experiments at IOTA. That results in more relaxed parameters for the injection. Figure 20 shows the trajectory of incoming particles and the closed orbit bump required for oscillation-free injection. Table 4 presents corresponding fields in the kicker and vertical correctors creating the bump. Proposed schema allows one to have a large clearance between the closed orbit and the physical aperture.

During IOTA Run-II (spring of 2020) IOTA was operating at 100 MeV. The closed orbit bump was used for injection of electrons. Typical injection efficiency was about 80-90%. Capabilities of the FAST linac [6] significantly exceed IOTA requirements, and therefore tuning for perfect injection was not necessary. During IOTA Run-I a special low-emittance lattice (similar to the OSC optics) was tested with up to 0.2 mA beam current. It demonstrated good performance for the beam injection.

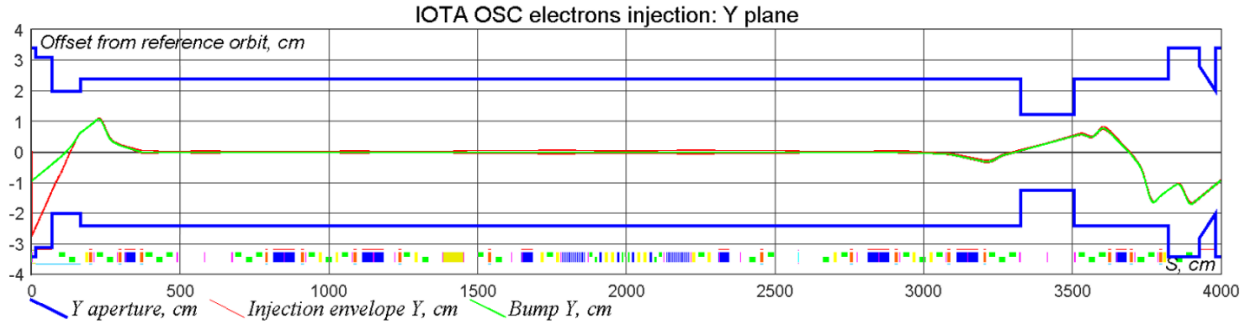

Figure 20: Beam centroid trajectories at injection for the IOTA ring; red line – trajectory of injected beam, green line – the orbit distortion required for injection, blue lines – aperture limitations.

Table 4: Fields in the kicker and correctors for OSC injection

| Element | Kicker | yBumpL | ySQA1R | ySQA2R | ySQB1R | ySQC1L | ySQB2L | ySQB1L | ySQA2L | ySQA1L | yBumpL |
|---|---|---|---|---|---|---|---|---|---|---|---|
| H [G] | 70 | -270 | 11 | 48 | 111 | -51 | 67 | -120 | -120 | -120 | -60 |
| L[cm] | 95 | 10 | 10 | 10 | 10 | 10 | 10 | 10 | 10 | 10 | 10 |

## 5.2. Orbit and Optics Correction

There are two major requirements for the closed orbit correction. First, the orbit should be within 20 μm from the centers of OSC chicane sextupoles to avoid gradient feeddown errors (see details in Section 4). Second, the beam in the kicker undulator should be aligned with light focused from pickup undulator. This means that the radiations radiated in both undulators should coincide in position and angle. We require that two light beams should coincide in coordinate not worse than 5% of the total size of light spot on the entire travel in the kicker undulator. The total light spot size in the kicker undulator is about 500 μm (see Table 7). Consequently, the coordinate errors of two-beam alignment have to be smaller than 25 μm. The corresponding angular error, which keeps beams together on the entire travel in the kicker undulator, error is 35 μrad. Note that the alignment of two light beams is affected and, consequently, can be partially corrected by alignment of light focusing structure.

The other requirement for the insertion is an ability to change the electron's delay. The only way to comply with both requirements is to have all 3 sextupoles on movable supports that will allow to adjust their horizontal positions. With a delay set by the physical location of sextupoles, the horizontal and vertical orbits require control over five degrees of freedom to align beam through 3 sextupoles and to match electrons trajectory in the kicker undulator with radiation from the pickup undulator. Ring trims allow control of 2 degrees of freedom (offset and angle in the pickup undulator). Additionally, there are 4 trims in each plane but two trims adjacent to the undulators effectively control only one degree of freedom. Vertical trims are standalone magnets. For horizontal orbit manipulations chicane dipoles (powered in pairs with additional coils enabling independent control of magnetic field in all four dipoles) and end trims of undulators will be used. Figure 21 shows an example schematic diagram of vertical orbit correction. Starting from the center of the middle sextupole, two inner trims are used to align the beam in the outer sextupoles, two outer trims are used to align electrons and photons and ring trims are used to: two correctors between sextupole doublets are used to center orbit in the sextupoles; other two, located between sextupoles and undulators, are used to match positions of electrons and radiation. Ring correctors are used to close the orbit. For the horizontal plane, where variations of individually powered chicane dipoles will be used for orbit correction, similar approach can be applied. For commissioning and operations a set of knobs will be prepared that independently changes orbit in the sextupoles and undulators.

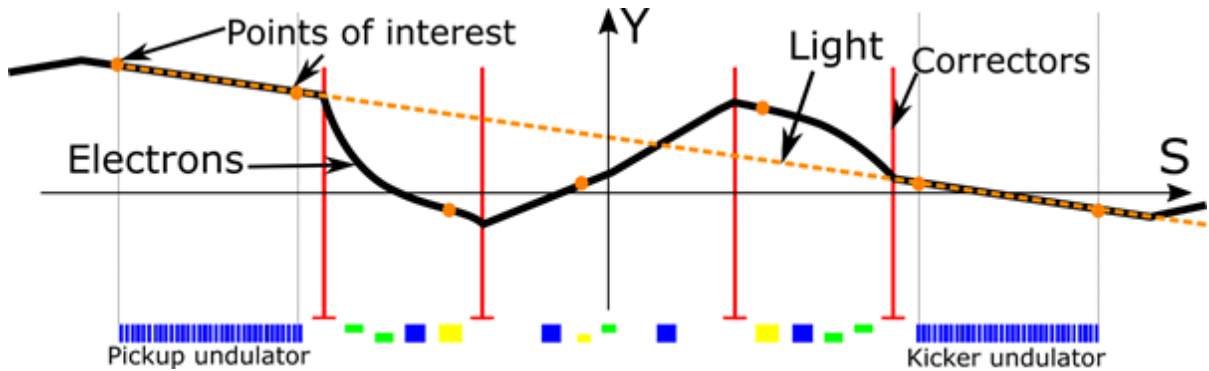

Figure 21: Schematic configuration of vertical orbit corrected for OSC experiment; blue – dipoles and undulators, green – quads, red – correctors, yellow – sextupoles, solid black line – electrons, dashed orange line – photons.

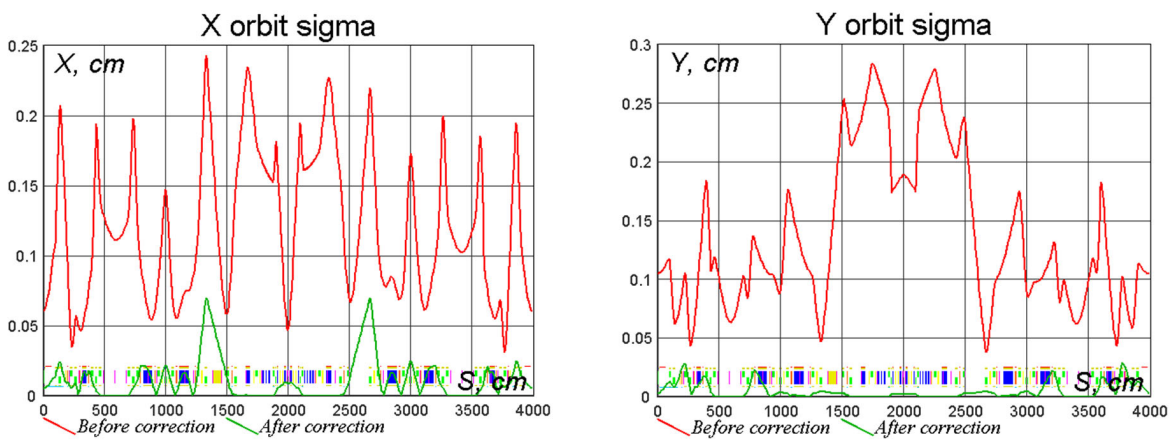

Figure 22: RMS orbit errors before and after corrections for horizontal (left) and vertical (right) planes.

Individually controlled chicane dipoles will easily handle horizontal orbit corrections. For vertical plane, inner and outer correctors must be capable of producing field integral of 2 kG·cm and 0.4 kG·cm, respectively.

Anticipated typical orbit errors and necessary corrector strengths were studied by statistical method. The following sequence was repeated multiple times. Random misalignments of elements were introduced with consequent attempt to correct resulted distortion of closed orbit. Introduced misalignments had normal distributions with σ's presented in Table 5. Figure 22 shows typical orbit errors before and after corrections for both transverse planes. Figure 23 shows corresponding typical field amplitudes in the correctors.

**Table 5: RMS misalignments of elements for statistical analysis of closed orbit correction**

| Quads, X and Y shifts | 0.1 mm |
|---|---|
| Main dipoles, X and Y shifts | 0.1mm |
| Main dipoles, X and Y tilts | 0.6 mm/m |

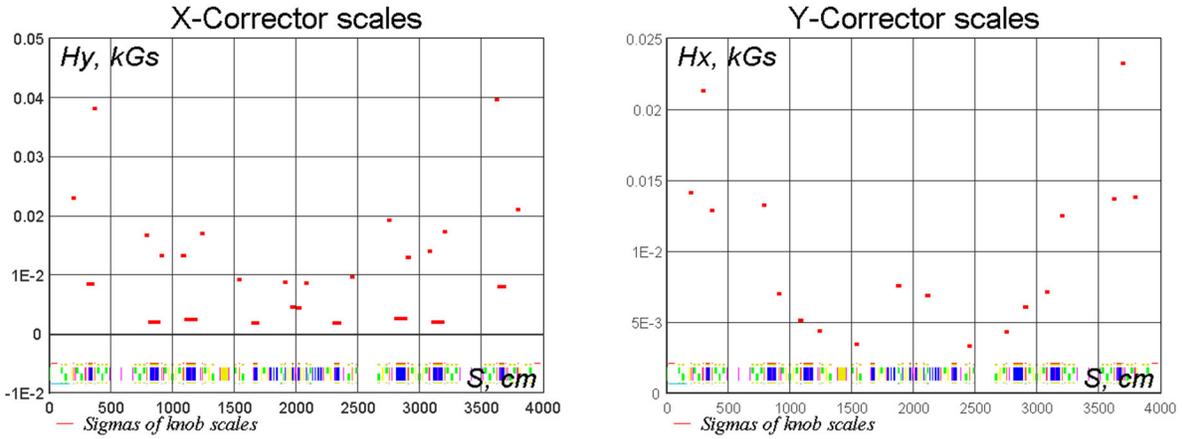

Figure 23: RMS corrector fields used for orbit corrections for horizontal (left) and vertical (right) planes.

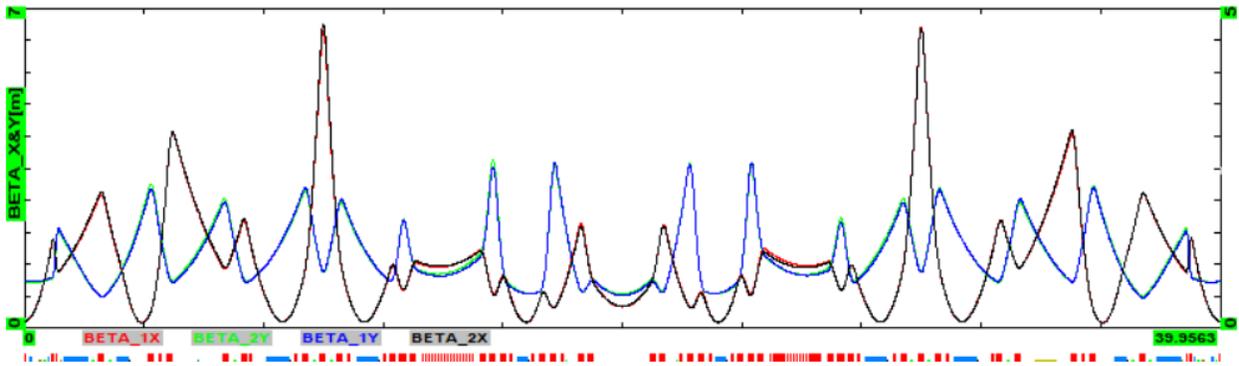

Figure 24: 4D beta-functions for coupled optics of passive cooling. Each pair of beta-functions $\beta_{1x}$ and $\beta_{2x}$ (red and black lines), and $\beta_{1y}$ and $\beta_{2y}$ (green and blue lines) are well overlapped at the entire length. Plot starts from the center of OSC straight. Tune split is 0.0027.

### 5.3. Operation on Coupling Resonance

To create optics with strong *x-y* coupling we initially decouple the lattice so that the tune split would be about or below $10^{-3}$. Then, we introduce *x-y* coupling with one or few skew-quads. To achieve stable and reproducible optics the skew quad(s) must create a tune split which is significantly larger than the minimum tune split achieved. In further consideration we consider an example where the coupling is introduced by a single skew-quadrupole SQA1L with integral field of 20 G. It makes the 4D beta-functions belonging to the same plane almost equal as one can see in Figure 24. To good accuracy each of this beta-function is equal to half of the corresponding plane normal uncoupled beta-functions. The horizontal dispersion at the location of SQA1L skew-quadrupole is equal to zero. It implies that the vertical dispersion is not excited and is equal to zero at the entire length of the ring. Equal values of horizontal 4D beta-functions for both betatron modes and unchanged dispersion result in that the horizontal cooling rate computed below in Section 6 is equally split between two betatron modes. Table 6 shows major parameters of the coupled optics for passive OSC.

**Table 6: Main parameters of passive OSC optics with strongly coupled optics**

| | |
|---|---|
| Betatron tunes, $\nu_1/\nu_2$ | 5.42/2.42 |
| Emittance cooling times in the absence of OSC, s ($\tau_1/\tau_2/\tau_s$) | 1.056 / 1.056 / 0.519 |
| Rms emittances of betatron modes set by SR, $\varepsilon_1 / \varepsilon_2$, nm | 0.426 / 0.423 |
| Rms relative momentum spread set by SR | $9.86 \cdot 10^{-5}$ |
| Tune chromaticities of betatron modes with OSC sextupoles on | 14 / -25 |
| Dynamic aperture with OSC straight sextupoles on, $\varepsilon_{1max}/\varepsilon_{1max}$, nm | 350/350 |
| Dynamic aperture in momentum with nominal RF | $1.2 \cdot 10^{-3}$ |

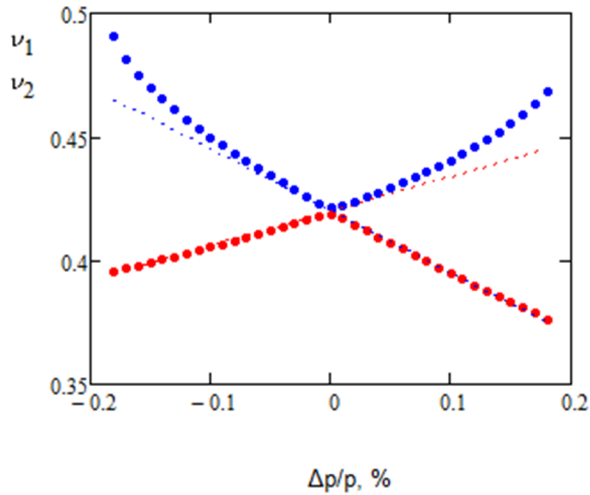

Figure 25: Dependence of betatron tunes on the momentum for coupled optics of passive OSC; dotted lines present a linear fit with chromaticities of 14 and -25.

Figure 25 shows dependence of betatron tunes on the momentum for the case when the OSC straight sextupoles are at nominal values and no other sextupoles engaged. As one can see at the momentum deviation of about -0.185% the tune approaches the half-integer resonance where the motion stability is lost. Such momentum acceptance is sufficient for OSC experiment, and therefore a compensation of chromaticity introduced by OSC sextupoles is not planned. Particle tracking performed with RF on yielded the dynamic aperture of about 350 nm with momentum acceptance of ±0.12%.

## 6. OSC Cooling Rates

To compute the OSC cooling rates we need to find a longitudinal kick which a particle receives in the kicker undulator from its own radiation radiated in the pickup undulator and then amplified and focused to the kicker undulator. We split this problem into the following steps: finding electric field of the radiation on the focusing lens surface, computing the electric field in the kicker undulator by integration of the field distribution on the lens, and, finally, finding the longitudinal kick in the kicker undulator. We assume that the distances from the pickup center to the lens and from the lens to the kicker center are equal and are much larger than the pickup and kicker lengths; so large that the depth of field would not result in a deterioration of the interaction. Applicability of such requirement will be discussed later. We also assume that the pickup and kicker undulators are flat, have the same length and the same number of periods.

The e.-m. radiation coming out from the pick-up undulator is determined by the Liénard-Wiechert formula [11]:

$$\mathbf{E}(\mathbf{r},t) = \frac{e}{c^2} \frac{(\mathbf{R}-\boldsymbol{\beta} R)(\mathbf{a}\cdot\mathbf{R}) - \mathbf{a}R(R-(\boldsymbol{\beta}\cdot\mathbf{R}))}{(R-(\boldsymbol{\beta}\cdot\mathbf{R}))^3}, \quad \mathbf{a} = \frac{d\mathbf{v}}{dt}, \quad \mathbf{R} = \mathbf{r}-\mathbf{r}' \ . \tag{23}$$

Here $e$ is the particle charge, $\boldsymbol{\beta}=\mathbf{v}/c$ is the dimensionless particle velocity, $\mathbf{R}$ is the vector from point of the radiation, $\mathbf{r}'$, to the point of observation, $\mathbf{r}$, and all values in the right-hand side are taken at the retarded time $t' = t - R/c$. Let the coordinates of a particle moving in a flat undulator depend on time as following:

$$v_x = -c\theta_e \sin\tau', \quad v_y = 0,$$
$$v_z = c\left(1 - \frac{1}{2\gamma^2} - \frac{\theta_e^2}{2}\sin^2\tau'\right), \quad \tau' = \omega_u t' + \psi, \tag{24}$$

where $\omega_u$ is the angular frequency of particle motion in the undulator, and $\theta_e$ is the amplitude of particle angle oscillations. Substituting velocities of Eq. (24) to Eq. (23) and simplifying the obtained equation one obtains the horizontal component of electric field in the far zone (at the lens):

$$E_x(r,t) = 4e\omega_u \gamma^4 \theta_e \cos\tau' \frac{1+\gamma^2\left(\theta^2(1-2\cos^2\phi) - 2\theta\theta_e \sin\tau'\cos\phi - \theta_e^2\sin^2\tau'\right)}{cR\left(1+\gamma^2\left(\theta^2 + 2\theta\theta_e\sin\tau'\cos\phi + \theta_e^2\sin^2\tau'\right)\right)^3}, \tag{25}$$

where $\theta$ and $\phi$ are the angles in the polar coordinate frame for the vector $\mathbf{R}$ ($x = R\sin\theta\cos\phi$, $y = R\sin\theta\sin\phi, z = R\cos\theta$), and we took into account that $a_x = c\theta_e\omega_u$. The vertical and longitudinal components of the electric field are averaged out at the focus (in kicker undulator) and therefore can be safely omitted from further consideration.

Only the first harmonic of the radiation interacts resonantly with the particle in the kicker undulator[8]. Therefore, we keep only the first harmonic of radiation in further calculations:

$$E_\omega(r,\theta) = \frac{\omega(\theta)}{\pi} \int_0^{2\pi/\omega(\theta)} E_x(r,t) e^{-i\omega(\theta)t} dt ,\qquad(26)$$

$$\omega(\theta) = 2\gamma^2 \omega_u / \left(1 + \gamma^2 \left(\theta^2 + \theta_e^2/2\right)\right) .\qquad(27)$$

To find the electric field in the kicker undulator, where the radiation is focused, we use a modified Kirchhoff formula[9]:

$$E(\mathbf{r}'') = \frac{1}{2\pi i c} \int_S \frac{\omega(\theta) E_\omega(\mathbf{r})}{|\mathbf{r}'' - \mathbf{r}|} e^{i\omega(\theta)|\mathbf{r}''-\mathbf{r}|/c} ds .\qquad(28)$$

Here $\mathbf{r}''$ is the coordinate in the observation point in the kicker undulator, the integration is performed over the lens surface $S$ where vector $\mathbf{r}$ is located, and the electric field at the lens is described by Eqs. (25) and (26). The focal length of the lens is equal to $R/2$. It results in that an increase of delay time related to path lengthening, $2(R\theta^2/2)$, is compensated in the lens. In computation of electric field at the axis it makes all waves arriving to the focal point having the same phase, and, consequently, the exponent in Eq. (28) accounting for these delays is reduced to a complex constant which is omitted in further calculations. Note that although the frequency of radiation coming out from the radiation point depends on $\theta$ this dependence disappears at the location of particle (in the moving image plane) due to longitudinal particle displacement in the kicker undulator.

## 6.1. Cooling Force for Small Undulator Parameter

The above equations can be significantly simplified in the case of small undulator parameter, $K=\gamma\theta_e \ll 1$. Then Eq. (25) can be simplified yielding the wave amplitude on the lens surface:

$$E_x = \frac{4e\gamma^4 \omega_u \theta_e}{cR} \frac{1 + (\gamma\theta)^2 (1 - 2\cos^2\phi)}{\left(1 + (\gamma\theta)^2\right)^3} ,\qquad(29)$$

Substituting Eq. (29) into Eq. (28) and neglecting dependence of frequency on $\theta_e$, which yields that $\omega(\theta) = 2\gamma^2 \omega_u / (1 + \gamma^2\theta^2)$, one obtains the field in the focal point (in the kicker undulator):

---

[8] We also note that in the IOTA case, as in majority of other possible practical cases, a dependence of group velocity in the lens material on the wavelength results in a longitudinal separation of wave packets belonging to different harmonics. That completely excludes an interaction with radiation of higher harmonics. Additionally, the radiation of higher harmonics can be absorbed in the lenses and/or is not amplified by optical amplifier (if present).

[9] This is an approximate formula which however yields the exact result for the case of small undulator parameter and infinite angular acceptance of the focusing system. Exact solution would require a Fourier transform of incoming wave packet both in time and all coordinates, followed by a usage of Kirchhoff formula for obtained flat waves and inverse Fourier transform. Such procedure is used in numerical simulations with SRW reported below. A comparison between analytical theory and numerical results was found to be well within practically required accuracy.

$$E_x = \frac{8e\gamma^6 \omega_u^2 \theta_e}{c^2} \int_0^{2\pi} \frac{d\phi}{2\pi} \int_0^{\theta_m} \frac{1+(\gamma\theta)^2(1-2\cos^2\phi)}{\left(1+(\gamma\theta)^2\right)^4} \theta d\theta = \frac{8e\gamma^6 \omega_u^2 \theta_e}{c^2} \int_0^{\theta_m} \frac{\theta d\theta}{\left(1+(\gamma\theta)^2\right)^4} \quad . \tag{30}$$

Here $\theta_m$ is the angle subtending the lens from the pickup undulator (or a kicker undulator) and we assume a round lens. An integration in Eq. (30) results in the amplitude of electric field in the kicker undulator:

$$E_x = \frac{4e\gamma^4 \omega_u^2 \theta_e}{3c^2} f_L(\gamma\theta_m), \quad f_L(x) = 1 - \frac{1}{(1+x^2)^3} \quad . \tag{31}$$

Averaging the energy transfer ($dE/dt = ev_x E_x$) over oscillations in the kicker undulator we finally obtain the amplitude of the energy change in the kicker undulator in the absence of optical amplification:

$$\Delta E = \frac{2e^4 B_0^2 \gamma^2}{3m^2 c^4} L_u f_L(\gamma\theta_m) \quad . \tag{32}$$

Here $B_0$ is the peak magnetic field in the undulator, $L_u$ is its total length, $m$ is the particle mass, and we also took into account that $\theta_e = eB_0 / (mc\gamma\omega_u)$.

Note that in the absence of optical amplification and $\gamma\theta_m \gg 1$ the amplitude of energy loss is equal to the total energy loss in both undulators: $\Delta E_{tot} = 2e^4 B_0^2 \gamma^2 L_u / (3m^2 c^4)$. In the above calculations we did not account this energy loss and the gain in amplifier. Including them one obtains:

$$\Delta E(s) = -\Delta E_{tot} \left(1 + Gf_L(\gamma\theta_m) \sin(k_0 s)\right) , \tag{33}$$

where $G$ is the optical amplifier gain (in amplitude), and the same as in Eq. (2) $s$ is the particle longitudinal displacement on the way from pickup to kicker relative to the reference particle. Comparing Eqs. Eq. (2) and (33), and accounting Eq. (27) one obtains for parameters $k_0$ and $\kappa$ introduced in Eq. (2):

$$\begin{aligned} k_0 &= \frac{2\gamma^2 \omega_u}{c}, \\ \kappa &= \frac{G\Delta E_{tot}}{cp} f_L(\gamma\theta_m) = \frac{2e^4 B_0^2 \gamma}{3m^3 c^6} L_u \, Gf_L(\gamma\theta_m). \end{aligned} \tag{34}$$

It is also assumed in Eq. (33) that, if the optical amplifier is present, its bandwidth is sufficiently large so that the spectrum widening due to finite number of undulator periods $\approx 2\gamma^2 \omega_u / n_w$ and the angular spread of radiation $2\gamma^4 \theta_m^2 \omega_u$ would be inside the amplifier bandwidth, where $n_w$ is the number of undulator periods. Otherwise one needs to average the cooling force with actual dependence of amplifier gain on the frequency.

Eq. (33) clearly shows that the OSC is effective even in the absence of optical amplifier. In this case an interference of radiation of two undulators results in the energy loss being modulated with the path length difference on the travel from pickup to kicker. Taking into account that the cooling rate is proportional to $d\Delta E/dp$ one obtains that the interference of SR coming from two undulators amplifies the cooling rate by the ratio of beam energy to it's the cooling range, $p/n_{\sigma p}\sigma_p$. The same statement is justified for the betatron motion. Note that an average energy loss presented in Eq. (33) is compensated by the RF system and does not affect the cooling dynamics.

Figure 26 presents dependence of relative cooling force on the angle subtending the lens from the undulator. One can see that for $\gamma\theta_m = 0.8$ the cooling force achieves about 80% of its theoretical maximum and that requires 40% bandwidth ($\omega_{max}/\omega_{min}=1.66$).

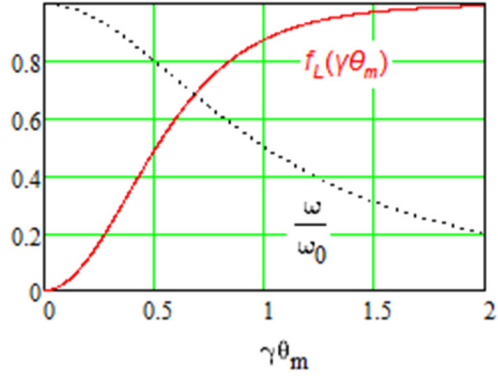

Figure 26: Dependence of relative cooling force on the angle subtending the lens from the undulator, $f_L(\gamma\theta_m)$, (red line), and a ratio of e.-m. radiation frequency at the aperture boundary to the frequency of forward radiation, $\omega_0=k_0c$, (black dashed line) for the OSC with small undulator parameter, $K \ll 1$.

To find the distribution of electric field in the focal point vicinity one needs to restore in Eq. (30) the phase advance correction which e.-m. wave obtains on the travel from the lens to the observation point located in the focal plane (in the kicker undulator):

$$E_x(\rho,\phi) = \frac{8e\gamma^6\omega_u^2\theta_e}{c^2}\int_0^{2\pi}\frac{d\phi'}{2\pi}\int_0^{\theta_m}\frac{1+(\gamma\theta)^2(1-2\cos^2\phi')}{(1+(\gamma\theta)^2)^4}\exp\left(\frac{i\rho k_0\theta\cos(\phi'-\phi)}{1+(\gamma\theta)^2}\right)\theta d\theta. \quad (35)$$

Here $\rho$ is the distance from the observation point to the system axis, and $\phi$ is the angle between the direction to the observation point and the horizontal plane. Integrating over $\phi'$ one obtains

$$E_x(\rho,\phi) = \frac{8e\gamma^6\omega_u^2\theta_e}{c^2}\int_0^{\theta_m}\left(J_0\left(\frac{\rho k_0\theta}{1+(\gamma\theta)^2}\right)+(\gamma\theta)^2 J_2\left(\frac{\rho k_0\theta}{1+(\gamma\theta)^2}\right)\cos(2\phi)\right)\frac{\theta d\theta}{(1+(\gamma\theta)^2)^4}. \quad (36)$$

Figure 27 presents a dependence of electric field on coordinates of the observation point in the focal plane obtained by numerical integration in Eq. (36). The point, where the electric field approaches zero the first time, determines the radiation half-size in the corresponding plane. For estimates the horizontal and vertical half-sizes can be approximated by the following equations:

$$x_0 \approx \lambda_0 \sqrt[3]{(1.51\gamma)^3 + (0.159 + 0.619/\theta_m)^3},$$
$$y_0 \approx \lambda_0 \sqrt[3]{(1.08\gamma)^3 + (0.619/\theta_m)^3}, \qquad \gamma\theta_m \geq 0.1. \tag{37}$$

Expressing the undulator magnetic field through the undulator parameter, $K = \gamma\theta_e = eB_0\lambda_w/(2\pi mc^2)$, one can rewrite Eq. (32) in the following form:

$$\Delta E = \frac{2\pi}{3} K^2 e^2 k_0 n_w f_L(\gamma\theta_m), \quad K \ll 1, \tag{38}$$

where $n_w$ is the number of undulator periods. As one can see for a chosen wavelength, $2\pi/k_0$, the energy kick amplitude is proportional to $K^2$ and a usage of maximum possible $K$ is desired. As will be seen below its increase is limited by a loss of wave-particle interaction with $K$ increase.

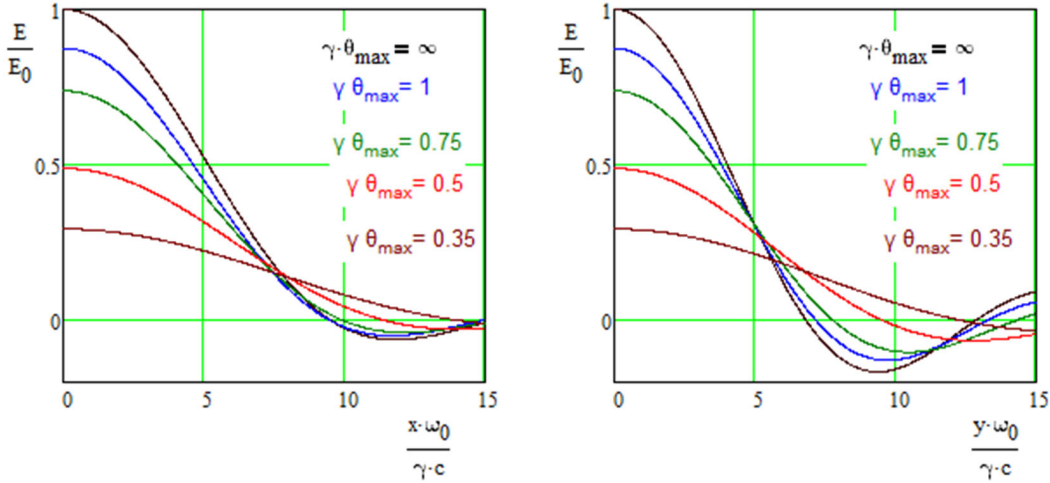

Figure 27: Dependence of electric field on the coordinates of the observation point in the focal plane: left - on the horizontal coordinate ($y=0$), right - on the vertical coordinate ($x=0$).

## 6.2. Cooling Force for Arbitrary Undulator Parameter

In the general case of arbitrary undulator parameter, the amplitude of electric field at the axis of kicker undulator can be expressed in the following form:

$$E_x = \frac{4e\omega_u^2 \gamma^4 \theta_e}{3c^2} F_h(\gamma\theta_e, \gamma\theta_m), \tag{39}$$

where

$$F_h(\Theta_e, \Theta_m) = \frac{3}{\pi^2} \int_0^{\Theta_m} d\Theta \int_0^{2\pi} d\phi \int_0^{2\pi} d\tau \frac{\Theta F_c(\Theta, \Theta_e, \tau, \phi)}{1+\Theta^2 + \frac{\Theta_e^2}{2}} \frac{1+\Theta^2(1-2\cos^2\phi) - 2\Theta\Theta_e \cos\phi \sin\tau - \Theta_e^2 \sin^2\tau}{\left(1+\Theta^2 + 2\Theta\Theta_e \cos\phi \sin\tau + \Theta_e^2 \sin^2\tau\right)^3},$$

$$F_c(\Theta, \Theta_e, \tau, \phi) = \left(1 + \frac{4\Theta\Theta_e \cos\phi \sin\tau - \Theta_e^2 \cos(2\tau)}{2\left(1+\Theta^2 + \frac{\Theta_e^2}{2}\right)}\right) \exp\left(-i\tau + i\frac{\Theta_e^2 \sin(2\tau) + 8\Theta\Theta_e \cos\phi \cos\tau}{4\left(1+\Theta^2 + \frac{\Theta_e^2}{2}\right)}\right) \cos\tau.$$

(40)

For the large acceptance lens, $\theta_m \geq \theta_e + 3/\gamma$ the function $F_h(\Theta_e, \theta_m)$ computed with numerical integration can be interpolated by the following equation:

$$F_h(K, \infty) \approx \frac{1}{1+1.13K^2 + 0.04K^3 + 0.37K^4}, \quad 0 \leq K \leq 4. \quad (41)$$

Integrating the force along the kicker length and neglecting electric field dependence on the horizontal coordinate but accounting a kick reduction due to longitudinal oscillation (associated with the particle transverse motion) one obtains the longitudinal kick amplitude in a flat undulator:

$$\Delta E = \frac{2\pi}{3} e^2 k_0 n_w K^2 \left(1 + \frac{K^2}{2}\right) F_h(K, \gamma\theta_m) F_u(\kappa_u), \quad \begin{aligned} F_u(\kappa_u) &= J_0(\kappa_u) - J_1(\kappa_u), \\ \kappa_u &= K^2 / \left(4(1+K^2/2)\right). \end{aligned} \quad (42)$$

Here $k_0 = 2\gamma^2 \omega_u / (c(1+K^2/2))$ is the wave-number for the base frequency of the radiation. For small $K$ this equation coincides with Eq.(38). Figure 28 presents a dependence of dimensionless kick, $F_t(K, \gamma\theta_m) = K^2(1+K^2/2) F_h(K, \gamma\theta_m) F_u(\kappa_u)$, on the undulator parameter for different values of $\gamma\theta_m$. The same as above we imply here that the optical amplifier gain is equal to one and the amplifier's bandwidth is larger than the bandwidth of the first harmonic radiation coming from the pickup undulator to the kicker undulator. Otherwise averaging over the optical amplifier bandwidth is additionally required.

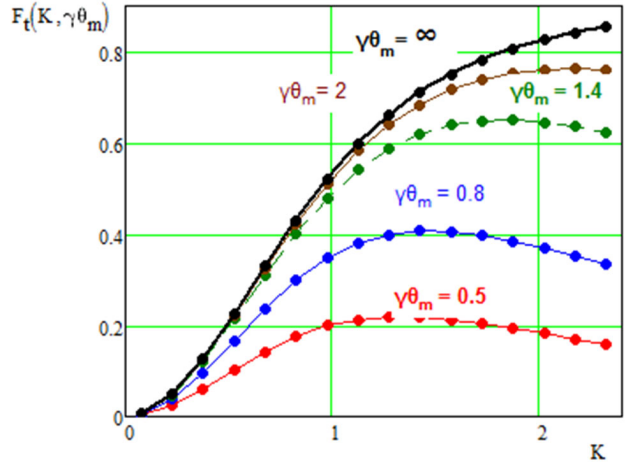

Figure 28: Dependence of dimensionless longitudinal kick on the undulator parameter. Dots are obtained by numerical integration of Eq. (40). The black line is built using Eq. (42) with accounting Eq. (41).

A reduction of cooling force with $K$ increase is related with increase of particle oscillation amplitude in the field of undulator resulting in a separation of electromagnetic wave and particle in the kicker undulator. This separation has two contributions. The first one comes from shifts of focal point from axis in the kicker undulator due to radiating particle transverse displacements in the pickup undulator. This is accounted in Eq. (42). The second contribution comes from transverse particle displacements in the kicker undulator and it is not accounted in this equation. In other words, we neglected a dependence of the electric field of the e.-m. wave on the transverse coordinates in the kicker undulator assuming that the field is equal to the field at the axis. Such choice was used to simplify analytic calculations and is justified by the following. The energy transfer in the kicker is most efficient at small deviations from the axis where the maximum horizontal velocity is achieved; and there is no energy transfer at maximum deflection where transverse velocity is equal to zero. The maximum value of the kick amplitude is achieved for moderate $K$ values ($K \leq 2$) where accounting for this effect does not produce significant correction. This effect, as well as many other effects reducing the kick amplitude, are accounted in the computer simulations reported in the following sections.

**Table 7: Cooling parameters for OSC**

|  | Passive cooling | Active cooling |
|---|---|---|
| Basic radiation wavelength, μm | 0.95 | 2.2 |
| Undulator period, $\lambda_w = 2\pi c / \omega_u$, mm | 48.40 | 110.63 |
| Number of undulator periods, $n_w$ | 16 | 7 |
| Undulator length, $L_u = \lambda_w n_w$, mm | 774.7 ||
| Peak magnetic field of the undulator, $B_0$, kG | 2.22 | 1.005 |
| Undulator parameter, $K=eB_0\lambda_w/(2\pi mc^2)$ | 1.003 | 1.038 |
| Angle subtending the lens from the pickup undulator, $\theta_m$ ($\gamma\theta_m = 0.8$), mrad | 4.07 ||
| Amplitude of energy transfer for the passive cooling, $\Delta E$, meV | 114 * | 21.9 * |
| Horizontal half-size in the focal plane, $x_0$, μm | 282 | 653 |
| Vertical half-size in the focal plane, $y_0$, μm | 202 | 467 |
| Amplitude of transverse particle motion in undulator, $x_w=\lambda_w K/(2\pi\gamma)$, μm | 40 | 93 |
| OSC horizontal emittance cooling rate in absence of $x$-$y$ coupling, $\lambda_x$, s$^{-1}$ | 38 * | 9.8 * |
| OSC longitudinal emittance cooling rate, $\lambda_s$, s$^{-1}$ | 36 * | 8.4 * |

* In computation of this value we ignored the dependence of e.-m. wave on the horizontal coordinate in the kicker undulator and an intensity loss in the lenses due to reflections and dispersion in lens material.

### 6.3. Colling Force for IOTA OSC

Table 7 presents major parameters of the passive and active OSC in IOTA. In both cases the number of periods is limited by available length for undulators and we assume that undulators have the same length. Cooling force for the active OSC is computed for the case when optical amplifier is not present. Accounting for amplifier is carried out below in Sections 8 and 10. The angle subtending the lens from the pickup undulator is chosen to be 4 mrad. Compared to the case without aperture limitations it results in a reduction of cooling force by ~1.5 times. Further increase of the subtending angle would not deliver significant increase of the kick amplitude but would further complicate OSC magnets (due to larger required aperture) and the dispersion correction in the optical telescope which focuses radiation. Note that a usage of such large bandwidth is unfeasible in the case of active system due to limited bandwidth of optical amplifier. A relatively moderate undulator parameter of 1.038 was chosen. As can be seen from Figure 28 it yields the kick amplitude within 15% of its theoretical maximum. Further increase of $K$ would result in an increase of beam heating due to SR radiation in the OSC undulators and, consequently, an increase of transverse emittance. That is undesirable due to limited cooling aperture.

## 7. Light Optics for Passive OSC

In this section we consider possible ways to focus radiation from the pickup undulator to the kicker undulator for both passive and active OSC. However, for active OSC, to be carried out at 2.2 μm wavelength, we consider here only passive focusing leaving discussion on the limitations coming from optical amplifier to Sections 8 and 10. In the rest of this section we will refer passive and active OSC as 0.95 and 2.2 μm cases.

### 7.1. Focusing with Lens Telescope

Above we assumed that the radiation emitted by a particle at some location in the pickup undulator is focused to the same location in the kicker undulator for the particle entire travel in the pickup undulator. That results in that the particle interacts with its optimally focused radiation on its entire travel in the kicker undulator. It is automatically achieved for the lens located at the infinity. In practical terms it means that the distance to the lens is much larger than the length of undulator. That condition is impossible to achieve in a real accelerator. A practical solution can be obtained with a lens telescope which has the transfer matrix, $M_T$, from the center of pickup to the center of kicker equal to $\pm I$, where $I$ is the identity matrix. In this case the transfer matrix between emitting and receiving points is $O(l) M_T O(-l) = \pm I$, i.e. coincides with the matrix for the system where the lens is located at infinity. Here $O(l)$ is the transfer matrix of a drift with length $l$. Taking into account that the pickup-to-kicker transfer matrix for the electron beam is close to -$I$ we choose the matrix for the radiation to be -$I$ as well. That reduces separation of particle and its focused radiation in the kicker undulator which appears due to betatron motion. The simplest telescope requires 3 lenses. Straightforward calculations yield the focal lengths of the lenses:

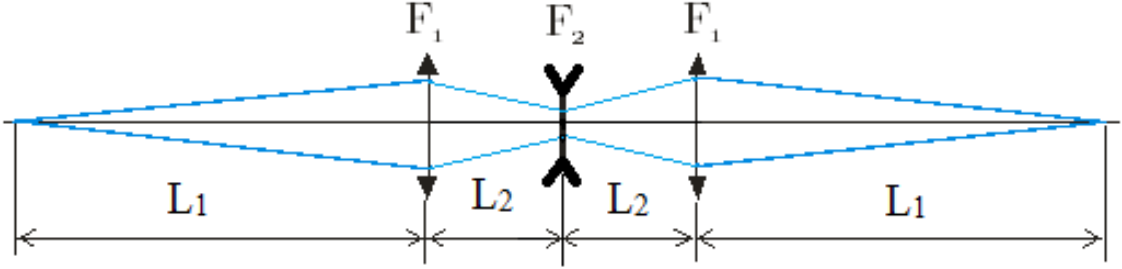

Figure 29: Light optics layout for passive cooling.

$$F_1 = L_2, \quad F_2 = -\frac{L_2^2}{2(L_1 - L_2)} \quad . \tag{43}$$

Here $F_1$ is the focal length for two outer lenses, and $F_2$ is the focal length for the central lens, $L_1$ is the distance from an undulator center to the outer lens, and $L_2$ from this outer lens to the central lens located in the center of OSC straight as shown in Figure 29. Figure 30 presents an example of ray propagation through such focusing system. As one can see the light focus is moving together with particle displacement in the kicker undulator. Table 8 presents tentative parameters of the three-lens telescope. Distances $L_1$ and $L_2$ are determined by requirement to avoid an interference of elements of beam and light optics. The focusing strengths of light optics lenses are chosen to be the same for 0.95 and 2.2 μm cases while lens materials and thicknesses are different.

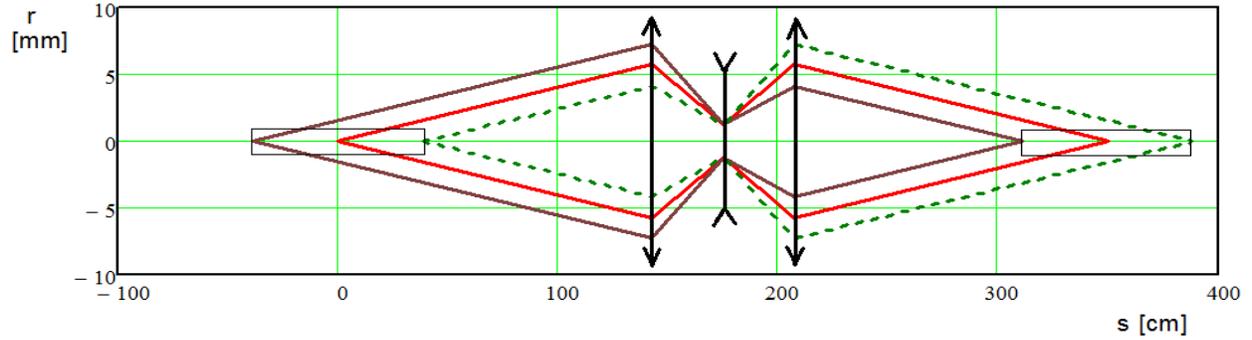

Figure 30: Trajectories of rays radiated at the beginning, in the middle and at the end of pickup undulator with initial angle of 0.407 mrad ($\gamma\theta_m=0.8$) representing acceptance angle of the optics.

**Table 8: Three-lens telescope parameters computed in the thin lens approximation**

|  | 0.95 μm | 2.2 μm |
|---|---|---|
| Distance from the OSC section center to an undulator center, $L_1+L_2$ | 172.765 cm | |
| Distance from an outer to the central lens, $L_2$ | 32 cm | |
| Focal length of outer lenses, $F_1$ | 32 cm | |
| Focal length of central lens, $F_2$ | -4.707 | |
| Angular acceptance of pickup undulator radiation, $\theta_m/\gamma\theta_m$ | 4.07 mrad / 0.8 | |
| Lens material | Quartz | BaF$_2$ |
| Total light delay in telescope material, $\Delta s$ | 0.648 mm | 2 mm |
| Total thickness, $L_g$ | 1.397 mm | 4.244 mm |

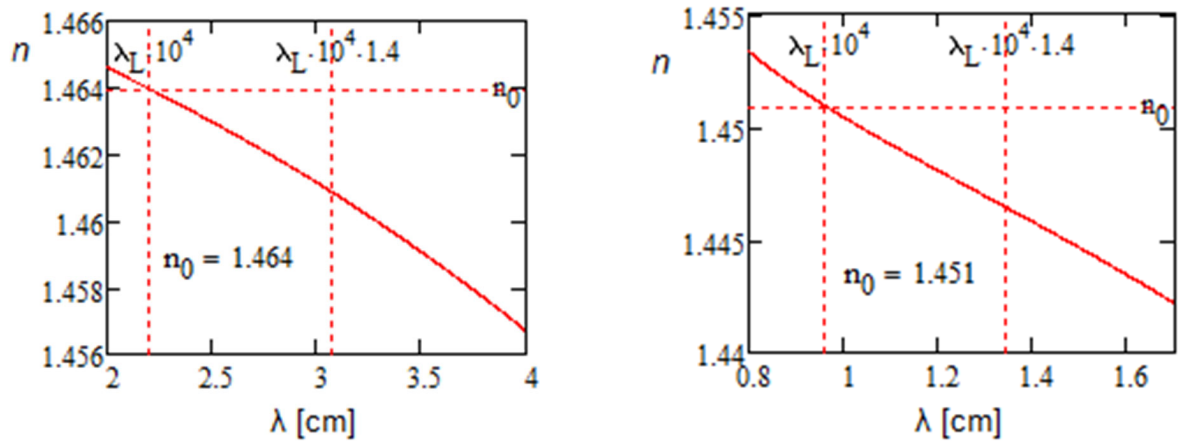

Figure 31: Dependence of the refraction indices for fused quartz (left) and Barium fluoride (right) on the wavelength for 0.95 and 2.2 μm cases.

Both the first and second order dispersions in the lens material result in a reduction of the energy transfer in the kicker undulator. The first order dispersion worsens light focusing and the second one results in lengthening of the wave packet and, consequently, a reduction of wave amplitude.

Choice of lens material is driven by minimization of effects of the second order dispersion. It is fused quartz for 0.95 μm and Barium fluoride (BaF$_2$) for 2.2 μm. In the corresponding wavelength ranges used for 0.95 and 2.2 μm cases they have sufficiently small GVD (group velocity dispersion) of 21 and -9.74 fs$^2$/mm, respectively. That is significantly smaller than the GVD for other optical glasses. The refraction indices of fused quartz and BaF$_2$ can be approximated by the following formulas [12]:

$$n(\lambda) = \begin{cases} \sqrt{1 + \dfrac{0.6961663\lambda^2}{\lambda^2 - 0.0684043^2} + \dfrac{0.4079426\lambda^2}{\lambda^2 - 0.1162414^2} + \dfrac{0.8974794\lambda^2}{\lambda^2 - 9.896161^2}} & - \text{Fused quartz}, \\ \sqrt{1 + \dfrac{0.643356\lambda^2}{\lambda^2 - 0.057789^2} + \dfrac{0.506762\lambda^2}{\lambda^2 - 0.10968^2} + \dfrac{3.8261\lambda^2}{\lambda^2 - 46.3864^2}}, & - \text{Barium fluoride}, \end{cases} \quad (44)$$

where $\lambda$ is the wavelength in μm. Corresponding plots are presented in Figure 31.

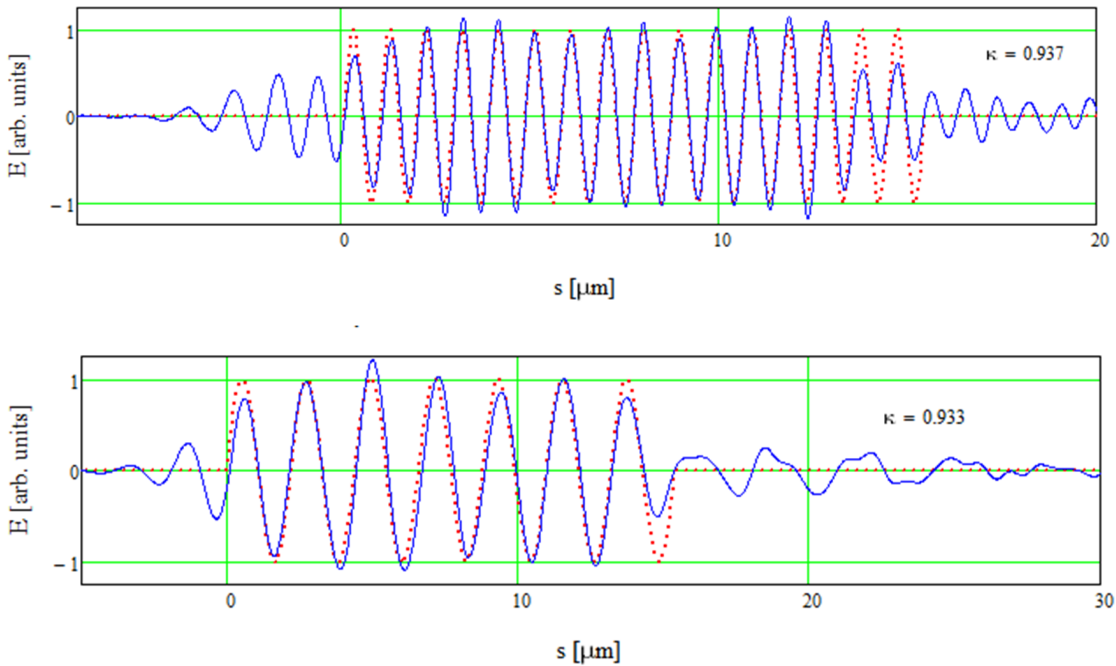

Figure 32: A distortion of the waveforms after passing lens material for 0.95 μm case (top) and 2.2 μm case (bottom); corresponding delays are 0.648 and 2 mm, thicknesses: fused quartz - 1.397 mm and BaF$_2$ - 4.244 mm; initial waveforms - red dashed lines, final waveforms – solid blue lines.

To see an effect of the second order dispersion we consider propagation of forward radiation through lens material. Figure 32 demonstrates a distortion of ideal 16-period and 7-period waveforms after passing fused quartz and BaF$_2$ for the 0.95 and 2.2 μm cases, respectively. The corresponding delays, $\Delta s$, are 0.648 mm and 2 mm. As one can see the distortions are relatively moderate. In both cases they result in a kick efficiency reduction of about 7%. Note that accounting for the difference between the group and phase velocities is important in determining the required glass thickness. In this case the total thickness of the glass is:

$$L_g = \frac{\Delta s}{n_0 - 1 - \lambda_0 \frac{dn}{d\lambda}}, \qquad (45)$$

where $n_0 = n(\lambda_0)$. Compared to the case without dispersion that results in a -2.8% and -1.5% (-39 µm, -64 µm) corrections to the glass thicknesses. Accounting the refraction index non-linearity yields additional -1.31 µm and -1.53 µm corrections, resulting in the total glass thickness of 1.397 mm and 4.244 mm for the 0.95 and 2.2 µm cases, respectively.

The dependence of group velocity on the wavelength also results in a time separation of the first and second harmonics of undulator radiation after their passage through the telescope (see Figure 33). Obviously, the radiations of the first and second harmonics radiated in an undulator coincide in time. Without the time separation of the harmonics that would result in a strong interference of the second harmonic radiations outgoing from the pickup and kicker undulators. Although the time separation of the second harmonics is significant it still leaves about 15 – 20% effective overlap. While only odd harmonics are present in the forward radiation, there is considerable power radiated at non-zero angles for the second harmonic. That may be used in optimization of OSC tuning. Forward radiations of the third harmonic are completely separated in time and therefore their interference hardly can be used as a diagnostic tool.

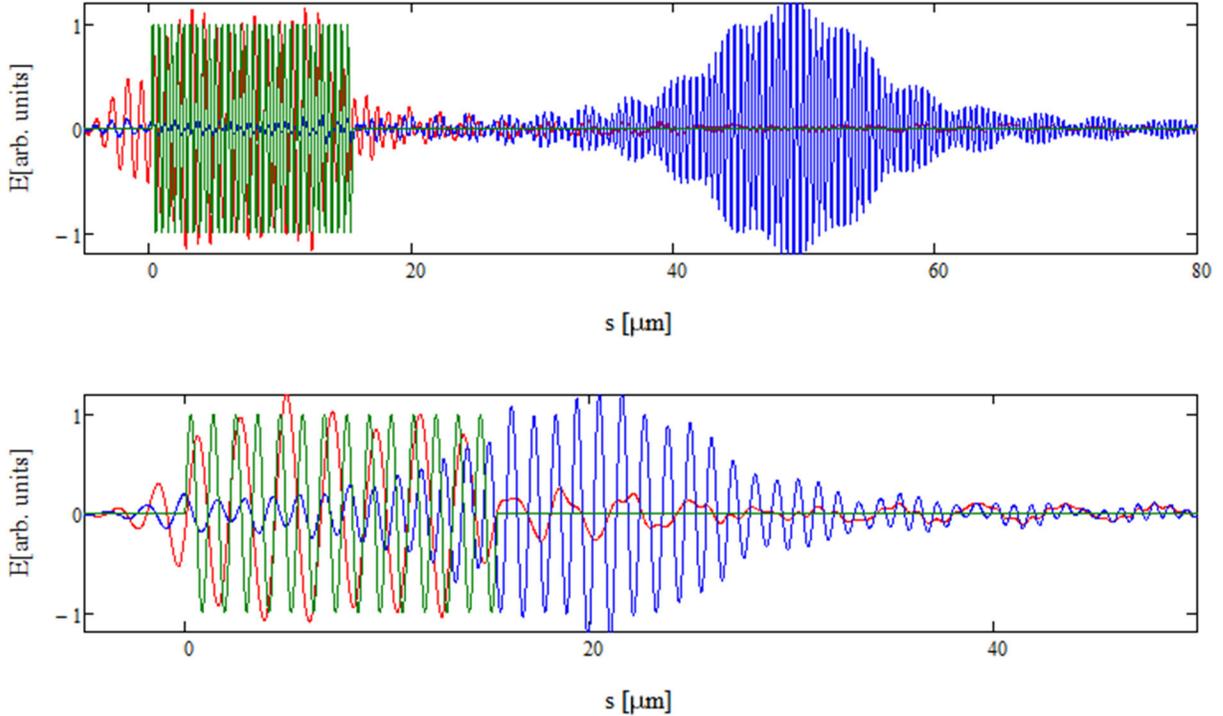

Figure 33: Separation of the first (red line) and second (blue line) harmonics of pickup undulator radiation after passing through the lens telescope for 0.95 µm case (top) and 2.2 µm case (bottom). The green line shows the second harmonic radiation of kicker undulator (or unperturbed radiation of pickup undulator). The amplitudes of the second and first harmonics were chosen to be equal for demonstration purpose.

## 7.2. Requirements to Lens Telescope Focusing

Next, we consider requirements to the required accuracy of lens focusing initially neglecting chromatic aberrations. In our consideration, we will assume axial symmetry and small undulator parameter but, as will be seen below, it does not limit the generality of conclusions on the lens focusing requirements.

Errors in lens focusing introduce an additional phase term, $\Phi(\theta)$, in Eqs. (30) and (40). In the case of small undulator parameter when Eq. (30) can be used it may be rewritten as following:

$$E_x = \frac{8e\gamma^6 \omega_u^2 \theta_e}{c^2} \left| \int_0^{\theta_m} \exp(i\Phi(\theta)) \frac{\theta d\theta}{\left(1+(\gamma\theta)^2\right)^4} \right| . \qquad (46)$$

Eq. (40) should be modified in the same way. As one can see, $\Phi(\theta) \ll 1$ is required to avoid a reduction of the cooling force.

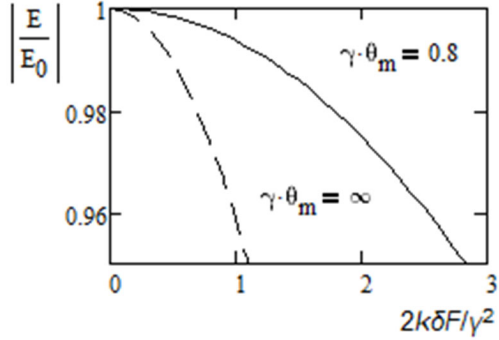

Figure 34: Dependence of cooling force reduction on the focusing error for single lens; solid line - the lens size corresponding to the IOTA OSC ($\gamma\theta_{max}$=0.8), dashed line – infinite transverse lens size.

To obtain a simple estimate for required accuracy of the focal length we initially assume a single lens telescope and small undulator parameter, $K \ll 1$. We also neglect the depth of field which implies that the lens is located in the center between the radiation point and the point where the radiation is focused. Assuming that the focal length, $F$, has an error $\delta F$ one obtains the phase term:

$$\Phi(\theta) = -2k(\theta)\theta^2 \delta F = -\frac{2k_0 \theta^2 \delta F}{1+(\gamma\theta)^2} , \qquad (47)$$

where $k_0$ is the wave number for forward radiation, $k(\theta)$ is the wave number for the radiation under angle $\theta$, and we accounted that the pickup and kicker are located at distance $2F$ from the lens. Figure 34 presents a reduction of cooling force due to focusing error obtained by numeric integration of Eq. (46) with the phase term of Eq. (47). A requirement of cooling force reduction below 2% results in that the focusing error should be below 1.26 cm. Note that in this estimate we used an approximation of $K \ll 1$ which, strictly speaking, is not justified for the IOTA OSC. However, it does not produce a significant error because the phase term makes a major contribution to the reduction of the cooling force. The same value for the focusing error can be also obtained if one will require the light spot size due to geometric misfocusing being about one quarter of diffraction beam size determined by Eq. (37).

To find requirements to focusing errors for the lenses of the lens telescope we require a focus displacement due to lens focusing errors, $\delta s$, being equal to the single lens case considered above where $\delta s = 4\delta F$. The light transport matrix between radiating and receiving points of the pickup and kicker undulators is:

$$\begin{bmatrix} 1 & L_1+s \\ 0 & 1 \end{bmatrix} \begin{bmatrix} 1 & 0 \\ -(1+f_3)/F_1 & 1 \end{bmatrix} \begin{bmatrix} 1 & L_2 \\ 0 & 1 \end{bmatrix} \begin{bmatrix} 1 & 0 \\ -(1+f_2)/F_2 & 1 \end{bmatrix} \begin{bmatrix} 1 & L_2 \\ 0 & 1 \end{bmatrix} \begin{bmatrix} 1 & 0 \\ -(1+f_1)/F_1 & 1 \end{bmatrix} \begin{bmatrix} 1 & L_1-s \\ 0 & 1 \end{bmatrix}, \quad (48)$$

where $F_1$ and $F_2$ are the unperturbed focusing strengths of the lenses of Eq. (43), $f_1$, $f_2$ and $f_3$ are the relative changes of lens focusing, and $s$ is the location of point of radiation in the pickup undulator relative to its center. Performing matrix multiplication and leaving only the first order terms in $f_k$ one obtains:

$$\begin{bmatrix} -1 + \frac{f_3-f_1}{L_2}L_1 + \frac{f_1+f_3}{L_2}s & M_{12} \\ \frac{f_3+f_1}{L_2^2} & -1 - \frac{f_3-f_1}{L_2}L_1 - \frac{f_1+f_3}{L_2}s \end{bmatrix}, \quad (49)$$

$$M_{12} = -2(L_1-L_2)f_2 + \frac{L_1^2}{L_2}(f_1+f_3) + 2\frac{L_1}{L_2}(f_3-f_1)s + \frac{f_3-f_1}{L_2}s^2.$$

The focal point displacement is: $\delta s = -M_{12}/M_{22} \approx -M_{12}$. Requiring that it would not exceed $\delta s = 4\delta F \approx 5.04$ cm one obtains requirements for relative focusing errors: $|f_1|=|f_3|\leq 0.8\%$, $|f_2|\leq 2.3\%$ or $|\delta F_1|=|\delta F_3|\leq 2.5$ mm, $|\delta F_2| \leq 1$ mm. Performing similar computations for errors in the distances between the central lens and the outer lenses, $L_2$, one obtains $|\delta L_2|\leq 2.3$ mm for each of two distances assuming that only one of them is changing at time.

### 7.3. Focusing Chromaticity and its Compensation

Above we considered waveform lengthening due to the second order chromaticity for the forward radiation (see Figure 32). Note that for the forward radiation, the first order chromaticity makes the phase and group velocities different but does not change the wave amplitude and does not reduce the cooling force[10]. However, the first order chromaticity affects the focal strength of the lenses, and consequently, it results in misfocusing and a reduction of cooling force. To minimize effects of the focusing chromaticity we need to choose the optimal wavelength, where the telescope transfer matrix is equal to -**I**, somewhere in the center of the cooling band. Figure 26 presents a dependence of cooling force on the angular acceptance of the telescope for a small undulator parameter, $K\ll 1$. Calculations with use of Eq. (39) show that for $K = 1$ this dependence is quite close. Figure 26 demonstrates that for chosen angular acceptance of $\gamma\theta_m = 0.8$ the major contribution to the cooling force comes from the angles $\approx 0.5/\gamma$ where the derivative is maximal. Taking this into account we choose the optimal angle to be $\theta_{opt} = 0.53/\gamma$, which corresponds to the optimal wavelength $\lambda_{opt}=$ equal to 1.13 μm and 2.6 μm for 0.95 μm and 2.2 μm cases, respectively.

---

[10] In the case of linear dispersion, $d\omega/dk$ = const, a wave packet described by the following equation $E(t) = E_0 e^{i\omega_0 t} f(t)$ is transformed to the following $E'(t) = E_0 e^{i\omega_0(t-d/v_p)} f(t-d/v_g)$ after passing through a plate. Here $d$ is the thickness of the plate, and $v_p = \omega_0/k_0$ and $v_g = d\omega/dk$ are the phase and group velocities, respectively.

For the case of 0.95 μm the total glass thickness is 1.397 mm. For the 3-lens telescope it is quite small to be split between 3 lenses and makes the lens production problematic. Here one also need to account a necessity of pathlength correction plate, which requires not negligible delay. In this case the thicknesses of lenses and the correction plate would be too small to support required mechanical stability. Therefore, we do not consider a three-lens telescope for the 0.95 μm case and further consideration of the 3-lens telescope will be carried out for the 2.2 μm case only. Table 9 presents tentative parameters of such telescope.

Table 9: Three-lens telescope main parameters for 2.2 μm case (no optical amplifier)

| | |
|---|---|
| $L_1$, mm | 1408.30♦ |
| $L_2$, mm | 319.35♦ |
| Focal length for 2.6 μm radiation, $F_1$, mm | 318.99 |
| Focal length for 2.6 μm radiation, $F_2$, mm | -45.814 |
| Thickness of the outer lenses (at axis) $t_1$, mm | 1.113 |
| Thickness of the central lens (at axis) $t_2$, mm | 1.019 |
| Thickness of the length adjustment plate, $t_r$, mm | 1.000 |
| Radius of lens surface for outer lenses*, mm | 294.984 |
| Radius of lens surface for central lens*, mm | -43.395 |
| Radius of outer lenses (half-aperture), mm | 8 |
| Radiation radius on the outer lenses at the angular acceptance, mm | 7.39 |

\* Radii of both surfaces are equal

♦ Distances are accounted as center-to-center for lenses, and as lens center to undulator center.

A dependence of the telescope transfer matrix on the wavelength results in a corresponding dependence of the focal point position in the kicker undulator, $\Delta s(\lambda)$. Taking into account that the focal length of a lens is scaled with wavelength as $F(\lambda) = F_0 \left( n(\lambda_{opt}) - 1 \right) / \left( n(\lambda) - 1 \right)$, substituting this scaling to Eq. (48) (where $f_k = 0$ and $s = 0$) and performing matrix multiplication one obtains the transfer matrix from the center of pickup undulation to the center of kicker undulator, $\mathbf{M}_{tot}(\lambda)$. The corresponding focal point displacement is: $\Delta s(\lambda) = M_{12}(\lambda) / M_{22}(\lambda)$. The left pane in Figure 35 shows results of computations for $\Delta s(\lambda)$. As one can see in the band of the IOTA OSC there is a considerable displacement ranging from -3 to 4 cm. This displacement results in an increase of the radiation size in the focal point and, consequently, a reduction of kick amplitude. To estimate an increase of the total spot size at the focal point the following expression was used:

$$\rho(\lambda) \approx \sqrt{\frac{x_0^2 + y_0^2}{2} + \left( \theta_m \Delta s(\lambda) \right)^2} \quad . \tag{50}$$

Here $\theta_m$ is the angular acceptance of the telescope, $x_0$ and $y_0$ are the horizontal and vertical beam sizes determined by Eq. (37), and we assume that the spot size increase due to misfocusing is equal to $\theta_m \Delta s(\lambda)$. As one can see in the right pane of Figure 35 an increase of the total spot size does not exceed about 4%. Consequently, a drop in the kick amplitude due to chromaticity of the focusing should not exceed few percent.

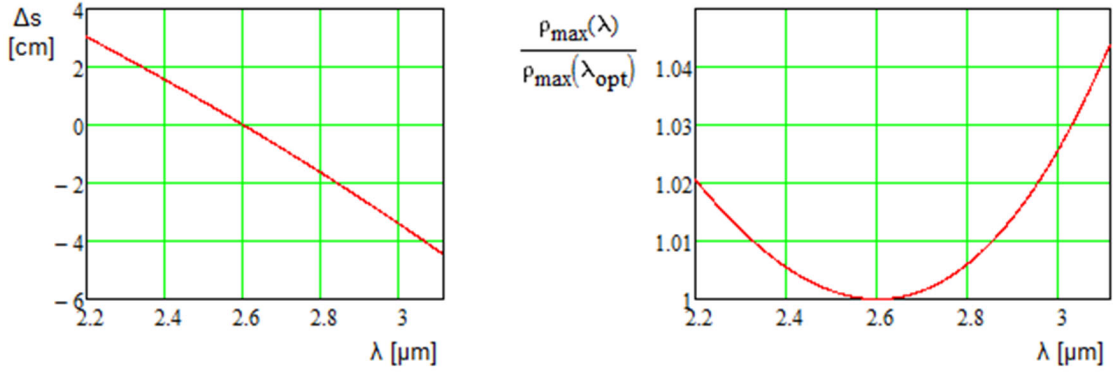

Figure 35: Dependences on the wavelength for the focal point displacement (left) and corresponding increase of the total spot size (right) for the three-lens telescope of 2.2 μm case.

We would like to note that the kick amplitude loss due to focusing chromaticity could be reduced with a high order correction to the lens shape. Taking into account that the radiation wavelength depends on the angle as[11]:

$$\lambda(\theta) = \lambda_0 \frac{1 + \gamma^2 \theta^2 + K^2/2}{1 + K^2/2}, \quad (51)$$

and assuming that the radiation comes from the center of the pickup undulator one can obtain a dependence of the radiation wavelength on the radial location for each lens. A ray outgoing from the center of pickup undulator with angle $\theta$ hits the central and outer lenses at the radii of $L_1\theta$ and $L_2\theta$, respectively. Substituting $\theta$ in Eq. (51) with $r_1/L_1$ and $r_2/L_2$ one obtains dependencies of the wavelength on the radii. To obtain an equation for the optimal lens shape we account that the bending angle of a lens is:

$$\Delta\theta_k \equiv \frac{r_k}{F_k} = (n(\lambda) - 1)\frac{dt_k(r_k)}{dr_k}; \quad (52)$$

where $t_k(r_k)$ describes the thickness of $k$-th lens at the ray radial position in the lens. Integrating the above equation and expending the result into Tailor series one obtains the dependence of lens thickness on the radius:

$$t_k(r_k) = t_k - \frac{r_k^2}{2(n_0-1)F_k} + \frac{1}{(n_0-1)^2 F_k}\int_0^{r_k} (n(\lambda(r_k')) - n_0) r_k' dr_k'. \quad (53)$$

---

[11] Here we neglect the spectrum widening due to finite duration of the waveform; and that the beam longitudinal displacement in the pickup undulator changes the e.-m. field distribution at surfaces of the lenses.

Figure 36 presents differences of the lens thicknesses for the lens described by Eq. (53) and for the spherical lens with optimal surface radius:

$$R_k = 2F_k \left( n(\lambda_{opt}) - 1 \right), \tag{54}$$

where we assume that the both surfaces of the lens have the same radius. As one can see the lens described by Eq. (53) is thinner at large radius for the outer lenses, and thicker for the central lens. The difference of the thicknesses does not exceed 0.5 μm in the working aperture of the lenses.

The lens surface of Eq. (53) can be approximated by the polynomial $t_k(r_k) = t_k + \sum_{n=1}^{4} S_{2 \cdot n}^{k} r_k^{2n}$. The polynomial coefficients for the inner and outer lenses are shown in Table 10.

**Table 10: Values of polynomial coefficient describing changes in the lenses thickness**

|  | Outer lenses | Central lens |
|---|---|---|
| Curvature radius at the axis (same for both surfaces), $R = 1/a_2$, cm | 29.694 | -4.368 |
| $a_2$, cm$^{-1}$ | -0.03368 | 0.2289 |
| $a_4$, cm$^{-3}$ | -3.209·10$^{-4}$ | 0.04146 |
| $a_6$, cm$^{-5}$ | -7.469·10$^{-5}$ | 0.2226 |
| $a_8$, cm$^{-7}$ | -1.897·10$^{-5}$ | 0.5134 |

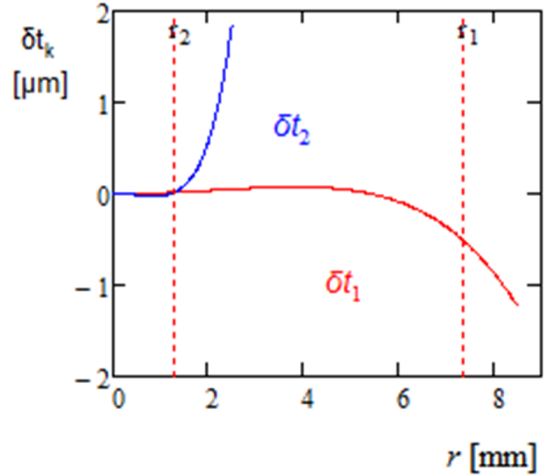

Figure 36: Differences of the lens thicknesses for the central ($\delta t_2$) and outer ($\delta t_1$) lenses for the case of lens surface described by Eq. (53) and the spherical lens with optimal radii of the surfaces; 2.2 μm case.

We need to note that the above consideration of chromatic effects misses few important details of the process. In particular, we did not consider the actual distribution of radiation fields at the lens surfaces and the second order effects for radiation outgoing at non-zero angles. The latter results in additional lengthening of the wave packet. Accurate calculations were carried out numerically with SRW (Synchrotron Radiation Workshop) [13]. They are discussed in Section 8 below.

## 7.4. Requirements to the Lens Telescope and its Transverse Alignment for 2.2 µm case

Above, if not specified, we assumed that the surfaces of the lenses are parabolic while a spherical shape would be preferred from a manufacturing point of view. However, for the chosen parameters of the lenses the difference in thicknesses for the parabolic and spherical lenses does not exceed 0.02 µm in the working aperture of each lens and 0.1 µm in their full aperture. That makes the parabolic and spherical lenses identical from any practical point of view.

The longtime accuracies of the telescope positioning and positioning of lenses inside telescope are determined by the longtime reproducibility of relative aiming of electron beam and its radiation in the kicker undulator. We require that the beams coincide with accuracy better that 10% of the total size of radiation in the kicker undulator. Splitting this error between displacements of the electron beam and its radiation in equal parts one obtains the stability of the radiation positioning of 5% of the total spot which corresponds or 25 µm. Taking into account that a displacement of telescope results in a twice larger displacement of the radiation in the kicker undulator (see right pane in Figure 37 one obtains the telescope longtime positioning of ~12 µm. As can be seen in the left pane the telescope rotation in a transverse plane produces a displacement of radiation spot linearly growing in the pickup undulator. The maximum amplification of ≈2.5 is achieved at the ends. That determines the maximum each-end displacement of ±10 µm corresponding to the telescope rotation of 31 µrad.

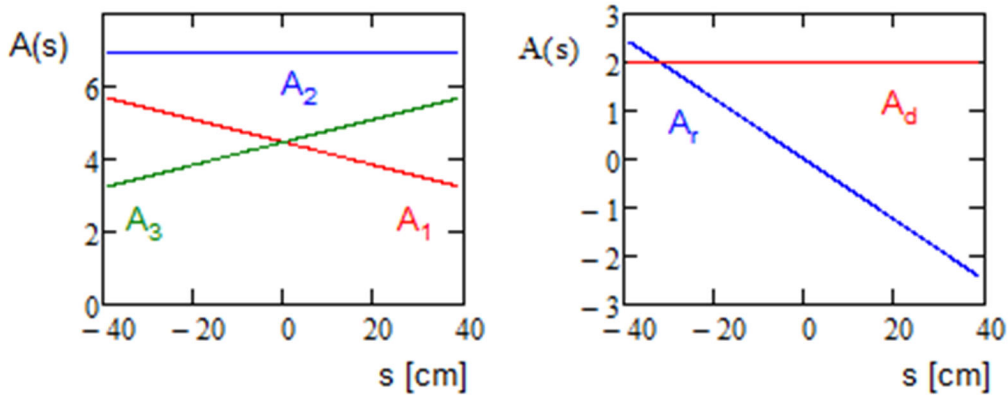

Figure 37: Dependence of the focal point displacement amplification along the kicker undulator: left – amplifications of lens displacements for each of three lenses; right - amplification of the entire telescope displacement in coordinate (red) and the telescope rotation relative its center expressed in the displacements of telescope ends, $\delta x_1 = -\delta x_3$.

Requirements for the longtime stability of lens positioning inside the telescope are tighter. They are determined by larger values of amplification for the case of separate lenses (see left pain in Figure 37). The maximum values of amplifications are about 7. That results in the required lens stability inside the telescope of about 4 µm.

To determine a relative alignment of lenses inside the telescope we assume that these errors have to be compensated by the telescope displacement within ±0.5 mm. It makes 1 mm displacement in the focal point. Accounting that the maximum displacement amplification is 7 one obtains the required accuracy of transverse lens positioning inside the telescope of about 150 μm. Table 11 summarizes the requirements for alignment and longtime stability of the telescope and its elements. It also includes requirements to accuracy of focal strengths discussed above. Note also that errors in the focusing strengths of the lenses can be partially compensated by adjustments of lens-to-lens distances inside the telescope.

Table 11: Requirements to accuracies of telescope parameters for 2.2 μm

|  | Outer lenses | Central lens | Telescope |
|---|---|---|---|
| Relative accuracy of focal length, $\Delta F/F$ | <0.8% | <2.3% |  |
| Accuracy of distance between central and outer lenses, mm | - | - | <2.3 |
| Long-time stability of telescope displacement, μm | - | - | <12 |
| Long-time stability of telescope rotation, μrad | - | - | <31 |
| Long-time stability of lenses position inside telescope, μm | - | - | <4 |

### 7.5. Single Lens Telescope

In the previous subsection we considered a three-lens telescope for focusing the pickup radiation to the kicker undulator. The telescope allows us to suppress the depth-of-field effects in the focusing and, consequently, results a uniform distribution of the kick along the kicker undulator, and thus maximizes it.

However, in the IOTA OSC both undulators are significantly shorter than the distance between them. That greatly reduces the depth of field effects making possible a usage a single lens telescope without significant loss in the cooling rate. To make an estimate, first, we add the squared spot sizes set by the diffraction and the errors of geometric focusing. That results in a dependence of the radiation effective radius on the position in the kicker undulator:

$$r_{\mathit{eff}}(n,s) \approx \sqrt{\frac{4\theta_m^2 L_{tot}^2}{(n_0-1)^2}\left((n-1)\frac{s^2}{L_{tot}^2}-(n-n_0)\right)^2 + x_0(s)y_0(s)} \ . \tag{55}$$

Here we assume that the lens focal length is chosen to focus the radiation from the center of the first undulator to the center of the second undulator, $s$ is the distance from the undulator center, $L_{tot}=L_1+L_2 = 172.765$ cm is the distance from an undulator center to the lens located in the center of the OSC chicane, $n_0$ and $n$ are the refraction indices of the lens material in the band center and at an arbitrary wavelength, and $x_0(s)$ and $y_0(s)$ are the horizontal and vertical radiation spot sizes set by diffraction and defined as radii where the electric field approaches zero the first time. The first term in Eq. (55) has two contributions: the first one due to depth of field misfocusing, and the second one due to chromatic aberration. Only the leading order terms are left. The dependence of $x_0(s)$ and $y_0(s)$ on $s$ comes from a dependence of diffraction on the lens subtending angle from the observation point located in the kicker undulator. Figure 38 presents a dependence of the radiation effective size along the kicker undulator for three different wavelengths located at the band center and at the band boundaries for the 2.2 μm case. One can see that the focusing chromaticity introduces significantly smaller correction than the depth of field.

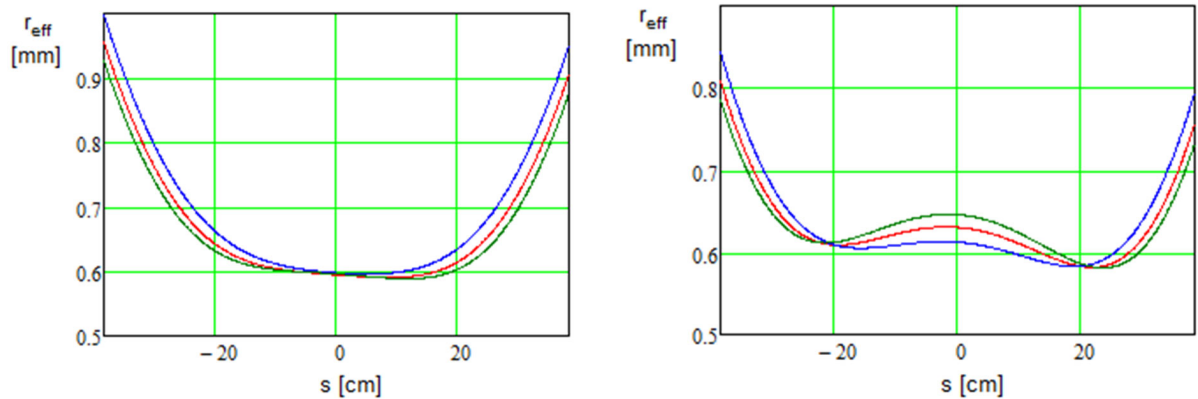

Figure 38: Dependence of the radiation effective spot size along the kicker undulator for the radiation in the band center (red, $\lambda$=2.6 μm), and the upper (green, $\lambda$=2.2 μm) and bottom (blue, $\lambda$=3.11 μm) boundaries of the band for the 2.2 μm case. The lens is located in the OSC chicane center. Right pane – the lens focusing strength is chosen to focus from center-to-center of undulators. Left pane – the lens focusing strength is chosen to maximize the kick value.

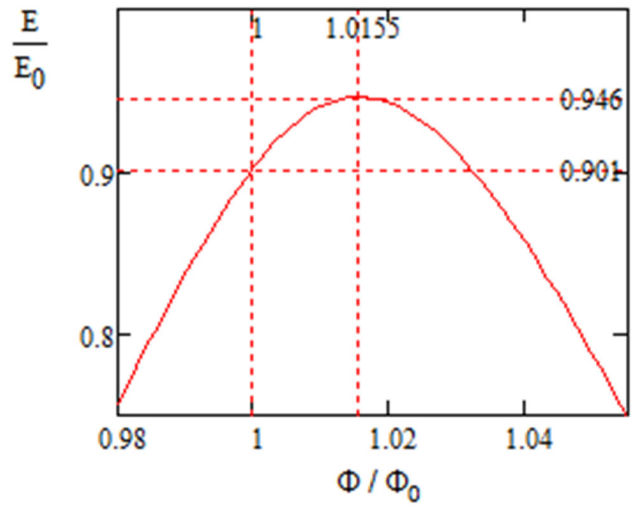

Figure 39: Dependence of the relative kick value on the relative focusing strength of the lens for a single-lens telescope for the 2.2 μm case. The lens is located in the OSC chicane center.

In the first approximation the electric field is inversely proportional to the effective radiation radius. Integrating the electric field along the undulator length and comparing it with the integral obtained in the absence of depth of field, we obtain about 9% decrease of the kicker efficiency due to an absence of depth of field cancelation. Almost half of the loss can be compensating by an increase of the lens focusing strength. The corresponding overfocusing, which reduces the diffraction contribution to the spot size in the first half of the undulator and decreases the depth of field effect at the undulator ends (see right pane in Figure 38), compensates the loss. Figure 39 presents the dependence of relative kick value on the relative focusing strength. The maximum is achieved at the focusing strength of about 1.5% above nominal. At this point only about 5% of kick value is lost. The difference with the three-lens telescope will be even smaller if one accounts its chromatic effects which are significantly larger for the three-lens telescope. A displacement of the lens downstream of the OSC section center yields additional 0.04% gain. However, this value is too small to be accounted in a further consideration.

Accurate accounting of the described above effects is presented in Section 8 below.

### 7.6. Separation of Particle and its Radiation due to Particle Oscillations

Betatron oscillations of a particle result in its transverse displacements in both undulators. To minimize a divergence of a particle and its radiation in the kicker undulator we targeted to make the transfer matrices for both the beam and the radiation to be close to **-I**. The separation is growing with betatron amplitude and when the value of the separation is approaching the transverse size of radiation spot the cooling force starts to decrease.

Let a particle with the single particle emittance of $\varepsilon$ experience betatron oscillations. Then, its position and angle in the kicker undulator are:

$$x_p = \sqrt{\varepsilon \beta_p} \cos\psi ,$$
$$\theta_{xp} = -\sqrt{\frac{\varepsilon}{\beta_p}} \left( \sin\psi + \alpha_p \cos\psi \right), \tag{56}$$

where $\psi$ is the betatron motion phase, and $\beta_p$ and $\alpha_p$ are the beta- and alpha-functions in the pickup undulator center. For a single lens telescope the transfer matrix of the radiation is

$$\begin{bmatrix} -1 & 0 \\ -2/L_c & -1 \end{bmatrix}, \tag{57}$$

where $L_c$ is the distance from the center of undulator to the OSC straight center. Consequently, the maximum separation between the particle and its radiation in the kicker undulator center is:

$$\Delta x = \sqrt{\varepsilon \beta_p} \sqrt{ \left( M_{11} + 1 - \frac{\alpha_p}{\beta_p} M_{12} \right)^2 + \frac{M_{12}^2}{\beta_p^2} } , \tag{58}$$

where $M_{11}$ and $M_{12}$ are the matrix elements of beam transport matrix between centers of pickup and kicker undulators, and we accounted that the position of radiation in the kicker undulator is $-\sqrt{\varepsilon \beta_p} \cos\psi$. For the beam transfer matrix of **-I** we have $M_{11} = -1$ and $M_{12} = 0$ and the separation is equal to zero, as expected.

Although obtaining the beam transfer matrix between undulators close to **-I** was one of the goals of the IOTA optics design the space limitations did not allow to achieve it for both planes. There is quite good compensation in the horizontal plane where the maximum separation is ~1% of betatron amplitude. That means that the horizontal beam separation is negligible for any betatron amplitude within cooling range. The situation is significantly worse for the vertical plane where $\Delta y/y_0 \approx 1.16$, *i.e.* the cooling rate will start to decrease at betatron amplitudes larger than about double rms vertical betatron size in the absence of OSC.

Synchrotron oscillations move particle locations in undulators proportionally to the dispersions which are equal in both undulators. For **-I** optical transfer matrix the separation is twice larger than a displacement in the undulators. However, for the passive OSC in IOTA the dispersion in the undulator centers is close to zero. That greatly suppresses this effect, making it negligible in practice.

## 8. Simulations with SRW for OSC Rates and Light Optics Tolerances

In this section, we describe simulation results obtained with Synchrotron Radiation Workshop (SRW) [13]. In SRW, the synchrotron radiation from an electron passing through an arbitrary magnetic field can be computed and propagated through an optical system. Computations are performed in an approximation of classical electrodynamics. SRW computes the radiation wave packet emitted by a single electron passing through the pickup undulator and propagates this radiation through an optical system to the location of the kicker undulator. The calculations are done in the frequency domain. The frequency components within the pickup's bandwidth are separately propagated to the kicker location and then Fourier transformed into the time domain. The found dependence of the electric field on time and coordinates is used to compute the value of the longitudinal kick in the kicker. The slippage between the electron and the wave packet and transverse and longitudinal variations of the field are accounted in the kick calculations.

### 8.1. Simulations for 2.2 μm Case.

We first consider the case for a single lens without dispersion as a benchmark of our simulations. The lens is placed 1.75 m downstream of the center of the pickup undulator corresponding to the midway point between pickup and kicker centers. A circular aperture is placed in front of the lens. The aperture's radius and distance from pickup undulator center determine $\theta_m$. Figure 40 presents a comparison of the kicks from electric field obtained with SRW and computed with Eq. (39) for different $K$ values and angular lens acceptances, $\gamma\theta_m$. Note that Eq. (39) implies an absence of depth-of-field effect which is significant in single lens focusing. For each point used in SRW calculations and presented in Figure 40 for a given $K$ the undulator period was adjusted so that to obtain the same wavelength of forward radiation (2.2 μm). As one can see from Figure 40 the single lens makes about 5% smaller kick. That is in a good coincidence with an estimate presented in the previous section.

Now we consider a radiation propagation through the three-lens telescope which corrects a waveform distortion due to the depth-of-field effect. The telescope dimensionless angular acceptance, $\gamma\theta_m$, is equal to 0.8, where $\theta_m$, is determined by circular masks installed near the first and the last lenses. The mask radius is set by the angle $\theta_m$ referenced to the pickup undulator center. In the absence of dispersion the wave packet should be symmetric relative to the kicker center, and should have as many wiggles as there are periods in the pickup undulator, $n_w$; in the 2.2 μm case $n_w=7$. As the wave packet propagates from the entrance to the exit of the kicker the waveform is changing. Figure 41 presents the waveform in the center and at both ends of kicker. The waveform change is expected since the telescope is designed to focus the radiation that originates at a specific location in the pickup to the same location in the kicker, thus maximizing the kick at each location. The slippage between an electron and the wave keeps the electron in vicinity of maximum amplitude at the entire motion in the kicker, thus compensating the depth of field effect and maximizing the kick. In other words, the telescope supresses the depth of field effect. In the center the field amplitude is 10.8 V/m. Using Eq. (39) one obtains the field amplitude of 11.8 V/m. The kick reduction at the kicker undulator ends is related to a reduction of angular acceptance for radiation coming from ends of pickup undulator (see Figure 30).

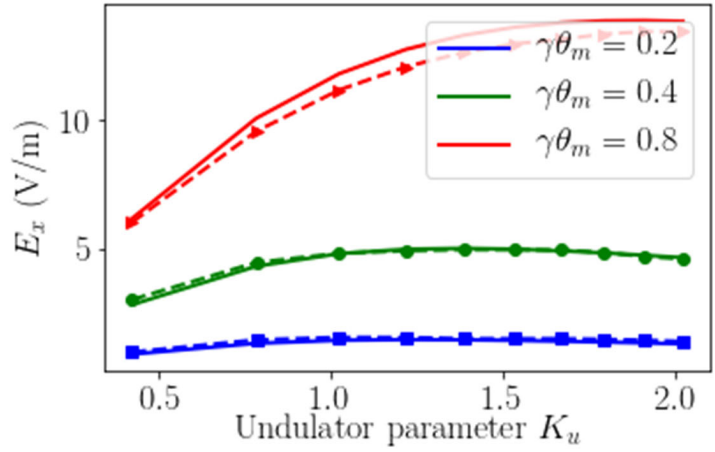

Figure 40: Comparison of the electric field amplitude computed using Eq. (39) (solid line) and SRW (symbols with dashed traces) in the absence of dispersion. The radiation is focused by a single lens located in the center of OSC chicane.

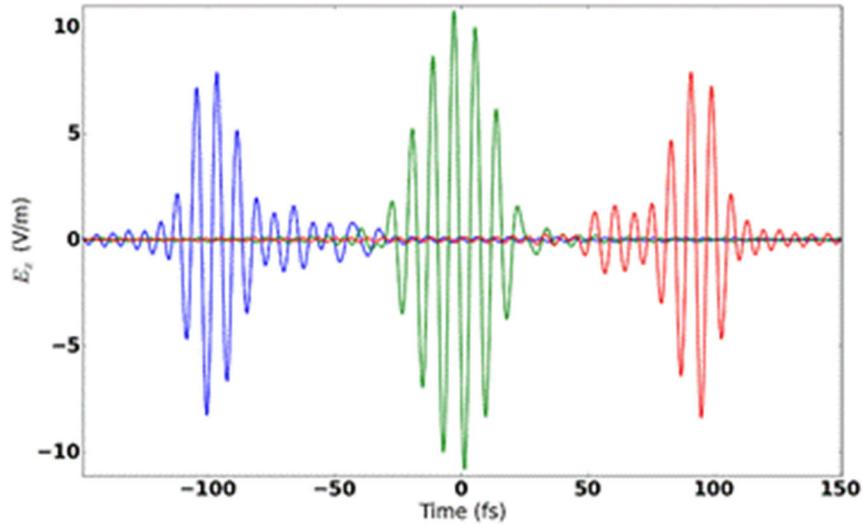

Figure 41: Wave packet generated in the pickup in the absence of dispersion is shown at the entrance (blue), center (green) and exit (red) of the kicker after passing through the dispersionless three-lens telescope. An artificial delay is separating the pulses for clarity.

To see how well the telescope addresses the depth of field, the electric field as seen by an electron is computed next. Within a given wave packet the electron is slipping back since it travels at an average longitudinal velocity $\overline{\beta} = \beta\left(1 - K^2/(4\gamma^2)\right)$ which is less than $c$. Thus, at a given location in the kicker, $s$, the particle is found at time, $t_p$,

$$t_p = \frac{1}{c}\left(s(1-\bar{\beta}) + \frac{K^2}{8k_p\gamma^2}\sin(2k_p s)\right) + t_0 .  \qquad (59)$$

Here $t_0$ is a constant chosen so that at the entrance of the kicker the electron is just ahead of the wave packet, and we account for the longitudinal oscillations that occur from the particle's transverse motion excited by kicker magnetic field. A plot of $E_x$ at the particle location over the kicker length is shown in Figure 42. The amplitude of the electric field is seen to be gradually decreasing moving away from the kicker center. This effect is not accounted for in the kick value given in Section 6; and it is the result of the effective $\gamma\theta_m$ decreasing from 0.8 to 0.63 at the kicker edges.

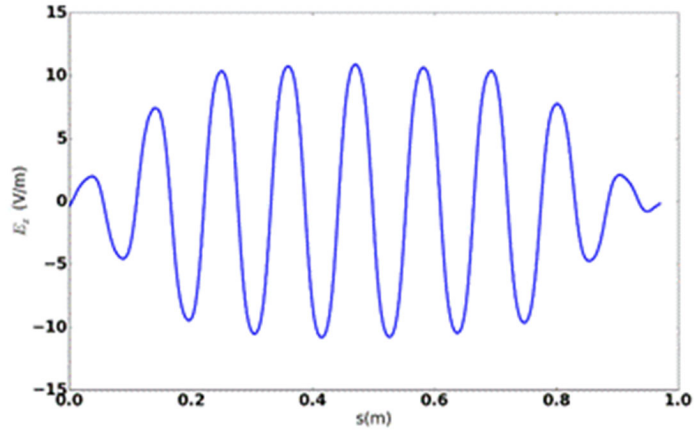

Figure 42: The electric field as seen by an electron as it slips back relative to its wave packet. The kicker begins at $s=0.1$ m and ends at $s=0.874$ m and the period of oscillation matches the undulator period.

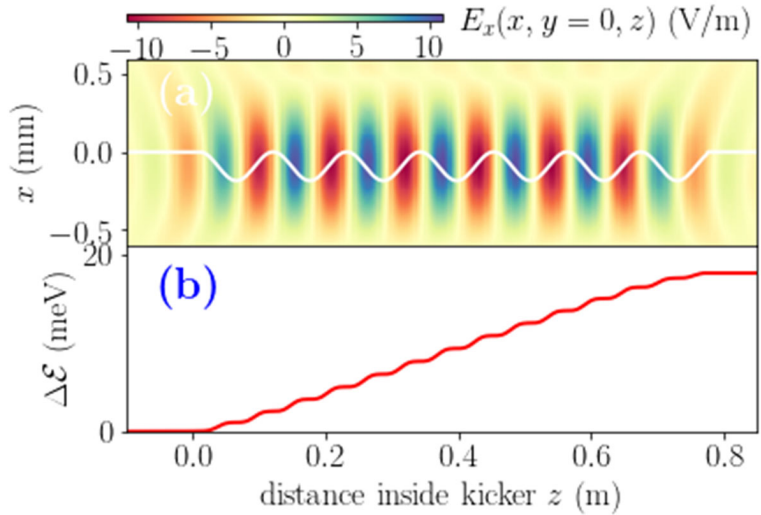

Figure 43: (a) The imaged electric field in the *x-s* plane when slippage is accounted. The white line corresponds to the particle trajectory in the kicker and the arrival time (electric field phase) was chosen to result in the maximum energy transfer. (b) The energy transfer as a function of position in the kicker.

The calculations of the kick value presented in Section 6 neglect the dependence of electric field on transverse coordinates. To some degree it is justified since the coupling between an electron and its field is strongest when the transverse velocity is at the maximum, or when the particle is on axis. To check this assumption the field is computed in the transverse plane as shown in Figure 43. The maximum extent of electron trajectory is 93 μm where 5.5% reduction of the field amplitude is observed. A correction to the kick value is significantly smaller since the electron horizontal velocity is small at the maximum extent.

We also note that the horizontal half-waist of the field from SRW is found to be 530 μm in the *x*-plane while Eq. (37) predicts 653 μm (see Table 7). It is quite good agreement, in particular, if one takes into account that the waist size of Eq. (37) was computed for K<<1.

Since $E_x(x,z)$ and $v_x$ are known the energy exchange can be computed by numerical integration:

$$\Delta E = e \int E_x(x,z) v_x dt , \qquad (60)$$

which results in a maximum kick value of 18 meV. A summary of values found from the semi-analytic theory presented in Section 6 and the results from SRW are summarized in Table 12. The largest effect is the longitudinal dependence of $E_x$ which reduces the kick by 10.4%, while the transverse field dependence reduces the kick by only 1.1%.

**Table 12: Summary of values obtained with theory developed in Section 6 and SRW simulations in the absence of dispersion in the lens material and ideal lenses**

|  | Theory | SRW |
|---|---|---|
| Field Amplitude (V/m) | 11.8 | 10.8 |
| Field half-waist in x-plane (μm) | 653 | 530 |
| Kick Amplitude (meV) | 22 | 18 |

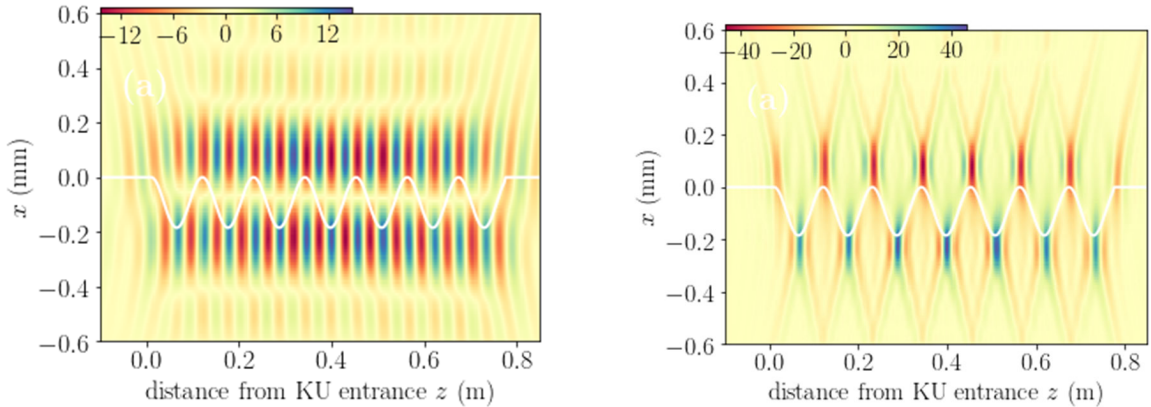

Figure 44: The imaged electric field for the isolated 2nd harmonic (left), and the imaged electric field including all radiation up to the 3rd harmonic (right).

All calculations so far have neglected the presence of higher harmonics in the undulator radiation. As already stated, the dispersion in the lens material yields a temporal separation of the first and higher harmonics. That results in an absence of energy exchange between an electron and the radiation of higher harmonics; but even in the absence of temporal separation there is no resonant energy transfer between electron motion mostly happening at the first harmonic and the higher harmonics of the radiation. Note also that the radiation of even harmonics radiation is strongly suppressed on the axis thus further reducing an interaction with the electron.

The right side of Figure 44 shows the imaged electric field where the radiation up to the 3$^{rd}$ harmonic is included. Electric field radiated from the pickup undulator is proportional to its acceleration as can be inferred from Eq. (23). Inside the pickup undulator the transverse acceleration achieves its maximum at the particles farthest extent off axis. Thus, it should be expected that, when the radiated electric field is imaged into the kicker, "hot spots" should be off axis. Such spots become visible when higher harmonics are included in the calculations.

The dispersion in the lens material is accounted with Sellmeier's formula (see Eq. (44)) describing a dependence of the index of refraction on the wavelength. It provides a straightforward way to account all orders of dispersion in SRW simulations. Dispersion reduces the kick value in two ways: (i) due to chromaticity in the lens focal length resulting misfocusing across the OSC band, (ii) and the packet lengthening due to Group Velocity Dispersion (GVD).

In Section 7.3 an estimate of lens chromaticity on OSC cooling rates was considered. It suggested that chromatic effects are minimized when the focal lengths of lenses are designed for 2.6 μm wavelength. This was confirmed in SRW simulations using the thin lens approximation with parameters presented in Table 9. The field amplitude as a function of lens design wavelength in the kicker center is shown in the left side of Figure 45. It is in good agreement with the results presented in Section 6. The right side of Figure 45 shows an energy exchange along the kicker length for a particle receiving the maximum kick. Chromaticity in the lenses results in a reduction in the kick amplitude to 17.2 meV or about 5%. It is in good agreement with results presented in Section 7.3.

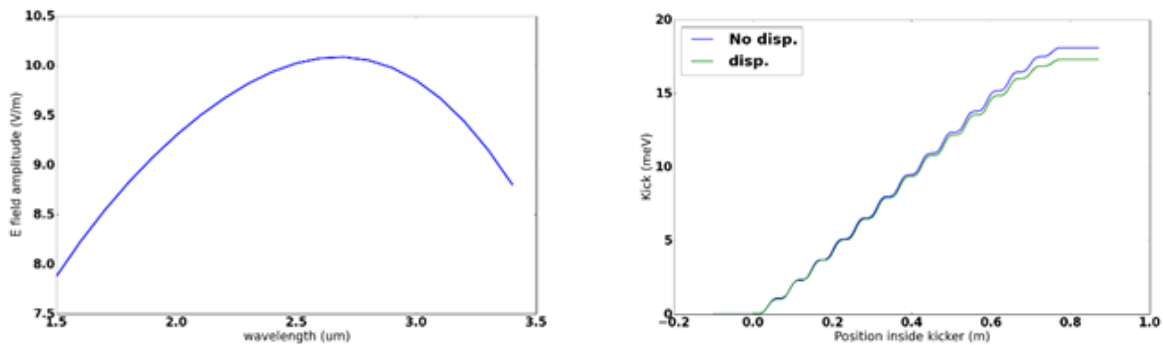

Figure 45: Left plot shows the dependence of field amplitude in the kicker center on the wavelength used to compute focal strengths of the telescope lenses. Right plot shows how the kick value is accumulated along the undulator length with and without dispersion accounting. Both plots use the thin lens approximation.

When the lens thickness is accounted the kick is further reduced through GVD. Using the thick lens parameters from Table 9 and including the delay plate in the optical system the kick amplitude is further reduced to 16.5 meV. Adjusting the lens profile as suggested in Eq. (53) results in an increase in the kick to 16.6 meV.

It was discussed in Section 7.5 that if the three-lens telescope is replaced by a single lens[12] with the focal length optimized for maximum kick, then the kick value is expected to be very close to the value achieved with the telescope. Simulations showed even larger kick value (16.7 meV versus 16.5 meV) for a single lens telescope. This surprising result can be explained as follows. The depth of field is not corrected for a single lens, however the larger strength of the lens (compared with the lens focusing radiation from center-to-center of undulators) considerably compensates the loss due to depth of field. There are two contributions. The first one results in an over-focusing and reduction of radiation spot size in the first half of undulator with subsequent increase of the electric field. The second one is determined by a reduction of depth of field effect at the undulator ends (see Figure 38). The SRW simulations show that the over-focusing results in an approximately 20% increase in the electric field amplitude for a portion of the kicker near the entrance. The results are summarized in Figure 46. It is seen that the single lens focusing results in a much less uniform energy exchange along the kicker but still results in approximately the same over all kick value. For the single lens case the lens thickness was adjusted to result in the same delay of the wave packet as it is for the telescope and for electrons. Summarizing we can state that the kick values computed with SRW where all effects are accounted are: 16.5 meV for the three-lens telescope with spherical lenses, and 16.7 meV for a single lens focusing. Note that the much larger chromaticity of the three-lens telescope removes its advantage in comparison with a single lens. Note that the reflections from the lens surfaces may result in a further reduction of the kick value.

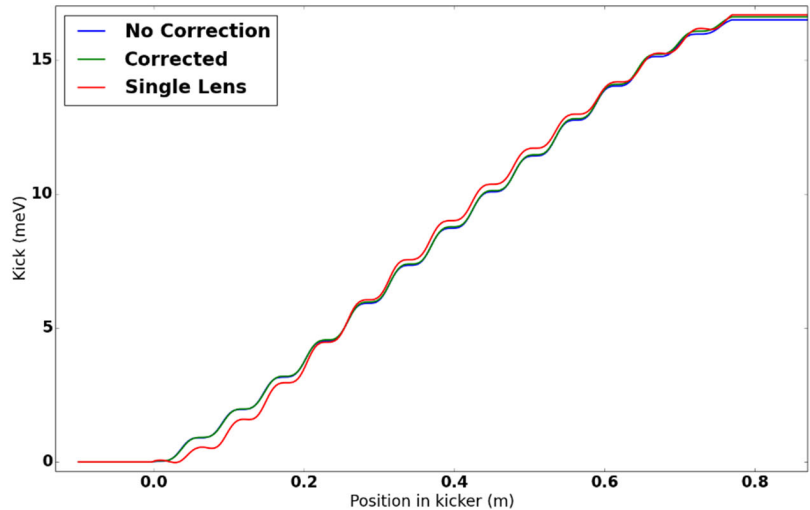

Figure 46: The energy exchange for the three-lens telescope designed for wavelength of 2.6 μm with spherical lenses (blue) and lens profiles adjusted to compensate dispersion (green) and a single lens (red).

---

[12] Note that both the single lens and the tree-lens telescope have the same angular acceptances implying a larger aperture radius for the single lens. In each case $\theta_m$ was referenced to the centers of undulators as described in the text.

## 8.2. Simulations for 0.95 μm Case

Due to small delay the passive cooling with the 0.95 μm basic wavelength can only be performed if the radiation is focused by a single lens. Therefore, we do not discuss simulations with the three-lens telescope in this subsection.

**Table 13: Results of SRW simulations for passive OSC at 0.95 μm**

| Maximum kick value computed with SRW no dispersion | 62.5 meV |
|---|---|
| Maximum kick value computed with SRW with dispersion | 60 meV |
| Horizontal HSHM* in the focal plane in the undulator center with dispersion | 126 μm |
| Vertical HSHM* in the focal plane in the undulator center with dispersion | 118 μm |

* half-size at half-maximum

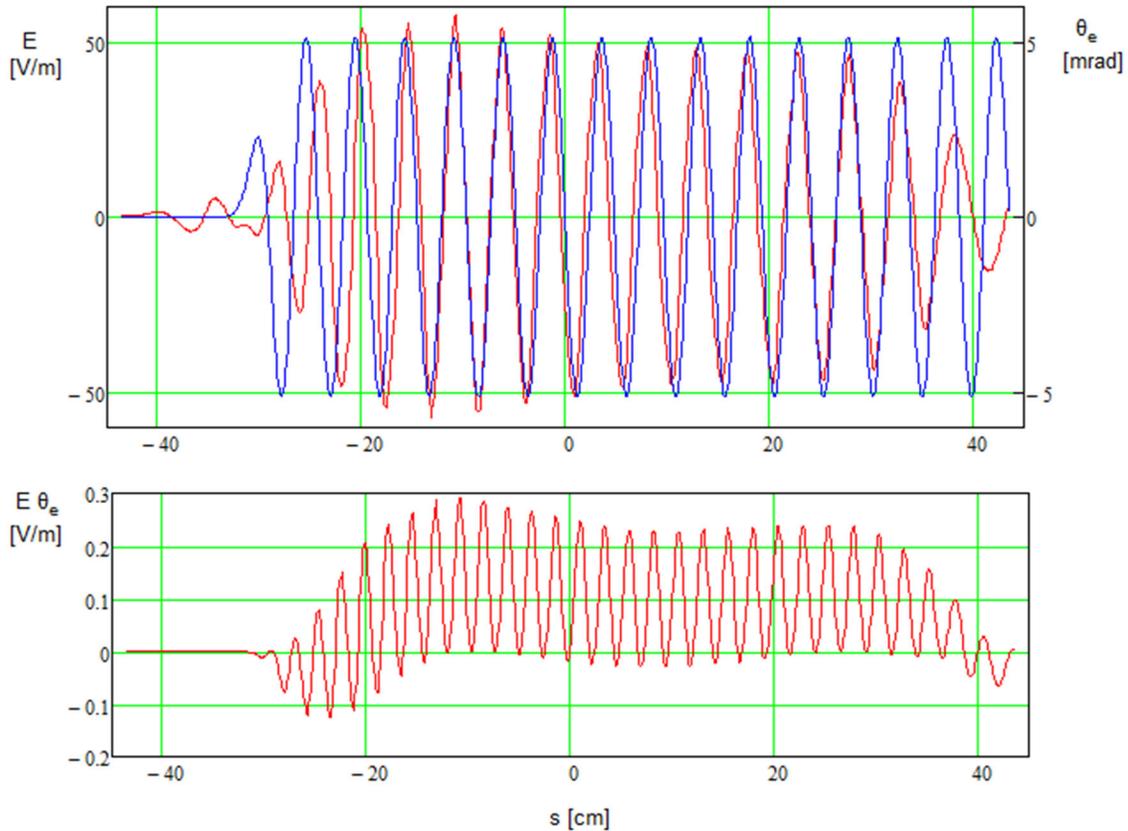

Figure 47: (top) Dependences of horizontal electric field of e.-m. wave (red line) and the particle angle excited by the undulator fields (blue line) on the particle coordinate along the kicker undulator for a particle phased for the maximum energy transfer. (bottom) The rate of energy transfer in the kicker undulator.

In general, the simulations for the 0.95 mm case yield results which are similar to the 2.2 μm case discussed above. However, there is an important new feature which was not discussed above, and which resulted in considerable loss in the cooling force. In the SRW simulations we use realistic fields of undulators which details are discussed below in Section 12.2. Table 13 presents main results of the simulations. It is assumed there, that the lens and its focal strength are optimized to maximize the cooling force.

Computations with Eq. (39) yield the amplitude of energy transfer for the passive cooling of 114 meV (see Table 7). Accounting for the loss due to wave packet lengthening (-6.3%, see Figure 32) and the depth of field (-5%, see Section 7.5) and accounting that the undulators effectively have 15 periods (-4%) due to the end effects one obtains the expected amplitude of energy transfer of 97 meV. However, as one can see from Table 13 the SRW calculations result in significantly smaller value of 60 meV. The major reason of this behavior is the Gouy phase which results in dephasing with the phase gradually acquired by the wave with deviation from the focal point. This phase achieves $\pm\pi/2$ for large upstream/downstream longitudinal displacements from the focal point. Figure 47 presents plots of the electric field of the e.-m. wave acting on a particle moving in the kicker undulator for the particle phased to obtain the maximum energy gain. It also shows the particle angle excited by the kicker undulator magnetic field, and the rate of particle energy gain, $E\theta_e$ (bottom pane). As on can see the optimal phasing results in that the particle angle and the electric field are not perfectly phased on the entire length of the kicker undulator. Dephasing achieves its maximum at the undulator ends. Comparing the total energy gain in the undulator and the energy gain in the undulator center scaled to 15 periods one obtains their ratio to be 0.65 which brings us close to the SRW value. Note that this dephasing was negligible for the 2.2 μm case because of larger wavelength and, consequently, larger Rayleigh range and smaller dephasing. Although the wavelength is changing by ~2.3 times only, the effect of detuning is quadratic with its value, $\cos(\Delta\phi) \approx 1-\Delta\phi^2/2$. That results in much larger effect of the detuning on the energy transfer for the shorter wavelength.

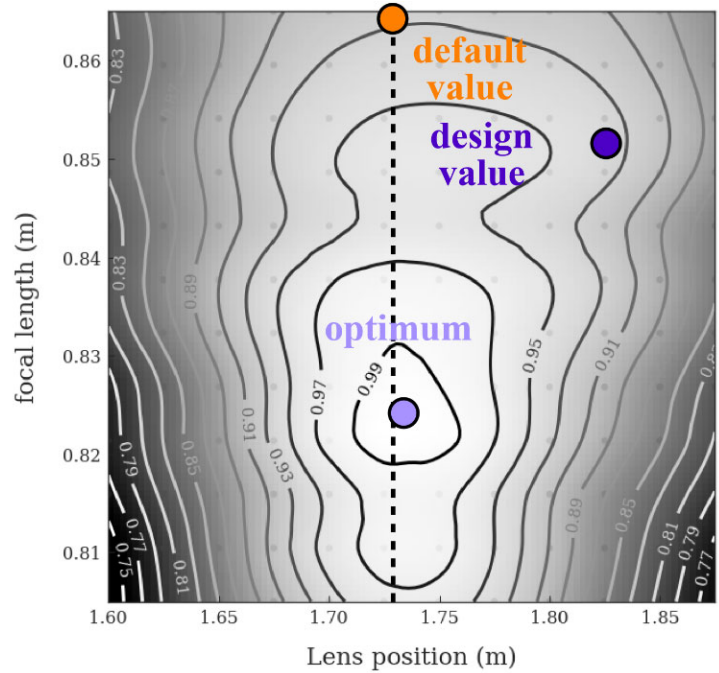

Figure 48: Dependence of relative cooling force on the lens position and focusing strength for the 0.95 μm case. The OSC center is located at 1.7276 m.

Figure 48 presents the dependence of cooling force (maximum energy gain) on the lens position and focusing strength. The maximum is achieved if the lens is shifted downstream of the OSC center by about 1 cm; and its focal length is reduced by about 4 cm. That increases the cooling strength by about ~7% as one can see in the figure. The lens is designed to obtain this optimal focusing at the wavelength of 1.13 µm located in the band center. Reflections from the lens and the path correction plate are not accounted in Table 14. The lens material is fused quartz. A necessity to separate the beam and light optics resulted in a shift of the optical lens downstream of the OSC center by about 10 cm. Optimization of lens focusing allowed us to reduce the loss in the kick value to ~6%. The corresponding point in Figure 48 marked as "design value".

There is good agreement between the beam sizes presented in Tables 7 and 13 if one takes into account that Table 13 presents the beam size at half maximum.

## 9. Beam Lifetime and Effects of Beam Intensity

In this Section we consider the beam lifetime and intensity related issues for the case of passive OSC at 0.95 μm basic wavelength. Due to smaller cooling acceptance the 0.95 μm case presents more challenges from the beam lifetime point of view. Limitations for the 2.2 μm case are driven by the same physics and will be considered elsewhere.

### 9.1. Residual Gas Scattering

The major mechanism determining the beam lifetime is associated with elastic scattering on the residual gas. The beam lifetime can be estimated with the following formula:

$$\tau_{rs} = \left( \frac{2\pi r_e^2 c}{\gamma^2 \beta^3} \left\langle \left( \frac{\beta_x}{\varepsilon_{mx}} + \frac{\beta_y}{\varepsilon_{my}} \right) \sum_k n_k Z_k (Z_k + 1) \right\rangle_s \right)^{-1}, \tag{61}$$

where $\beta$ and $\gamma$ are the relativistic factors, $c$ is the speed of light, $r_e$ is the electron classical radius, $n_k$ is the atomic density of $k$-th gas species and $Z_k$ is its nuclear charge, $\beta_x$ and $\beta_y$ are the horizontal and vertical beta-functions, $\langle \ \rangle_s$ denotes averaging over ring circumference, and $\varepsilon_{mx}$ and $\varepsilon_{my}$ are the ring horizontal and vertical acceptances, respectively.

Two other mechanisms affecting the beam lifetime are the inelastic scattering on atomic electrons and the bremsstrahlung. They result in mostly a loss in the electron momentum, while in the absence of dispersion excite the betatron motion relatively small. The bremsstrahlung contribution to the lifetime is determined by the total cross-section for radiation of a photon with energy above $\gamma m_e c^2 \theta_s$ [14]:

$$(\sigma_{BS})_k = 4\alpha_{FS} r_e^2 Z_k (Z_k + 1) \int_{\theta_s}^{1} \frac{1-\theta}{\theta} \left( \frac{1}{1-\theta} - \theta + \frac{1}{3} \right) \left( \ln\left( \frac{2(1-\theta)\gamma}{\theta} \right) - \frac{1}{2} \right), \tag{62}$$

where $\alpha_{FS}$ is the fine-structure constant, and we accounted for the contribution of atomic electrons. The cross-section of inelastic scattering at a residual gas atom is:

$$(\sigma_{ie})_k = \frac{2\pi r_e^2 Z_k}{\gamma \theta_s}. \tag{63}$$

Then, the total lifetime is determined by the summation of all contributions:

$$\tau_{tot} = \left( \frac{1}{\tau_{rs}} + c\beta \left\langle \sum_k n_k \left( (\sigma_{BS})_k + (\sigma_{ie})_k \right) \right\rangle_s \right)^{-1} \tag{64}$$

The beam lifetime measurements at small beam current, when the Touschek scattering can be neglected, were carried out in the IOTA Run II in the spring of 2020. They exhibited the beam lifetime of ~170 min. At that time the IOTA vacuum chamber was not baked and therefore, as will be seen below, the vacuum was not sufficiently good for the OSC studies. To address this problem, we will bake the IOTA vacuum chamber and expect an order of magnitude improvement in the ring vacuum. This improvement is used in the below estimate.

As one can see from the above formulas the particle loss depends on the nuclear charge, $Z$, and therefore is sensitive to the residual gas composition. In the below estimate of beam lifetime at the different stages of OSC we assume the relative gas composition presented in Table 14. To find overall pressure we used the IOTA Run II measurements. In the analysis we assume the relative partial pressures presented in Table 14. All other relevant parameters were measured or calibrated in the course of IOTA Run II. They are: acceptances of $\varepsilon_x=22$ μm and $\varepsilon_y=40$ μm, the average beta-functions of $\bar{\beta}_x=2.16$ m and $\bar{\beta}_y=1.94$ m, and the RF bucket height expressed in the relative momentum $\theta_s\equiv(\Delta p/p)_{max}=2.7\cdot10^{-3}$ ($V_{RF}=350$ V). Using these parameters, one scales the pressure to obtain the experimentally measured beam lifetime. That yields the following partial beam lifetimes:

- Elastic scattering – 3.3 hour,
- Bremsstrahlung – 46 hours,
- Inelastic scattering – 50 hours,
- Total lifetime – 2.9 hour.

That corresponds to 10 times larger density for each of the species than presented in Table 14. As one can see the elastic scattering strongly dominates. For the elastic scattering we introduce the effective pressure as the pressure of atomic hydrogen making the same scattering: $n_{eff} = \sum_k Z_k(Z_k+1)n_k$. For the partial pressures presented in Table 14 the effective pressure is equal to $6.2\cdot10^{-9}$ Torr of atomic hydrogen equivalent with main contribution coming from atoms of oxygen.

**Table 14: Expected partial pressures of different gases averaged over ring**

|      | P [$10^{-11}$ Torr] |
|------|---------------------|
| $H_2$ | 2.3 |
| CO   | 0.25 |
| $H_2O$ | 1.5 |
| $N_2$ | 0.34 |
| $CH_4$ | 1.5 |
| $CO_2$ | 0.6 |
| Ar   | 0.2 |

**Table 15: IOTA acceptances and beam lifetimes at different stages of OSC studies**

|                                | $\varepsilon_x$ [μm] | $\varepsilon_y$ [μm] | $\theta_s$ [$10^{-3}$] | Lifetime [min] |
|--------------------------------|----------------------|----------------------|------------------------|----------------|
| Acceptance limited by apertures | 10.5                 | 6.6                  | 7                      | 390            |
| Dynamic aperture               | 0.35                 | 0.35                 | 1.2                    | 17             |
| Cooling acceptance             | 0.072                | 0.072                | 0.57                   | 3.5            |

Compared to the IOTA Run II optics the OSC optics has significantly smaller momentum compaction and therefore significantly larger momentum acceptance even in spite of significantly smaller design RF voltage of 70 V. Therefore, the contributions coming from the bremsstrahlung and the inelastic scattering have negligible effect on the beam lifetime. Table 15 presents the IOTA acceptances at different stages of OSC experiments and corresponding beam lifetimes where we accounted that the average beta-functions are: $\overline{\beta}_x$=2.8 m, $\overline{\beta}_y$=3.1 m. The beam lifetime looks satisfactory if the OSC sextupoles are off (top line in Table 15). In this case the longitudinal acceptance is set by the RF bucket height and the transverse acceptances are limited by the vacuum chamber apertures. Switching on the OSC sextupoles greatly reduces the dynamic aperture both transversely and longitudinally. That results in a reduction of beam lifetime from ~6 hour to 17 minutes. Finally, we need to note that the cooling acceptance is even smaller than the dynamic aperture. In this case scattering on the residual gas may result in a particle being knocked out of cooling area and being stuck at large amplitude by OSC (see Figure 2). To return these particles back to the core we will need to switch off the OSC. Then within few seconds the beam will be cooled back to the core by synchrotron radiation (SR) cooling.

The gas scattering also results in an emittance growth. Usually the rms beam emittances are obtained in the logarithmic approximation. In this case they are equal to:

$$\begin{bmatrix} \varepsilon_x \\ \varepsilon_y \end{bmatrix} = \frac{2\pi c r_e^2}{\gamma^2 \beta^3} \sum_k n_k Z_k (Z_k+1) \ln\left(\frac{\theta_k^{\max}}{\theta_k^{\min}}\right) \begin{bmatrix} \overline{\beta}_x / \lambda_x \\ \overline{\beta}_y / \lambda_y \end{bmatrix}, \tag{65}$$

where

$$\theta_k^{\min} \approx \frac{\sqrt[3]{Z_k}}{192\gamma}, \quad \theta_k^{\max} \approx \min\left(\frac{274}{\sqrt[3]{A_k}\gamma}, \sqrt{\frac{\varepsilon_{x,y}^{\max}}{\overline{\beta}_{x,y}}}\right),$$

are the minimum and maximum scattering angles [15], $\varepsilon_x^{\max}, \varepsilon_y^{\max}$ are the ring acceptances for the horizontal and vertical planes, respectively, $\lambda_x$ and $\lambda_y$ are the emittance cooling rates due to SR for $x$ or $y$ planes, respectively, and $A_k$ is the atomic mass. We also accounted that in difference to the scattering in a medium, in accelerators $\theta_k^{\max}$ is typically determined by the ring acceptance.

More accurate analysis based on a solution of integrodifferential equation [16] shows that in the negligence of SR heating the equilibrium distribution function for scattering at $k$-th species is characterized by the dimensionless parameter

$$\left(\hat{B}_k\right)_{x,y} = \frac{2\pi c r_e^2}{\gamma^2 \beta^3 \lambda_{x,y}} \frac{n_k Z_k (Z_k+1)}{\left(\theta_k^{\min}\right)^2}, \tag{66}$$

which is equal to the average number of collisions during one SR emittance cooling time. For IOTA $\hat{B}_k \ll 1$. In this case the distribution function becomes non-Gaussian with sharp peak at zero amplitude and long non-Gaussian tails. The corresponding rms value of the emittance can be estimated as:

$$\varepsilon_{x,y} \approx \frac{2\pi c r_e^2}{\gamma^2 \beta^3 \lambda_{x,y}} \bar{\beta}_{x,y} \sum_k n_k Z_k (Z_k + 1) \left( \ln\left( \frac{\varepsilon_b}{1.35 \hat{B}_k \bar{\beta}_{x,y} \left(\theta_k^{\min}\right)^2} \right) - 1 \right). \quad (67)$$

Eq. (67) yields about half of the value determined by Eq. (65). The emittances growth rates following from Eq. (67) are equal to ~0.5 nm/s for both planes which are about ~1.25 times larger than the emittance growth rates due to quantum nature of SR when the ring operates at coupling resonance.

Figure 49 shows a numerically obtained distribution function for the vacuum observed in the IOTA Run II (right pane) and for 10 times better vacuum expected in the OSC studies (left pane). The calculations are performed for the small intensity beam when the IBS can be neglected and in the absence of OSC. The blue lines show the Gaussian fitting to the distribution function. For the left pane this Gaussian distribution contains 97% of the particles and has width 1.015 times larger compared to the distribution in the absence of gas scattering. For the right pane these number are 78% and 1.3 times. As one can see the 10 times worse vacuum (see the right pane) weakly affects the core but greatly affects the tails. The total rms emittance for the left pane is about twice larger than for the core, while for the right pane the total rms emittance exceeds the core emittance by almost an order of magnitude.

Concluding we can state that an achievement of the best possible vacuum in the IOTA ring is one of highest priority goals.

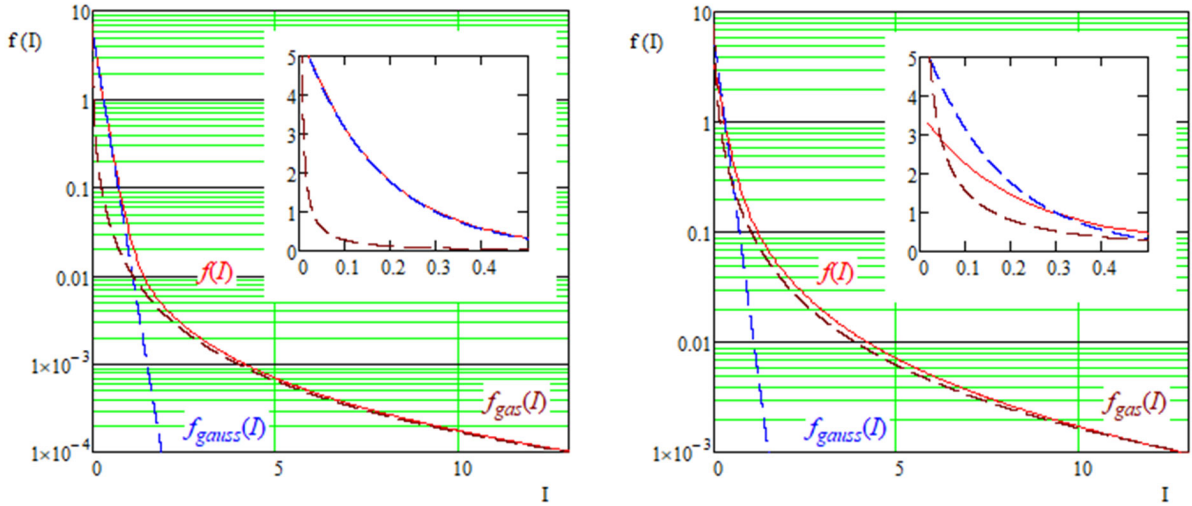

Figure 49: Dependence of distribution function on the dimensionless action $I = \left(\theta^2 + x^2/\bar{\beta}\right)/\left(\theta_k^{\min}\right)^2$ obtained numerically for the rms emittance set by SR of 0.42 nm for single species gas with $\hat{B}_k$ =0.035 (left pane) and $\hat{B}_k$ =0.35 (right pane). Insets show the same data for small actions. Red lines show the distribution functions. Brown lines show the distributions in the absence of SR diffusion. Blue lines show Gaussian fitting to the distribution cores.

### 9.2. Intrabeam Scattering

The RF harmonic number for IOTA is chosen to be equal to 4. The RF voltage choice was determined as a compromise between the multiple intrabeam scattering (IBS) which heating rate increases inversely proportional to the bunch length and the beam loss rate determined by Touschek scattering which increases fast with decrease of RF voltage. Table 2 shows the RF related parameters where we assume that the OSC is not present.

High beam intensity is desirable because it accelerates measurements and simplifies the beam instrumentation. For the IOTA OSC the major limitation on the beam intensity is determined by IBS. The beam equilibrium emittances are set by equalization of cooling and heating rates. The cooling rates are determined by SR cooling and the OSC. The beam heating rates are determined by quantum fluctuations of SR, scattering on the residual gas and intrabeam scattering. Note that the diffusion contribution of OSC is small and can be neglected. In calculations we accounted for strong $x$-$y$ coupling due to operation on the coupling resonance and controlled introduction of coupling by a skew-quadrupole as described in Section 5.3. Calculations of multiple IBS are based the IBS model which correctly accounts for coupling [17]. The Touschek scattering model also fully accounts for the coupling and, to simplify formulas, assumes non-relativistic particle velocities in the beam frame. That is satisfied for all beam parameters to be used in the OSC studies. In calculations of the gas scattering we use parameters presented in the previous section for the acceptance limitations presented in the top line of Table 15. The model was benchmarked on the measurements performed in the course of the IOTA Run II [18]. In computation of beam parameters with OSC on we use the following emittance cooling rates: the cooling rate for both transverse modes - 35 s$^{-1}$, the longitudinal cooling rate - 33 s$^{-1}$,

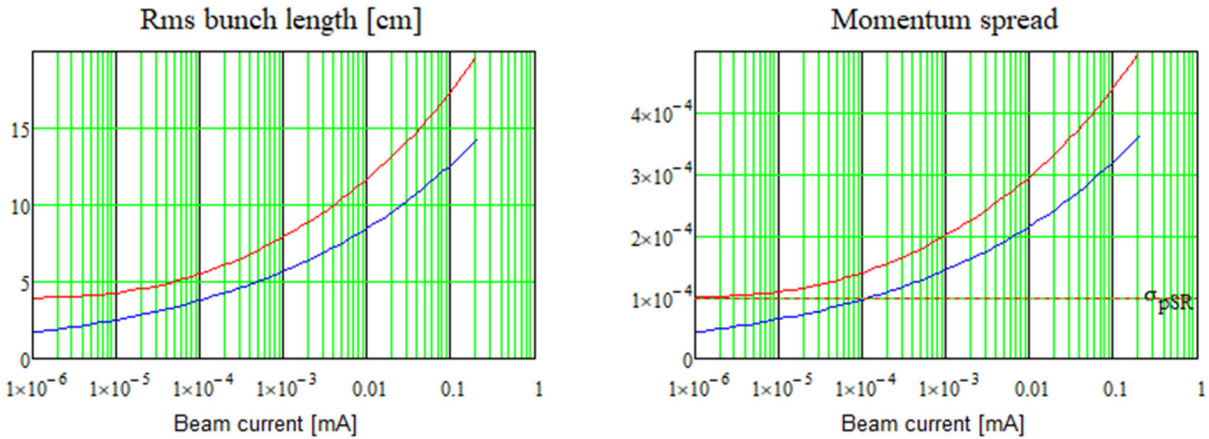

Figure 50: Dependences of rms bunch length (left) and rms momentum spread (right) on the beam current with (blue) and without (red) OSC. The horizontal dashed line shows momentum spread set by SR cooling in the absence of OSC.

As one can see from Figure 50 and Figure 51 the IBS starts to dominate for the beam current above 0.1 µA which corresponds to 8.3·10$^4$ particle per bunch. With further beam current increase and subsequent increase of transverse emittances the beam lifetime within cooling area starts to deteriorate fast when the ratio of transverse cooling acceptance to the rms emittance achieves 25 (or 5$\sigma$ in amplitude). That happens when the rms emittance achieves ~3 nm at the beam current of ~300 µA (see blue curve, OSC on, in Figure 51). That sets the maximum current for the OSC operation to ~300 µA.

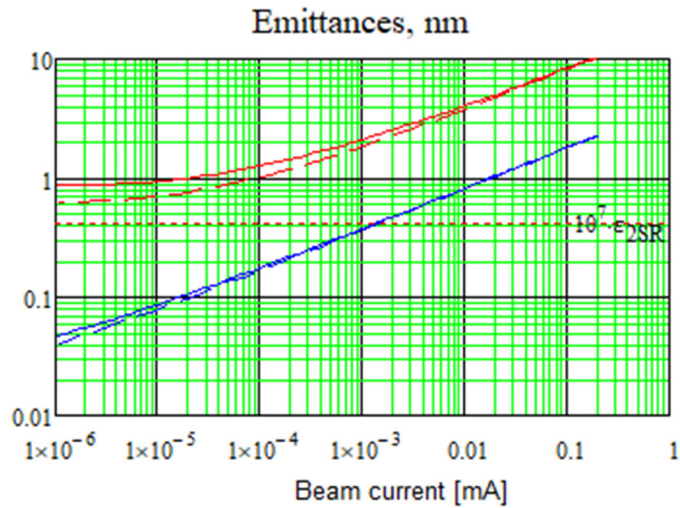

Figure 51: Dependences of the mode emittances on the beam current with (blue) and without (red) OSC. Horizontal dashed line shows equilibrium emittance set by SR in the absence of IBS, gas scattering and OSC. Dashed lines show the emittances of the second mode.

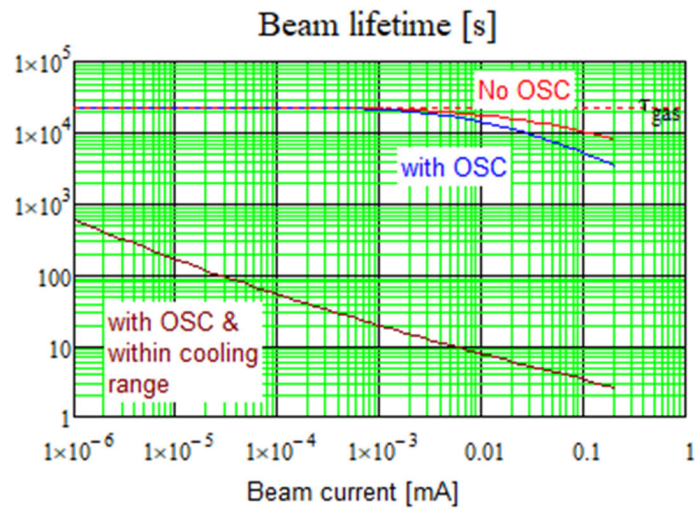

Figure 52: Dependences of beam lifetime on the beam current with (blue and brown) and without (red) OSC. The brown (bottom) curve shows the beam lifetime inside the cooling acceptance ($\Delta p/p=5.7\cdot 10^{-4}$), while red and blue curves show the total beam lifetime in the ring determined by the RF bucket height ($\Delta p/p=6.9\cdot 10^{-3}$)

Another important limitation comes from the Touschek scattering. The corresponding results are shown in Figure 52. As one can see the Touschek scattering starts to dominate the beam loss at the residual gas at the beam current above ~30 µA. The situation is much worse if one considers the beam lifetime within the cooling area. Similar to the single scattering on the residual gas a single Touschek scattering can knock out a particle out of the longitudinal cooling range but still leave it in the ring acceptance. Then, the particle will be cooled to a quasi-stable point with large synchrotron amplitude. Consequently, such scatterings lead to a storage of particles at large synchrotron amplitudes and a reduction of core intensity. The brown line in Figure 52 shows corresponding lifetime for the core. To bring these particles back to the core one needs to switch off the OSC and switch it on after few seconds when the entire beam is cooled to the core by SR cooling. Requiring 1 minute lifetime in the OSC cooling area one obtains that the beam current should be $\leq 0.1$ μA.

Concluding we can state that the single and multiple scattering set the nominal operating current for the OSC studies to be ~0.1 μA or to $8.3 \cdot 10^4$ particle per bunch. Some operational improvement may be achieved by reduction of the RF voltage to about 25 V which will make the bunch somewhat longer and Touschek scattering smaller. Note that the IBS effects will be much more pronounced in the absence of coupling due to much larger particle density. Therefore, the majority of experiments with OSC will be carried out at the coupling resonance where both transverse emittances are close to be equal.

# 10. Optical Amplifier for IOTA OSC

The Optical Amplifier (OA) to be used for demonstration of the active OSC is discussed in this section. An ideal amplifier for OSC would have the following characteristics: (i) high gain, (ii) an optical delay not exceeding a few millimetres, (iii) an operating bandwidth on the order of 50 THz, and (iv) the capability of amplifying pulses with length larger or equal to the particle bunch length, *i.e.* at minimum hundreds of picoseconds at a repetition rate of 7.5 MHz in the case of IOTA. An Optical Parametric Amplifier (OPA) would make a superb amplifier for the OSC if it were not for this last requirement. An OPA creates signal gain through a non-linear mixing process which requires pump intensities on the order of a GW/cm$^2$. Considering the bunch length and repetition rate in IOTA, and that the pickup light could be focused only to about a 100 µm radius inside the OPA crystal, the required average power of the pump laser would exceed a kW.

Therefore, instead of an OPA we consider amplification from a solid-state lasing medium which, due to the small intensity originating from the pickup, operates essentially independent of the bunch length and repetition rate of the accelerator. Here we note that points (i) and (ii), *i.e.* high gain and small delay, are mutually contradictory for a single-pass amplifier. Ultimately the delay is determined by the cooling beam optics and, consequently, by the chicane design. For the active OSC in IOTA the delay has been set to 2 mm. A requirement to obtain sufficiently large cooling ranges results in that the OSC must take place in the mid-IR regime. Considering criteria (i) – (iv) we identified Chromium-doped zinc-selenide (Cr:ZnSe) as a candidate medium for the OSC.

Table 16: Characteristics for Cr:ZnSe Amplifier in IOTA

| Parameter | Value | Reference |
|---|---|---|
| Peak emission wavelength, $\lambda_a$ | 2.4 µm | [19] |
| Index of refraction, $n$ | 2.45 | [20] |
| $dn/dT$ | $70 \cdot 10^{-6}$ K$^{-1}$ | [21] |
| Thermal conductivity, $\kappa_T$ | 1 W/(cm K$^*$) | [22] |
| Crystal length, $L$ | 1 mm | - |
| Pump wavelength, $\lambda_p$ | 1908 nm | |
| Absorption cross-section at the pump wavelength, $\sigma_{pa}$ | $1.0 \cdot 10^{-18}$ cm$^{-2}$ | - |
| Emission cross-section at the pump wavelength, $\sigma_{pe}$ | $0.4 \cdot 10^{-18}$ cm$^{-2}$ | [23] |
| Maximum emission cross-section in signal band ($\lambda$=2.4 µm), $\sigma_s$ | $1.3 \cdot 10^{-18}$ cm$^{-2}$ | - |
| Fluorescence lifetime, $\tau$ | 5.5 µs | [23] |
| Emission band | 1.8 – 3.3 µm | - |

$^*$ Thermal conductivity achieves it maximum of about 1 W/(cm K) in the temperature range 7-50 K. It decreases by about 6 times at the room temperature (300 K).

Cr:ZnSe has a central wavelength $\lambda_a$ = 2.45 µm and an operating band that covers the major fraction of the pickup undulator radiation bandwidth (2.2-2.9 µm). The high index of refraction ($n$ = 2.45) limits the length of the crystal. For this report we consider a 1 mm crystal yielding 1.45 mm delay, and thus, leaving 0.55 mm delay for the lenses. Table 16 summarizes the crystal parameters relevant to the OSC.

## 10.1. Gain for a Single-pass Cr:ZnSe Amplifier

Operation of Cr:ZnSe amplifier is well approximated by a model with 4 energy levels. Its detailed analysis is carried out in Ref. [24] which results are used below. In this analysis we also assume Continuous Wave (CW) pumping.

The emitted radiation from the pickup arrives at the entrance of the amplifier in short bunches at the revolution frequency of accelerator. For parameters presented in Table 7 one obtains the average total power radiated by one electron in one undulator to be 60 fW. Less than half of this power is radiated into the first harmonic which should be amplified by OA. Taking into account that the expected maximum number of electrons per bunch is $\approx 10^7$ we obtain the maximum average input power for the amplifier to be less than 600 nW. We will later determine the pump laser power to be 40 W. Taking into account that both the input and output powers of the OA are much smaller than the pump power we can neglect the effect of the stimulated emission induced by the incoming signal in computation of the population inversion in a steady state solution. The choice of pumping wavelength is determined by the availability of suitable Thulium laser. Its wavelength of 1908 nm is quite close to the optimum pumping wavelength of about 1750 nm [24]. Such choice results in an overlap between the absorption and emission cross-sections at the pumping wavelength. That, consequently, results in a possibility of stimulated emission induced by the pump laser. Accounting this one obtains the derivative of the pump laser intensity (power per unit square), $I_p$, on coordinate at a passage through the crystal [24]:

$$\frac{dI_p}{dz} = -I_p N_t \left( \frac{(1 + AI_p \sigma_{pe})(\sigma_{pa} + 2\sigma_{pe})}{1 + AI_p (\sigma_{pa} + 2\sigma_{pe})} - \sigma_{pe} \right), \quad A = \frac{\tau}{h\nu_p}, \tag{68}$$

where $\sigma_{pe}$ and $\sigma_{pa}$ are the emission and absorption cross-sections at the pump frequency, $\nu_p$ is the pump photon frequency, $\tau$ is the fluorescence lifetime, $N_t$ is the total ion-doping concentration, which we assume to be $2.0 \cdot 10^{19}$ cm$^{-3}$. This equation can be integrated numerically and used to compute the OA gain (in power):

$$G_P(\omega) = \exp\left( \Delta I_p \frac{\sigma_e(\omega)\tau}{h\nu_p} \right) \tag{69}$$

where $\Delta I_p = I_{p|(z=0)} - I_{p|(z=L)}$ is the change of pump laser intensity after passage through crystal, $L$ is the crystal length and $\sigma_e(\omega)$ is the emission cross-section at the signal frequency. As one can see the gain does not directly depend on doping concentration or pump power, but directly depends on the power left in the crystal by the pump laser. Consequently, for a chosen crystal the overheating represents the fundamental limitation on the achievable gain.

The gain and transmission for a single-pass through a Cr:ZnSe amplifier are shown in Figure 53. The transmission increases with pump intensity. Levelling-off of the absorbed pump intensity, $\Delta I_p$, with increasing pump intensity eventually limits the gain. At 125 kW/cm$^2$ the gain is 7 dB and cannot be increased significantly more. For crystals with a longer length the onset of the saturation is delayed, and the corresponding gain is higher. Unfortunately, a longer crystal cannot be used in the IOTA experiment given the limited optical delay set by the particle delay in the chicane bypass.

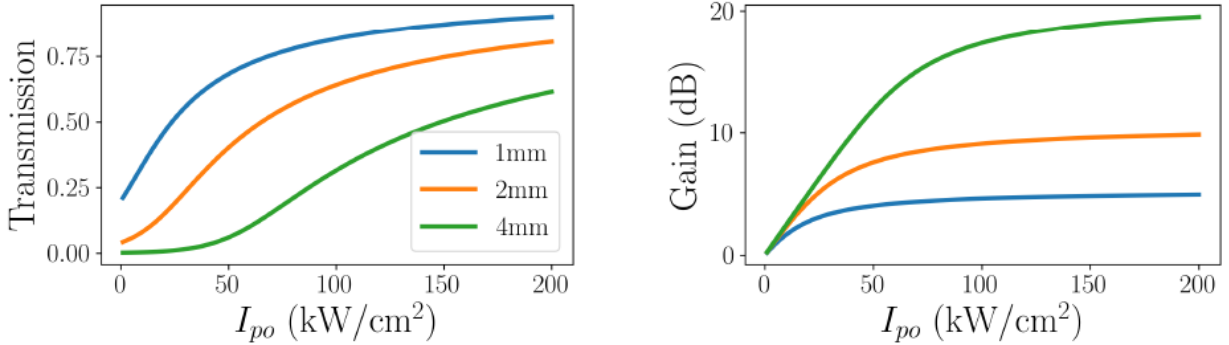

Figure 53: The transmission and gain curves for a single-pass through a Cr:ZnSe amplifier as a function of incident pump laser intensity for 3 different crystal lengths. Transmission is for 1908 nm light and amplification is at 2.45 μm.

To minimize heating of the crystal the radiation of pump laser is collinear with the amplified radiation and, consequently, with the electron beam in the OSC straight line. Due to aperture limitations it coincides transversely with the electron beam in the undulators. If the laser radiation coincides in time with the electron bunch it causes random longitudinal kicks which result in additional diffusion in both horizontal and longitudinal phase spaces. Although the frequency of pump laser is not in resonance with the base frequency set by the undulator period the effect is quite large. To keep the laser induced diffusion below the diffusion set by quantum fluctuations of synchrotron radiation the laser power density inside the undulators should not exceed 10 W/cm². That, in its turn, requires avoiding time coincidence of pump laser radiation and electron beam. This can be addressed by operating the pump laser in a pulsed regime at either a harmonic or subharmonic of the revolution frequency so that the electron beam and laser pulse do not overlap in time. The above analysis of OA operation was carried out for CW pumping. However, this analysis is still valid for pulsed pumping if the time between pumping pulses is sufficiently short,

$$T_{pump} \ll \tau \ln(G_P)/G_P \;, \tag{70}$$

and a time averaged value is used to compute the pumping laser intensity. This condition is well justified if pumping is performed at the revolution frequency or its first few subharmonics. Another possible option to avoid this laser induced diffusion might be a usage of off-axis mirrors or mirrors with holes in the middle to introduce and remove the pump radiation inside the chicane so that the pump radiation would never overlap with the electron beam.

### 10.2. Active OSC in IOTA

The simulation methods based on SRW as described in Section 8 can be extended to include an optical amplifier in order to address two separate issues concerning the amplifier:
- To suppress the depth-of-field the optical transport line should have a transfer matrix equal to ±**I**, where **I** is the identity matrix. This determines the relationship between the focal strengths of optical telescope lenses for given distance between them. The distance also determines the radiation spot-size in the amplifier which value can be accurately computed with SRW.

- The 7 dB of gain computed in the previous section corresponds to a wavelength where the maximum amplification is achieved. Since the OSC operates over a large spectral band a distortion of the undulator wave-packet from host-dispersion and finite amplifier bandwidth is expected. Because SRW computations are initially done in the Fourier-domain it is straight forward to account for both of these effects. That results in a realistic calculation of the kick amplitude achieved with the optical amplifier.

*Spot radius of pickup undulator radiation in the amplifier*

The laser pump intensity determines the gain. For the foreseen 7 dB of amplification the pump intensity should be 125 kW/cm$^2$. To keep the laser power reasonable, we need the spot radius of the pump laser to be as small as possible. It is also highly desirable to have an optics scheme which would suppress the depth of field effects related to the finite length of the undulators. These requirements may be satisfied with a three-lens telescope but in difference to the telescope considered in Section 7.1 this telescope should have +**I** matrix. Such a choice has an advantage that the radiation from the pickup undulator is focused to a much smaller size. In this case, in the geometric optics approximation, the radiation is focused to zero size for radiation coming out of the undulator center. For the radiation coming out of other locations in the undulator the spot size does not exceed ~150 μm if the radiation is within an angular acceptance of 4 mrad. The required focal lengths (see Figure 29) are:

$$F_1 = \frac{L_1 L_2}{L_1 + L_2}, \quad F_2 = \frac{L_2^2}{2(L_1 + L_2)}. \tag{71}$$

The focusing for the central lens, $F_2$, may be introduced to the profile of OA crystal or can be done with a nearby focusing lens. Table 17 presents parameters of the lens telescope for the OA.

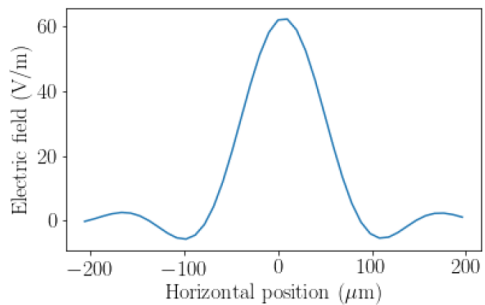

Figure 54: $E_x(x)$ in the plane of the amplifier.

Figure 54 presents SRW simulation of the radiation spot size in the plane of the OA. The size value is in a good agreement with Eq. (37) which results in the spot size in the focal point to be equal to $x_0 L_2/L_1 \approx 100$ μm; *i.e.* for the +**I** telescope the spot size at the OA is mainly determined by diffraction. It is impossible to achieve a same small spot size with the -**I** telescope. To achieve the same spot size with its use one needs the lenses to be of an order of magnitude stronger. In a geometric optics approximation an achievement of 200 µm spot size would require: $F_1$=50 mm, and $F_2$=0.76 mm. The second value is well outside of practical limit.

**Table 17: Geometrical parameters of lens telescope for active test of the OSC in IOTA.**

| $L_1$ | 154 cm |
|---|---|
| $L_2$ | 21 cm |
| $F_1$ | 18.48 cm |
| $F_2$ | 1.26 cm |
| $\gamma\theta_m$ | 0.8 |
| Outside lens radius | 6.3 mm |

Taking the above into account we choose the spot size of pumping laser to be $\rho_p \approx$ 100 µm and assume that the laser radiation has a flat-top distribution in the radial direction. That yields the total power of about 40 W. The corresponding angular divergence of pump radiation near the crystal is 10 mrad. Its propagation through the entire OSC straight will require 6 mm aperture (diameter) which is well within the vacuum chamber aperture.

Most of the laser power absorbed by the crystal is radiated through spontaneous emission. Using energy conservation and accounting difference in the wavelengths of the pump and emission we obtain the power left by the pump laser in the crystal in the form of heat:

$$P_h = \Delta I_p \left(1 - \frac{\lambda_p}{\lambda_s}\right) \pi \rho_p^2 . \tag{72}$$

For 40 W of pumping power it yields 1.7 W going into the crystal. The corresponding temperature dependence on the radius is given by:

$$T(\rho) = \begin{cases} \dfrac{P_h}{4\pi\kappa_T L \rho_b^2}\left(\rho_p^2 - \rho^2\right) + T(\rho_p), & \rho \leq \rho_p , \\ \dfrac{P_h}{2\pi\kappa_T L}\ln\left(\dfrac{\rho_{xtal}}{\rho}\right) + T(\rho_{xtal}), & \rho \leq \rho_p , \end{cases} \tag{73}$$

where $\rho_{xtal}$ is the radius of the crystal assumed to be cylindrical in shape, $\kappa_T$ is the crystal's thermal conductivity, and T($\rho_{xtal}$) is the temperature at the crystals surface assumed to be equal to the cooling fluid. The surface where the heat exchange occurs is assumed to be the lateral surface of the cylindrical crystal with 1 mm length and 5 mm radius. Assuming liquid nitrogen cooling one obtains $\kappa_T$ = 1.0 W/(cm·K), and the total temperature change from crystal center to surface of 12K.

The signal radiation is focused inside of $\rho_p$ where a temperature dependence on the radius is parabolic. The index of refraction of the host medium is temperature dependent. For small variations in temperature, the change in $n$ goes as $\Delta n(T) = T\, dn/dT$. Employing it one obtains a parabolic dependence of $n$ on the radius. Consequently, the crystal acts like a lens with a focal length given by [25]:

$$f_{th} = \frac{2\pi \kappa_T \rho_p^2}{P_h (dn/dT)} \quad . \tag{74}$$

Therefore, the pumping intensity uniquely determines both the gain and thermal lensing focal length. For the 7 dB amplifier expected to be used in IOTA, $f_{th}$ = 5 cm which is about 4 times larger than $F_2$ in Table 17. Therefore, additional focusing is required. This can be accomplished by curving the surface of the crystal or adding an additional lens near the amplifier. Correction of the focusing can be performed through minor change to $\rho_p$.

The radiation coming out of the crystal due to spontaneous emission makes random kicks to particles in the kicker undulator. The value of corresponding diffusion is very small, $\sqrt{\Delta p^2}/p \approx 1.6 \cdot 10^{-10}$ per turn, and can be neglected in comparison with diffusion due to SR in the IOTA dipoles, $\sqrt{\Delta p^2}/p \approx 7.5 \cdot 10^{-8}$ per turn. In this estimate we accounted an amplification of spontaneous emission by OA resulting in an increase of radiated power by $(G-1)/\ln(G) \approx 3$ times.

The beam transfer matrix between undulators is closer to -**I** then to +**I**. Therefore, a usage of +**I** transfer matrix for radiation, as required for the OA, increases the separation between particle and its radiation in the kicker undulator. Compared to the 0.95 µm case this increase is partially compensated by an increase of radiation spot size due to longer basic wavelength. Modifying Eq. (56) for +**I** radiation transfer matrix one obtains for the betatron motion: $\Delta x/x_0 \approx 0.38$, $\Delta y/y_0 \approx 0.2$; and for the synchrotron motion: $\Delta x/x_0 \approx 10^{-3}$. Although, the separation is significant for large betatron motion amplitudes it is still within an acceptable level.

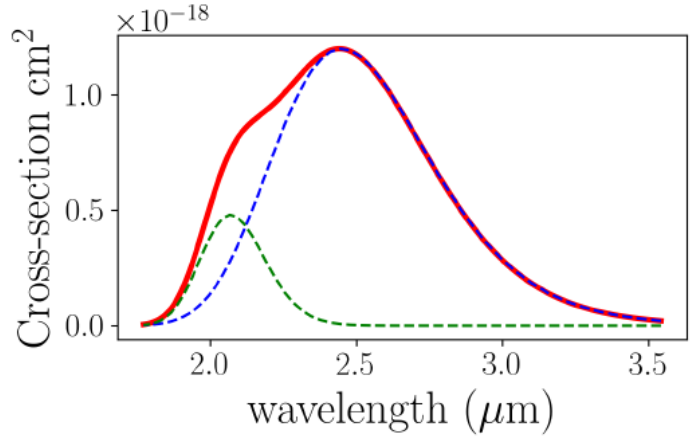

Figure 55: The modelled cross-section for Cr:ZnSe used for gain calculations (red solid line) with the individual Gaussian distributions comprising it (blue and green dashed lines).

### *Pulse amplification with SRW*

SRW allows one to account for both the host dispersion of Cr:ZnSe and OA bandwidth. For host dispersion the Sellmieir formula with coefficients for ZnSe was used [26]. To compute the dependence of amplifier gain on the frequency the emission cross-section was modelled as the sum of two Gaussian distributions centered at 2.07 and 2.45 µm and the gain was set to be 7 dB at 2.45 µm. The corresponding emission cross-section is plotted in Figure 55.

The left side of Figure 56 shows examples of the imaged pickup pulse in the center of the kicker with varying effects accounted. The green pulse is with no amplification and is seen to have an amplitude of 10.9 V/m. Using 7 dB of gain for the central wavelength of the amplifier, the orange pulse accounts the bandwidth but not the host dispersion. Its amplitude increases to 20.9 V/m. For comparison, an ideal (finite bandwidth, no host dispersion) 7 dB amplifier would increase the amplitude to 24.3 V/m. Finally, the blue pulse accounts for both the bandwidth and the host dispersion, and the amplitude is reduced to 19.1 V/m.

On the right side of Figure 56 the cumulative energy exchange between the co-propagating test particle (located at maximum acceleration phase) and the wave-packet in the kicker is shown. When both the host dispersion and amplifier bandwidth are accounted the kick amplitude increases by a factor of 1.7 from 17.4 to 29.7 meV over the un-amplified pulse.

Preliminary data indicates both the absorption and emission cross-sections of Cr:ZnSe may increase at 77K. Reliable values of the cross-sections are not currently available but we estimate the gain can improve to 15 dB yielding a kick amplitude of 40 meV [27].

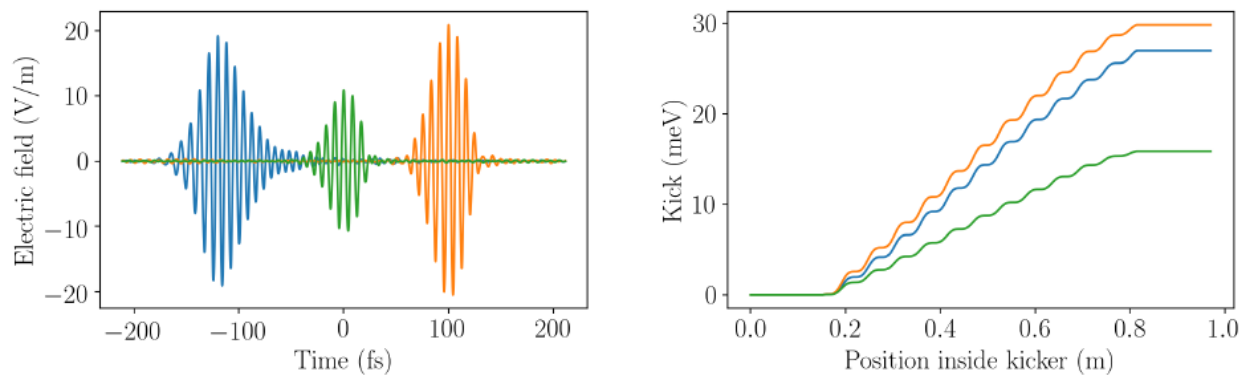

Figure 56: Left: Examples of the imaged pickup pulse in the kicker center. The green pulse is with no amplifier. The orange pulse includes the amplifier, with the bandwidth shown in Figure 55 but does not account host dispersion and finally the blue pulse includes both amplifier bandwidth and host dispersion. Right: The corresponding cumulative energy exchange.

In conclusion of this section we consider at which bunch population the particle interaction through OSC system will start reducing the cooling rates. In general, the theory developed for description of the stochastic cooling [28] is applicable to the OSC however its application requires accounting factors which are specific to the OSC. The diffusion due to OSC has two contributions. The first one is related to the amplified noise coming from nearby particles and the second one to the noise of amplifier induced by spontaneous emission. As it was mentioned at the end of the previous subsection the latter is small and can be neglected.

The cooling dynamics is determined by the gain and bandwidth of the cooling system which determine how the cooling of a particle is affected by the presence of nearby particles. In difference to the microwave stochastic cooling the OSC has two bandwidths. The first one is determined by angular divergence of undulator radiation which after focusing to the particles moving in the kicker undulator is recombined back to the base frequency (corresponding to forward radiation). The second frequency band is related to the finite number of undulator periods. Both bandwidths may be additionally limited by the bandwidth of OA. As one can see in Figure 56 (blue line) the dependence of kick value on the longitudinal coordinate can be approximated by a response of an 8-period undulator which indicates that the amplifier bandwidth weakly affects the overall bandwidth of the system. With an increase in the number of particles in the bunch the particle interaction through OSC system results in additional OSC driven diffusion which reduces the cooling rates. To see an effect of such particle interaction on the OSC rates we rewrite Eq. (92) of Appendix A so as to express the number of particles in the bunch which, for a fixed OSC gain, puts OSC to operate on the optimal gain:

$$N_{max} = \frac{3\mu_{01}^2 k_0 \sigma_s}{2\sqrt{\pi} n_w n_{\sigma s}^2 \lambda_{s\_opt}} . \tag{75}$$

Substituting $n_w$=8, $2\pi/k_0$=2.2 μm, and the values corresponding to the equilibrium achieved with active OSC: $\sigma_s$ =7.7 cm, $n_{\sigma s}$=9.4, $\lambda_{s\_opt}$=5.6·10$^{-7}$ per turn, we obtain $N_{max}$=2.7·10$^9$. Here we also accounted a factor-of-two reduction in the cooling rate due to particle interaction through the OSC system at the optimal gain, *i.e.* the cooling rate computed for small number of particles and equal to $\lambda_s$=8.4 s$^{-1}$ needs to be reduced by 2 times. Note that the obtained number of particles greatly exceeds the number of particles to be used in the OSC measurements in IOTA, which has a maximum value set by IBS. Consequently, we can conclude that the particle interaction through the cooling system is negligible for both the active and passive OSC to be tested in IOTA.

# 11. Effects of Quantum Nature of Radiation and Single Electron Cooling

As it was shown in Refs. [29, 30] the quantum nature of the light does not play a role in the OSC if cooling rates per turn are significantly smaller than $\alpha_{FS}$, where $\alpha_{FS}$ is the fine-structure constant. This condition is always justified for OSC in IOTA. Although the quantum nature of the light does not play a role in the IOTA OSC it still can be observed if only one electron is cooled [30]. Below we consider how that can be achieved for the case of passive cooling which looks preferable because it has a larger cooling rate and does not have noise present in an optical amplifier.

We start our consideration of quantum effects from Ref. [30] quote: "*Particles do not radiate in a series of stochastic quantum jumps, only once every $O(\alpha^{-1})$ passes through the pickup wiggler. The radiation from each particle is present on every pass, even though its average energy may correspond to much less than one photon, and is available to be amplified by a quantum mechanical interaction, provided no intervening measurement first projects the state by trying to count photons. Moreover the wiggler radiation possesses, on average, exactly the amplitude and phase it would possess classically, so the "coherent signal" information necessary for transit-time cooling is effectively always present, only corrupted by the equivalent of some additional additive (not multiplicative) noise needed to satisfy the uncertainty principle, noise that physically can be traced back to "vacuum fluctuations" and amplified spontaneous emission within the gain medium.*"

In the case of the passive OSC we can rewrite the above statement in the following way. Each undulator radiates a probability wave. The probability waves coming out of pickup and kicker undulators interfere creating a photon, or in other words the photon is not radiated before electron comes through both undulators. Interference of probability waves radiated from pickup and kicker undulators results in that a probability of photon radiation is oscillating together with synchrotron and betatron oscillations.

First, we find a probability of photon radiation by a single electron coming through an undulator. Similar to Eqs. (39) and (40) an expression for the first harmonic of horizontal electric field at the lens surface follows from Eqs. (25) and (26) which yield:

$$E_{\omega x}(\theta,\phi) = \frac{4e\omega_u \gamma^3 K}{\pi c R} \int_0^{2\pi} d\tau F_c(\Theta,K,\tau,\phi) \frac{1+\Theta^2(1-2\cos^2\phi)-2\Theta K \cos\phi \sin\tau - K^2 \sin^2\tau}{\left(1+\Theta^2+2\Theta K \cos\phi \sin\tau + K^2 \sin^2\tau\right)^3} \quad . \tag{76}$$

Here function $F_c(\Theta,K,\tau,\phi)$ is determined by Eq. (40), and $\Theta = \gamma\theta$. At a distance much larger than the undulator length the total energy of the first harmonic radiation into a solid angle is:

$$\frac{d\mathrm{E}}{d\Omega} \equiv \frac{d^2\mathrm{E}}{\theta d\theta d\phi} = \frac{c|E_{\omega x}(\theta,\phi)|^2}{8\pi} \frac{2\pi n_w}{\omega(\theta)} R^2 \quad , \tag{77}$$

where $n_w$ is the number of undulator periods, and $\omega(\theta)$ is the radiation angular frequency at angle $\theta$ determined by Eq. (27). Using Eq. (27) to bind derivatives over $\theta$ and $\omega$, dividing the energy by $\hbar\omega(\theta)$ and integrating over $\phi$ one finally obtains the probability to find a horizontally polarized photon in the dimensionless frequency band:

$$\omega_0 \frac{dW_\gamma}{d\omega} = \alpha \frac{\left(1+\Theta^2+K^2/2\right)^4}{2\left(1+K^2/2\right)} K^2 n_w \int_0^{2\pi} |I(\Theta,K,\phi)|^2 d\phi,$$

$$I(\Theta,K,\phi) = \frac{1}{\pi} \int_0^{2\pi} d\tau F_c(\Theta,K,\tau,\phi) \frac{1+\Theta^2\left(1-2\cos^2\phi\right)-2\Theta K\cos\phi\sin\tau - K^2\sin^2\tau}{\left(1+\Theta^2+2\Theta K\cos\phi\sin\tau + K^2\sin^2\tau\right)^3}.$$

(78)

Here frequency of radiation $\omega$ and the angle of observation $\Theta/\gamma$ are related by the following equation:

$$\Theta = \sqrt{\frac{2\gamma^2 \omega_u}{\omega} - 1 - \frac{K^2}{2}},$$

(79)

and $\omega_0 = 2\gamma^2 \omega_u$ is the frequency of forward radiation at the first harmonic.

Figure 57 shows results of computations obtained with Eq. (78) and computed with SRW for the same 16-period undulator at the nominal OSC parameters for the 0.95 µm case. In computations with SRW only photons which come through the optical system with acceptance $\gamma\theta = 0.8$ are accounted. That results in that the radiation at small frequencies is strongly suppressed. Contrary, Eq. (78) does not account acceptance limitations and therefore yields much higher intensity at small frequencies. This equation also does not account spectrum smearing due to finite number of undulator periods which reduces the peak intensity at $\omega/\omega_0 = 1$ in the SRW simulations. Note also that accounting of vertical polarization would increase the Eq. (78) result by ~15%. With this corrections accounted we conclude that the results of SRW and analytical calculations with Eq. (78) are in satisfactory agreement which supports both SRW and Eq. (78) results. An integration of the SRW spectrum in vicinity of the first harmonic $\omega/\omega_0 \in [0.55, 1.07]$ yields the probability of photon radiation into the first harmonic of $W_1 \approx 5\%$.

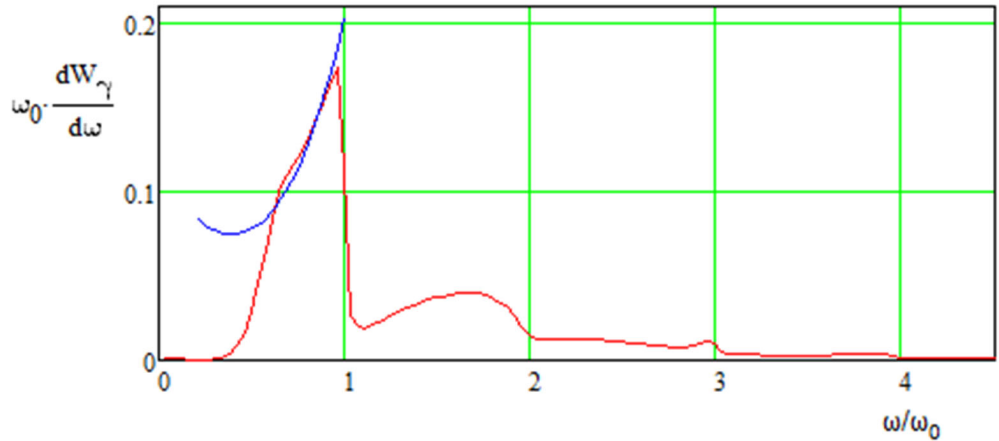

Figure 57: Dependence of probability density of photon radiation into the dimensionless bandwidth on the relative photon energy for the 16-period pickup undulator at nominal OSC parameters computed with SRW (red line). The SRW simulations account for both polarizations and the aperture limitations in the lens ($\gamma\theta = 0.8$). Blue line shows calculations with Eq. (78).

Now we find how the probability of photon radiation oscillates with betatron and synchrotron oscillations of an electron in the ring. For small undulator parameter the wave radiated from the pickup undulator after focusing in the ideal lens telescope has the same time and space structure as the wave radiated from the kicker undulator, and therefore in the absence of aperture limitations the waves interfere ideally resulting in that the probability of the photon radiation depends on a delay in the chicane, $s$, and is equal to:

$$W(s) = 2(1-\sin(k_0 s))W_1 \ . \tag{80}$$

Here $s$ is determined by Eq. (3), and $W_1$ is the probability of photon radiation in one undulator. Consequently, large amplitude betatron or synchrotron oscillations will result in close to 100% modulation depth of photon radiation probability at the betatron or synchrotron frequency, respectively.

In the general case of large undulator parameter the energy conservation binds the modulation of radiated energy with the energy transfer to a particle. Neglecting the dependence of photon energy on its frequency and accounting Eq. (3) we obtain the probability of photon radiation:

$$W(s) \approx W_0 + \frac{\Delta E}{\hbar \omega_c} \sin\left( k_0 \left( M_{51}x + M_{52}\theta_x + \left( M_{56} - \frac{L_{pk}}{\gamma^2} \right) \frac{\Delta p}{p} \right) \right), \tag{81}$$

where $\omega_c$ is the radiation frequency in the band center, and $\Delta E$ is the amplitude of the OSC kick, and $W_0$ is the average probability of the photon radiation. $\Delta E$ can be computed using Eq. (42) or more accurately with SRW code which can account all details of the OSC process. Assuming $\Delta E=114$ meV and $\lambda_c = 2\pi c / \omega_c = 1.13\ \mu\mathrm{m}$ (see Table 7) one obtains the amplitude: $\Delta W_{\max} = \Delta E / \hbar \omega_c \approx 0.1$. That coincides with modulation amplitude of Eq. (80) equal to $2W_1 \approx 0.1$.

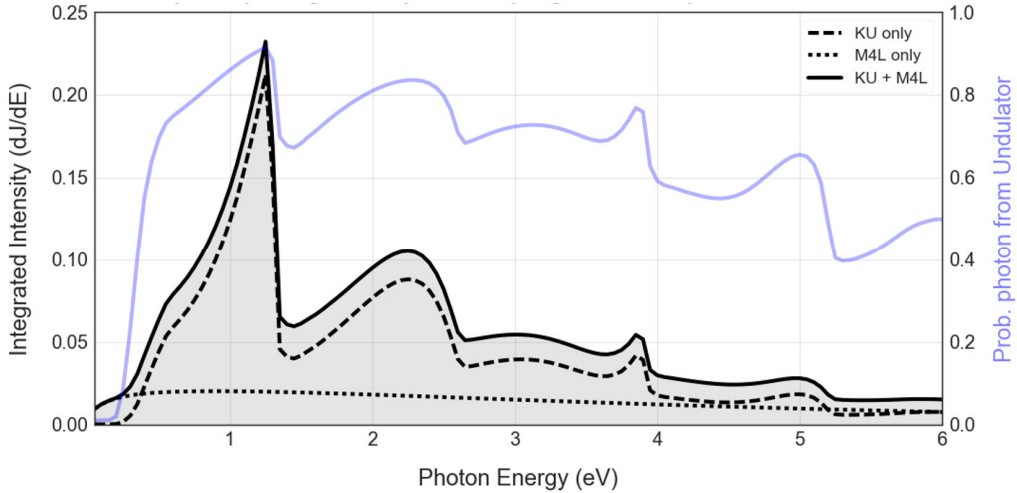

Figure 58: Spatially integrated spectrum of single electron radiation at ML4 window for the radiation from the kicker undulator and ML4 dipole (left scale), and the ratio of probability of photon radiation from the kicker undulator to the total probability of photon radiation in the second half of OSC straight (right scale, lilac color).

The average probability, $W_0$, also includes the radiation coming from the dipoles. The major contribution comes from the main IOTA dipole located downstream of the OSC straight. Figure 58 shows the spatially integrated spectrum, $dJ/dE$, of single electron radiation at ML4 window located directly downstream of the OSC straight, where $J$ is the radiated energy and $E$ is the photon energy. As one can see the dipole radiation adds ~20% to the radiation coming out from the kicker undulator or about 10% to the average radiation of both OSC undulators. Accounting this we conclude the amplitude of the photon probability oscillates at ~90% of the average probability if the radiation of higher harmonics ($\lambda \leq 0.95$ μm) is filtered out. The exact result will depend on the photodetector quantum efficiency response on the wavelength.

The maximum variation of $W(s)$ is achieved when the phase in Eq. (81) oscillates between $\pm \pi/2$. That corresponds to the betatron or synchrotron amplitudes equal to $\pi/(2\mu_{01}) \approx 0.653$ of the corresponding cooling range.

### *Variation of Photon Radiation Probability with Variation of Electron Momentum*

The work with a single electron in a storage ring was pioneered by scientists of BINP, Novosibirsk [31]. Recently single electron studies we also carried out in IOTA [32]. That allowed us to obtain the first-hand experience in such operations.

For the longitudinal motion the amplitude of momentum oscillations corresponding to the maximum oscillation of probability of photon radiation is equal to $3.7 \cdot 10^{-4}$. This amplitude of momentum oscillations can be observed at synchrotron radiation (SR) monitors. It corresponds to 100 μm amplitude of transverse displacement at the sync-light monitor M4L which has maximum modulo of the horizontal dispersion (26 cm). This is quite small displacement to measure with good accuracy. It will be easier to observe the oscillations in the longitudinal displacement of electron. For the nominal RF voltage of 70 V the corresponding amplitude of longitudinal motion is ~0.5 ns.

In the absence of external excitation, the OSC damps the longitudinal motion of an electron to the rms momentum of ~$3 \cdot 10^{-5}$ and an observation of probability variation is going to be extremely difficult. Therefore, we plan to excite the longitudinal motion with excitation of RF phase near the synchrotron frequency. A harmonic excitation of RF phase with 0.5 deg. amplitude and 5% detuning from the synchrotron frequency (from 204 to 194 Hz) will result in the desired longitudinal oscillation with 5 deg. amplitude required to maximize oscillations of the radiation probability.

The amplitude of the longitudinal oscillations can be controlled with photon arrival time referenced to the revolution marker. Similar measurements were already performed in the IOTA Run II for radiation coming from one of the IOTA dipoles.

A registration of photons arriving from the OSC straight could be done with infrared photomultiplier (PMT). Hamamatsu R5509-43 near infrared PMT is one of the best. It has peak sensitivity in the required band 900 - 1300 nm, but its quantum efficiency (QE) is only about 2%. Therefore, we preferred to use an infrared avalanche diode - Excelitas SPCM-AQRH-14-TR-BR2. It has a smaller wavelength band but significantly larger QE. In the infrared end of its sensitivity its QE drops approximately linearly from 25% to 0 in the wavelength range 950 – 1050 nm. Using a longpass filter (e.g. Thorlabs FEL0900) allows to remove radiation of higher harmonics. Averaging over the radiation band yields ~35,000 registered photons per second of the first harmonic radiation coming from both undulators. That is about twice larger than the number of photons which could be registered with R5509-43 near infrared PMT. Excelitas SPCM has excellent dark count of 100 s$^{-1}$ which is also much smaller than 15,800 s$^{-1}$ for R5509-43 operating at -80 C$^{\circ}$. SPCM does not require cooling. In normal OSC studies only half of the photon flux will be directed to the SPCM (see Section 13.1). Note also that the number of background events can be additionally reduced by an order of magnitude with gating referenced to the revolution marker.

### *Variation of Photon Radiation Probability with Betatron Oscillations*

For the horizontal motion the Courant-Snyder invariant corresponding to the probability oscillations at the maximum amplitude is 30 nm. It corresponds to the betatron amplitude of ~300 μm observed at the sync light (M4L) with maximum horizontal beta-function of 3 m. Present resolution of IOTA sync-light monitors allows one to observe betatron oscillations of single electron with such amplitude with good accuracy. An excitation can be performed by the existing kicker. This kicker has been used for the transverse damper/antidamper in the IOTA Run II.

Note that both the longitudinal and transverse degrees of freedom will be excited by the scattering at the residual gas expected happening every few tens of seconds.

## 12. Conceptual Design of the OSC Straight

The OSC apparatus, pictured in Figure 59, will occupy IOTA's "E-straight" and has approximately six meters available for its implementation. The design of the integrated system provides the necessary functionality and flexibility required for a broad OSC science program while minimizing overall cost and complexity. In this section, we describe the mechanical design of the OSC apparatus including vacuum chambers and pumping systems, magnets, mechanical supports and precision-motion and power systems. We will summarize the major components, design choices and their impacts on the OSC and system performance.

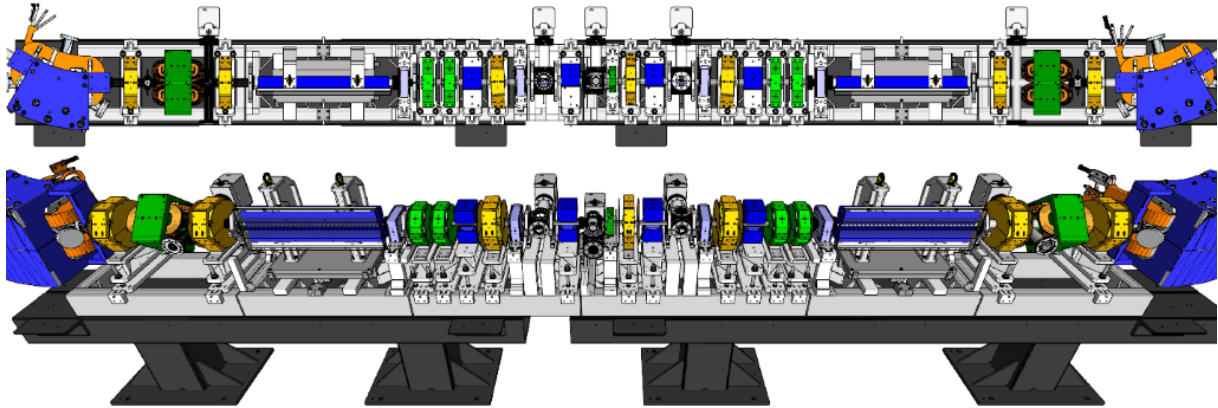

Figure 59: Integrated model of the OSC insertion: OSC dipoles (short and dark blue), quads (green), sextupoles (gold), vertical correctors (light blue) and undulators (long and dark blue). The beam moves from right to left.

The electron beam after passing the pick-up undulator is bent into the bypass section and then bent back into the IOTA ring passing through the kicker undulator. The light generated in the pick-up undulator is focused by an optical lens; then it comes through a computer controlled optical delay stage and finally is directed to the kicker undulator where it recombines with the electron beam.

### 12.1. Vacuum Envelope and Systems

In discussion of the IOTA vacuum system we separate it into two parts: the IOTA vacuum system and the vacuum system of the OSC straight which is described in more details.

The vacuum chamber for the OSC straight consists of three main elements: the bypass section, undulator chambers and end chambers. We consider the structure of these elements, any special processing issues , and finally the expected vacuum performance for the IOTA ring in the OSC configuration.

#### IOTA Vacuum System

The IOTA ring is 40 m long. The required overall vacuum level is in the low $10^{-11}$ Torr range. Because of strict requirements on the magnetic permeability the vacuum chamber is constructed of 316L stainless steel or silver-plated aluminium. 316LN flanges and components are used wherever possible, and hardware and heater tapes are nonmagnetic.

The beam-tube is 1.87" ID 316L stainless steel for most of the ring. Local aperture variations for the injection area and certain devices vary from 0.93" to 2.99" ID. Aluminium dipole chambers are coated with silver to minimize the possibility of charge build-up on the surface which can adversely affect the stability of beam orbit.

Vacuum levels are maintained with ion/NEG pumps combination (45 L/s ion pump and 200 L/s NEG). Pump spacing varies, but is typically 184 cm. Larger pumps, or closer pump spacing will be used in some areas depending upon outgassing rates of specific accelerator components (nonlinear magnet, Lambertson, etc.). Certain locations, such as the octupole string, will have slightly higher operating pressures due to smaller beam-tube size, lack of space for pumping, or higher outgassing. To meet the vacuum requirement, stainless steel tubes are electropolished and hydrogen degassed prior to installation. To transit from the present $2 \cdot 10^{-8}$ Torr (molecular hydrogen equivalent) to the OSC required vacuum the system will undergo an in-situ bake at 100 -150°C prior to operation. Ion pumps, cold cathode gauges and convection gauges will provide vacuum readbacks. Three gate valves separate the vacuum chamber into three sectors for ease of maintenance and bake-out. The valves are located in the upstream ends of BR and DR straights and in the middle DL straight.

To transit from the transport line vacuum to the IOTA required pressure, a differential pumping section separates the IOTA ring from the rest of the FAST beamline. This station consists of three combination pumps, each with 75 L/s of ion pumping and 200 L/s of NEG pumping.

### **OSC Bypass**

The OSC bypass section, pictured in Figure 60, comprises four flat-oval beam tubes connected by small chambers for optics and diagnostics. The small delay of the non-amplified OSC experiments corresponds to a horizontal offset (the radiation relative to the beam) of ~20 mm and allows for transport of the undulator radiation and electron beam within the same beam tube while remaining compatible with small, inexpensive magnets in a high-density configuration. The bypass chambers (316L tubing; 316LN flanges) have the cross-sectional geometry shown in Figure 61, which satisfies all required constraints while providing for the required radiation acceptance, $\gamma\theta \approx 0.7$. In the two innermost chambers, at maximum separation, the centers of electron beam and radiation are located near the two centers of curvature for the chamber profile.

A compact 6-way cross (316L tubing; 316LN flanges), which houses the in-vacuum optical lens and includes an inline formed bellows (316L), joins the innermost bypass chambers. The 6-dimensional in-vacuum precision-motion system for the lens is located beneath this cross in a 4.5" spherical-cube chamber (316L; Kimball Physics). The 6.3-mm, vertical half-aperture of the bypass chamber at the entrance of the optics cross establishes the angular acceptance for the pickup-undulator radiation of ~3.5 mrad. Two additional spherical-cube chambers (316L) house the optical-delay stage (left) and pickoff mirror (right) are situated symmetrically about geometric center of the bypass.

Each spherical cube is directly pumped by a Gamma Vacuum Titan 45S-CV Ion pump with an integrated 200 l/s NEG pump, which pumping power is sufficient for the expected gas loads in these areas.

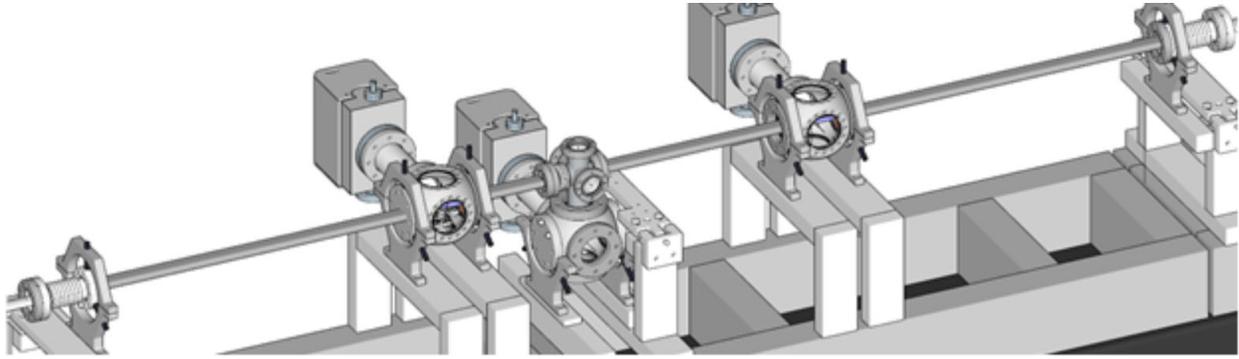

Figure 60: Vacuum envelope for the OSC-bypass section.

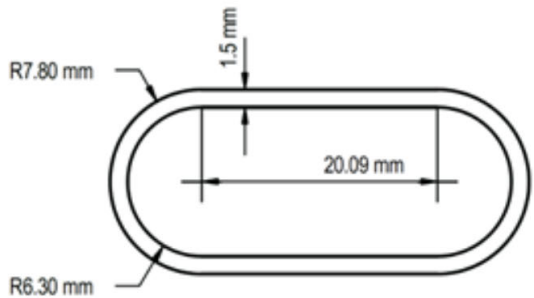
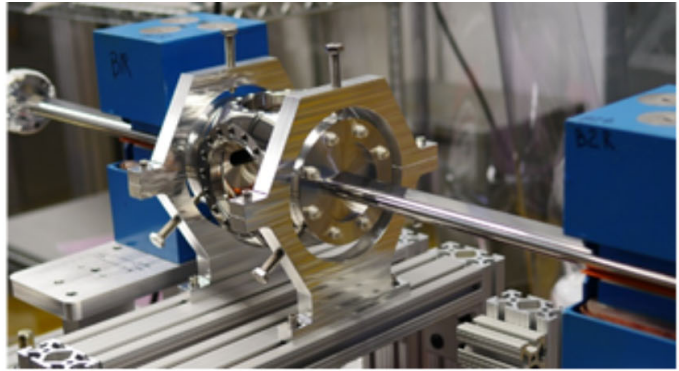

Figure 61: (left) The cross-sectional geometry of the four bypass chambers in the OSC insertion; (right) two of the completed bypass chambers integrated with a spherical cube.

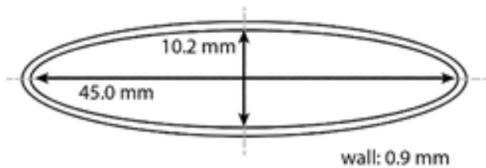
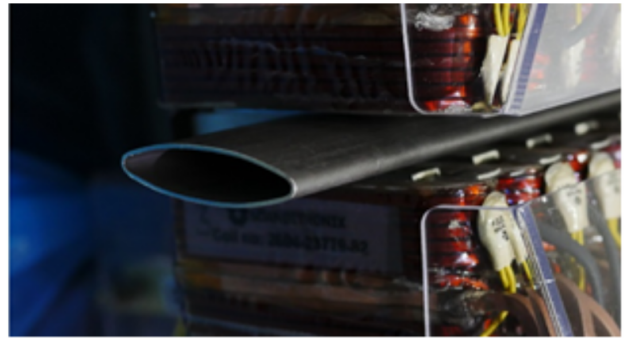

Figure 62: OSC undulator chamber cross-section (left) and beam tube during fit checks in undulator (right).

## **Undulator Chambers**

Preliminary simulations of the OSC electromagnetic undulators highlighted the need for a small vertical aperture to achieve the desired magnetic field (2.22 kG). Balancing the considerations of aperture, conductance, field strength and pole saturation, a value of 12.6 mm was chosen for the pole gap. The undulator chambers have an elliptical cross section (semi-major|semi-minor = 23.37|6.00 mm) and were fabricated by compression of seamless, 32-mm diameter, 316LN tubing (0.9-mm wall) in an aluminum mold. One of the undulator-chamber tubes and its cross-sectional geometry are shown in Figure 62. A 0.75"-ID formed bellows (316L tubing; 316LN flanges) joins each undulator chamber to the bypass section. The narrow diameter of these bellows provides compatibility with compact, high-strength vertical correctors. As shown in a subsequent section, the combination of the long, small-aperture bypass and undulator chambers without interstitial pumping results in the largest pressure rise of any region of the ring.

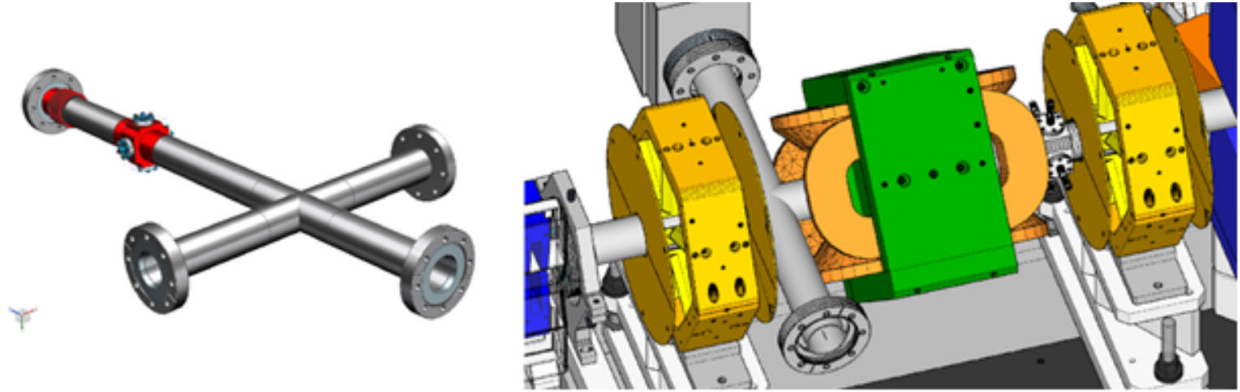

Figure 63: OSC end chambers and their integration in the E straight.

### End Chambers

The OSC end chambers are made from 1.87"-ID, 316L SS tubing and include a four-way cross for pumping, an integrated bellows (316L) and the E-straight's only electrostatic BPMs (one per side). The completed chambers are shown in Figure 63. The pumping crosses are situated as close to the undulator chambers as the magnetic optics would allow, and each chamber is pumped by the same 45S-CV ion/NEG pumps.

### Magnetic Permeability

To reduce magnetic errors in IOTA and the OSC insertion, all chambers are constructed from 316L and/or 316LN stainless steel (SS) and non-magnetic hardware is used throughout. The low magnetic permeability of the completed chambers was verified using a Stefan-Mayer FEROMASTER [33]. These measurements indicated $\mu_r \approx 1.001\text{-}1.002$ throughout the bulk of the tubing and $\mu_r \approx 1.005\text{-}1.007$ in localized areas near some of the welds. The spherical cubes, which were also built from 316L, but extensively machined, were measured at $\mu_r \approx 1.003\text{-}1.004$.

### Vacuum Processing

Due to the stringent vacuum requirements set by beam lifetime in the OSC experiments, special attention is given to reduction of the outgassing rate for the IOTA and OSC chambers. The outgassing of chemically cleaned, unbaked SS is set by $H_2O$ molecules at a typical rate of $\sim 3 \cdot 10^{-10}$ Torr·l/s·cm$^2$ and by $H_2$ molecules at a rate of $\sim 7 \cdot 10^{-12}$ Torr·l/s·cm$^2$. Strong baking under vacuum is required to achieve outgassing rates in the mid $10^{-13}$ Torr l/s·cm$^2$ range [34, 35]. As example, a chemically cleaned SS surface under vacuum for 2-3 days followed by 24 hours bake-out at $\sim 300°C$ can have an $H_2$ outgassing rate down to $\sim 5 \cdot 10^{-13}$ Torr·l/s·cm$^2$ with all other species (CO, $CO_2$ and $CH_4$) adding up to about 5% of the hydrogen level [34]. Baking at $\sim 400°C$ may reduce the $H_2$ outgassing rate to $\sim 2\text{-}3 \cdot 10^{-13}$ Torr·l/s·cm$^2$ [35]. Vacuum firing of SS at $\sim 950°C$ enables a hydrogen outgassing rate at or below $10^{-13}$ Torr·l/s·cm$^2$ [35].

The bypass, undulator and end chambers in the OSC section and all other beam pipes in the IOTA ring were electropolished and vacuum fired at 950°C. All chambers, bellows and other envelope components, e.g. BPM housings, have been processed with standard UHV-cleaning techniques.

Once all E-straight elements are installed, the entire IOTA ring, with minor exceptions, will be baked at ~120°C to remove adsorbed water from the surfaces of the vacuum envelope. IOTA will be segmented into three sections by gate valves to reduce the scale of the required bakeout system.

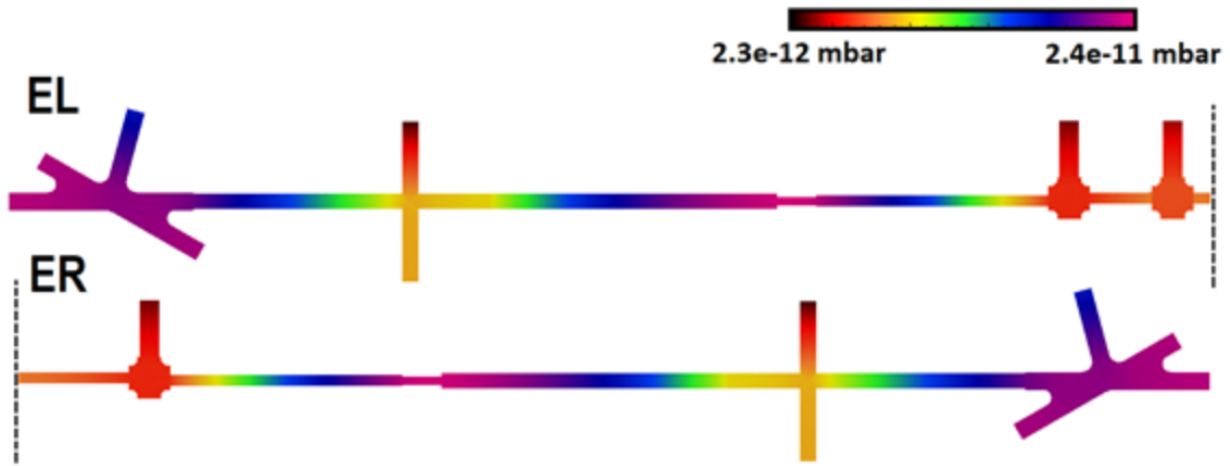

Figure 64: Molflow simulation of the two halves of the OSC insertion (EL and ER), including the IOTA main dipole chambers.

**Simulated UHV Performance**

The vacuum performance for the OSC section has been simulated using the software package Molflow [36]. The exact internal geometry of the OSC insertion, including the adjacent main-dipole chambers was imported and validated for proper surface description, opacity of surfaces and absence of leaks in its mesh. It is assumed that after baking the system will be dominated by $H_2$ outgassing, and cosine-type desorption was applied with a variety of outgassing rates to investigate different possible scenarios. Figure 64 shows the results of one such simulation. The chamber surfaces that were vacuum fired are set to an outgassing rate of $10^{-13}$ Torr·l/s·cm$^2$ in both simulations and the unfired and main-dipole-chamber surfaces are set to $10^{-12}$ Torr·l/s·cm$^2$. Note that the main dipole chambers are of UHV aluminum construction and may have significantly lower $H_2$ outgassing; however, a more conservative rate of $10^{-12}$ Torr·l/s·cm$^2$ is applied here. Each dipole chamber is pumped by a 45 l/s ($N_2$), 45S-CV ion pump; $H_2$ pumping speed is set at 80 l/s for the dipole pumps and 280 l/s for the ion/NEG combos on all other ports. Additional simulations were performed to estimate the impact of the higher outgassing rates that are expected from the precision motion systems in the central chambers. It is difficult to estimate an appropriate rate based on the composition and structure of these motion systems; however, as a simple test, the rate for one 4.5" flange on each of these chambers was increased by 100 times. This resulted in an increase of only ~50% in the equilibrium pressure for the central region of the bypass.

The OSC section is the lowest conductance section of the IOTA ring. Most of the remainder of the ring uses the same 1.87"-ID 316L tubing as the OSC end chambers. The expected equilibrium pressure around the ring should average to the low $10^{-11}$ Torr.

### 12.2. Hardware Systems

**Magnets**

In this section we describe the design and performance of the various magnets that have been fabricated for the initial OSC experiments. We also address considerations of field errors and stray fields in the OSC insertion. For reference, Figure 65 shows examples of the OSC dipoles, quadrupoles, sextupoles, coupling quadrupole and vertical corrector, and Table 18 presents the nominal performance specifications from 3D magnetic simulations of each magnet type. Performance details of the undulators are given separately in the section below. The integrated field quality in each case is taken at a reference radius of $R = 6.3$ mm, which is the typical physical aperture in the OSC bypass section.

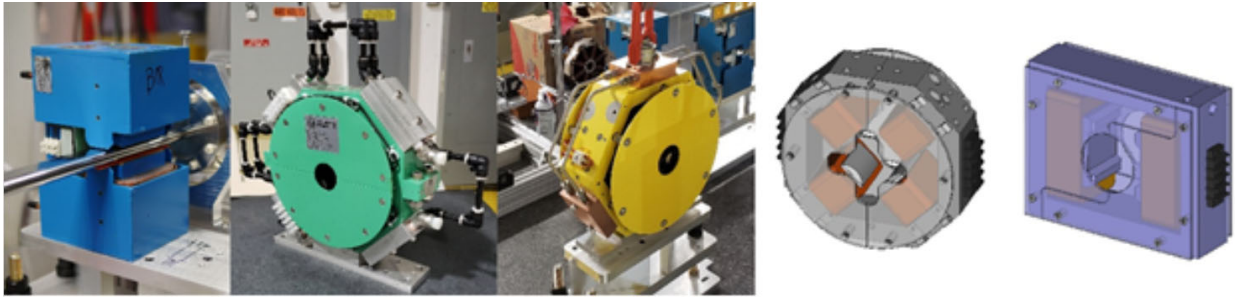

Figure 65: OSC magnets (from left to right): chicane dipole, quadrupole, sextupole, coupling quad, vertical corrector.

Table 18: Nominal properties of OSC magnets (from 3D simulations)

|  | Dipoles | Quads | Sextupole (L/S) | C quad | Y corr. |
|---|---|---|---|---|---|
| Physical length (cm) | 11.00 | 10.00 | 13.48/11.48 | 7.20 | 6.50 |
| Magnetic length (cm) | 8.76 | 8.30 | 10.00/8.00 | 4.97 | 5.68 |
| Nom. excitation (kA-turns) | 1.06 | 1.27 | 1.00 | 0.17 | 0.76 |
| Nom. strength (kG, kG/cm, kG/cm$^2$) | 1.42 | 1.21 | 0.336 | 0.09 | 0.31 |
| Integ. strength (kG-cm, kG, kG/cm) | 12.38 | 10.04 | 3.36/2.69 | 0.45 | 1.76 |
| Integ. field quality (at $R$=0.63cm) | $1.1 \cdot 10^{-4}$ | $2.0 \cdot 10^{-4}$ | $2.5 \cdot 10^{-7}/1.1 \cdot 10^{-7}$ | $7.0 \cdot 10^{-5}$ | $2.9 \cdot 10^{-4}$ |
| Yoke and screen material | 1006 | ARMCO | ARMCO | 1006 | 1006 |
| Cooling/thermal stabilization | Indirect | Indirect | Indirect | None | None |

**Field Screens and Stray Fields**

Possible sources of magnetic-field errors must be considered due to the relatively low energy of the electron beam (100 MeV) and the required precision of the OSC bypass.

As seen in Figure 59 the OSC straight has a large number of magnets and small inter-element separations; therefore, with the exception of the undulators, field screens are included to reduce interference with adjacent magnets. The screens were optimized to improve integrated field quality while reducing stray fields at neighboring elements to the order of 1 G.

The beam will experience non-negligible deflection by the Earth's magnetic field, the largest component of which is ~0.5 G in the vertical plane. At 100 MeV, the beam is deflected by about 0.15 mrad over a distance of one meter. This deflection is well within acceptable value due to the availability of dipole correctors (chicane dipoles and undulator trim coils) throughout the OSC straight.

Field maps for unshielded 45S-CV ion pumps indicate stray-field magnitudes at the beamline on the order of the Earth's field or less. Field data for the HPS/MKS cold-cathode gauges used in IOTA indicate the potential for stray fields at the beamline that are significantly stronger than the Earth's field; however, shielding these devices is straightforward and has been done for similar gauges on the FAST-linac beamline.

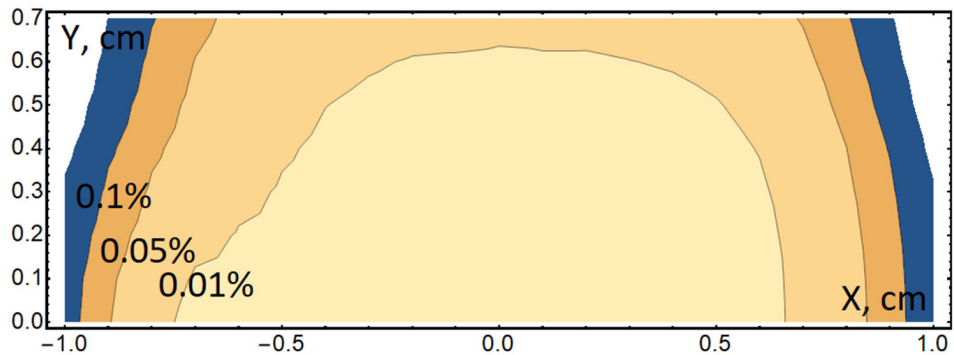

Figure 66: Contour-map of the integrated field error for the OSC chicane dipole at 2.4 kG field.

**Chicane Dipoles**

The OSC chicane dipoles have a C-shaped geometry to accommodate side loading of the vacuum chamber during assembly and magnet removal during bakeout. Asymmetric field-shaping bosses are used to compensate the quadrupole and sextupole components induced by the open geometry and finite width of the poles. Requirements on the accuracy of the dipole fields are established by the need to minimize the quadrupole and sextupole components so that the optics difference from the ideal chicane (built with ideal rectangular dipoles) would be determined by the coupling quad and the sextupoles correcting pathlength non-linearity. In particular, the closed orbit should be centered in the bypass sextupoles to within ~20 μm to minimize the quadrupole feed-down errors which should be an order of magnitude weaker than the strength of the coupling quad. For the beam offset in the bypass of ~2 cm that sets the field accuracy of the dipoles to be better than 0.1%. Similarly, the integrated focusing strength at the maximum aperture (0.63 cm) should be ~$6 \cdot 10^{-4}$ or lower. Figure 66 presents the relative accuracy of integrated dipole field computed with 3D model of the dipole.

To achieve the desired integrated field error a coils-on, multi-pass EDM of the critical field-shaping surfaces was performed. The narrow aperture of these magnets, which was set primarily by field strength requirements in the 2-mm OSC design, establishes the maximum aperture of the vacuum envelope in the bypass section and therefore the maximum angular acceptance for the pickup undulator's radiation (~3.5-mrad half angle). Field screens are used to clamp the field down to the ~1G at ~3.5 cm beyond the faces of the magnet. The shape and location of the screens were adjusted to minimize non-linearities of the edge field.

### OSC Quadrupoles

The insertion contains six main quadrupole magnets (see Figure 17): two of the IOTA-ring quadrupoles and four OSC quadrupoles (pictured in Figure 65) that establish the lattice in the bypass. Indirect cooling/thermal stabilization is provided via a sequence of four water blocks on the exterior of the yoke, and plumbing is routed such that alignment nests and spheres can be accommodated without disconnection of the water system. The field screens reduce the stray fields at the adjacent elements (3-cm separation) to ~1.3 G.

### OSC Coupling Quadrupole

The coupling quadrupole (see Section 3) enables longitudinal-horizontal coupling of the OSC cooling rate and must provide an integrated field strength of ~0.45 kG to split the cooling rates equally. The excitation can safely be increased by a factor of three or four without any significant thermal-management considerations. That offers flexibility in exploring OSC with different degrees of coupling. The quadrupolar field-screen geometry avoids interference with the flat-oval bypass chambers while maintaining excellent field quality and the desired level of field clamping at the adjacent sextupole's field screen (~1.5 G). Additionally, extra margin is included between the chamber wall and the screen to enable a quad displacement. That allows its use as a permanent field corrector in the center of the bypass.

### IOTA/OSC Sextupoles

For the 0.95-μm experiment, requirements for the compensation of nonlinear path lengthening are less substantial than the 2.2-μm case and can be met using three sextupoles in the electron bypass. Space constraints guided the design towards an asymmetric configuration with long and short versions of the sextupoles (marked by L and S in Table 18). In particular, the two outer sextupoles have the same design and construction as the IOTA main-ring sextupoles (90-mm core; 135-mm total length), and the inner sextupole has a shorter core but is otherwise identical (70-mm core; 115-mm total length). As discussed in the dipole section, quadrupole feed-down errors in the sextupoles should accumulate to less than 10% of the coupling quad's nominal integrated strength (~0.45 kG). While relatively strong vertical correctors have been included in the system design to accommodate various sextupole-misalignment scenarios, the transverse-positioning errors will be determined during lattice correction and, if necessary, alignment corrections will be made to minimize the required field of the correctors.

### Undulators

The OSC undulators, pictured in Figure 67, are electromagnetic with 14 full periods (4.84 cm/ea)

and include [1/4, -3/4] terminations for correction of the first and second field integrals. Each coil has indirect water cooling for thermal stabilization and an integrated thermal switch.

Requirements on the magnetic field are determined by the angular divergence of the undulator radiation ($1/\gamma \approx 5$ mrad) and transverse size of the pickup-undulator radiation when refocused into the kicker undulator ($y_0 \approx 200$ μm). We set the requirements for deviations of particle angle relative to the particle in the ideal undulator to be smaller than $0.1/\gamma \approx 0.5$ mrad, and for particle deviations of $0.1 y_0 \approx 20$ μm. Transformed to the magnetic-field requirements, it requires the imbalance of positive and negative fields to be $\overline{\Delta B}/B \leq 10^{-3}$. To have an effective resonant interaction in the second undulator, we require that the nominal wavelengths radiated from both undulators coincide with relative accuracy of $0.1/n_w$. This requires the undulators' fields and mechanical sizes to coincide within a relative accuracy of ~0.5%.

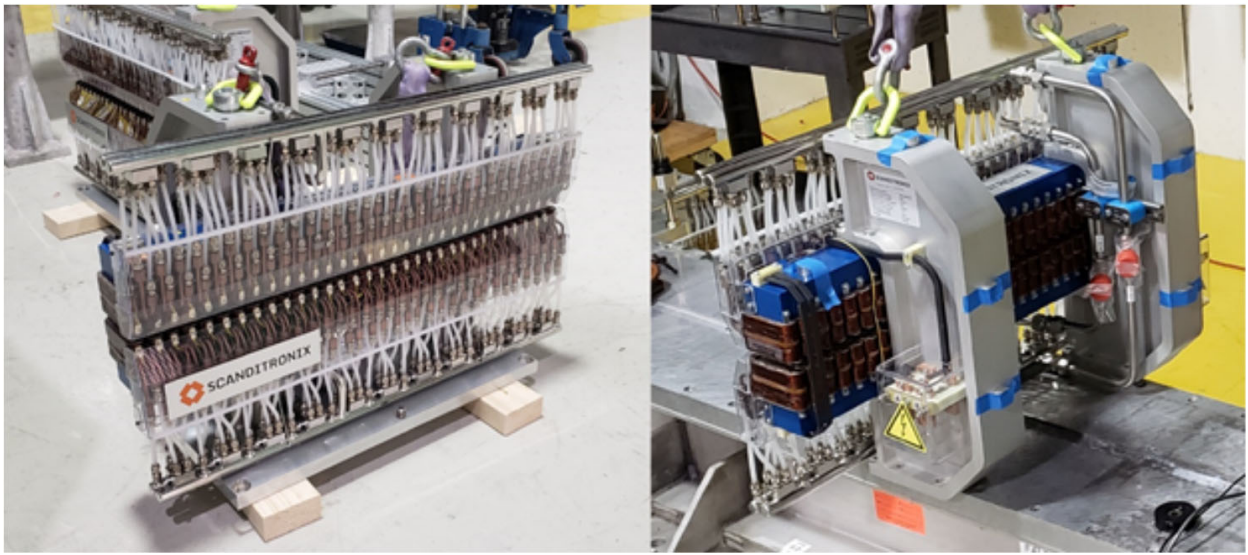

Figure 67: OSC Undulators being prepared for referencing and integration.

### OSC Dipole Correctors

Horizontal correction of the orbit trajectory in the OSC straight can be accomplished through a combination of the chicane dipoles, corrector coils in the undulators and, if necessary, horizontal offset of the coupling quad. Vertical correction requires the use of additional corrector magnets. The nominal design parameters shown in Table 18 were set by accounting for the most severe bypass-sextupole misalignment scenarios, while balancing against a number of considerations, including space, stray fields, field quality, yoke saturation, thermal management and system-integration procedures.

## 12.3. Power Systems

### Requirements on Field Stability of Chicane Dipoles

The most critical field stability requirements come from the chicane dipoles in the bypass. Ripple and noise in the chicane-dipole power supplies will feed down into magnetic-field errors and, with respect to the fixed optical delay, will appear as an effective error in the beam energy on a given pass through the system. Calculations were performed to examine the sensitivity of the bypass mapping to noise in the field strength of the individual chicane dipoles. It is assumed that the field errors at each dipole are independent and normally distributed with relative rms value of $0.5 \cdot 10^{-4}$. It is also assumed that the noise + ripple is dominated by spectral content significantly above the frequency of the OSC cooling rate, i.e. >~40 Hz. In this way, the beam randomly samples many effective energy errors during an OSC cooling time and an average reduction in the OSC cooling rate can be computed. The synchrotron-radiation loss and cooling rate, as modified by OSC (*xy*-uncoupled case), are shown in Figure 68 and indicate that a normalized rms field stability of $0.5 \cdot 10^{-4}$ limits the overall reduction in cooling rate to ~5% of the maximum value.

Energy stability due to variations in IOTA's main bending dipoles must be considered as well. The stability is estimated using beam jitter at the synchrotron-radiation BPMs in combination with the approximate dispersion function at those locations. Measurements in IOTA's first science run indicated an energy stability on the order of $10^{-5}$ for short timescales <1 ms. Significant variations (~$10^{-4}$) were observed for longer timescales (~1 s), but subsequent upgrades of IOTA's main bend bus and power systems have eliminated this noise. Modification of the OSC physics by energy variation in IOTA should be negligible.

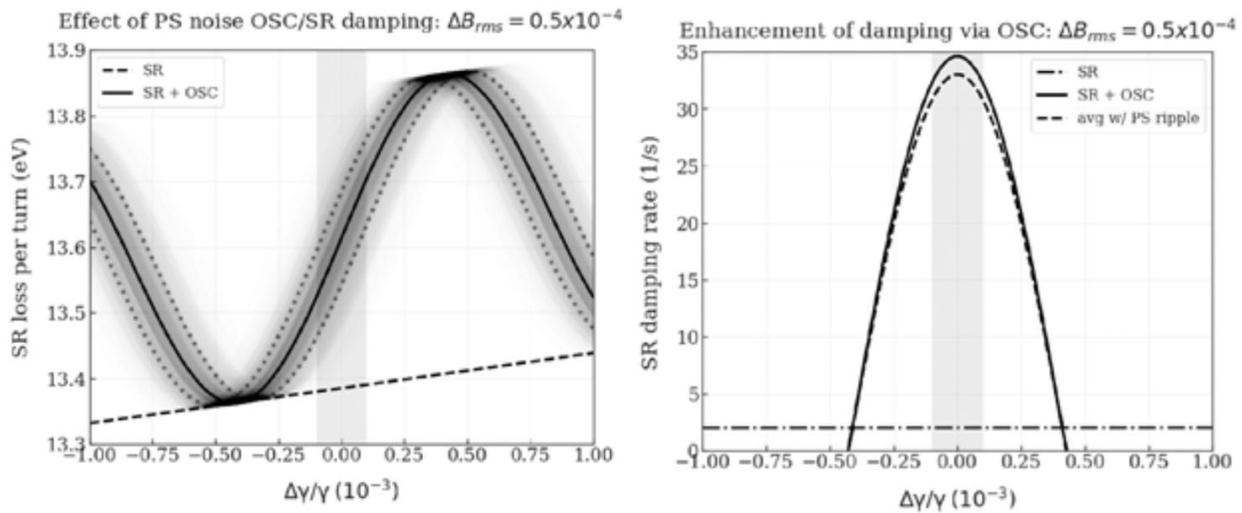

Figure 68: Dependence of synchrotron radiation loss (left) and cooling rate (right) on the relative particle energy with OSC on. Dotted lines surrounding the curves show the rms bounds when the field noise in the chicane is included.

### BiRa Systems PCRC and MCOR

To meet the field-stability requirements described above, the chicane dipoles will be regulated by a pair of Precision Current Regulator Controller (PCRC) systems from BiRa Systems [37]. Using a high-precision DCCT, the PCRC system regulates the ripple+noise of a standard bulk supply (~$10^{-3}$) to the $10^{-5}$ level and provides long-term stability of a few parts per million. Ideally, each dipole would be regulated by a separate PCRC; however, in the initial experiments, each PCRC unit will regulate a pair of the dipoles.

Other elements including the OSC quadrupoles, bypass sextupoles, undulator trim coils and vertical correctors will be powered by 12-A bipolar power modules in an MCOR system from BiRa Systems [38]. Each MCOR power module provides a regulated output to the load with ripple+noise and long-term stability of ~+/-50 parts per million.

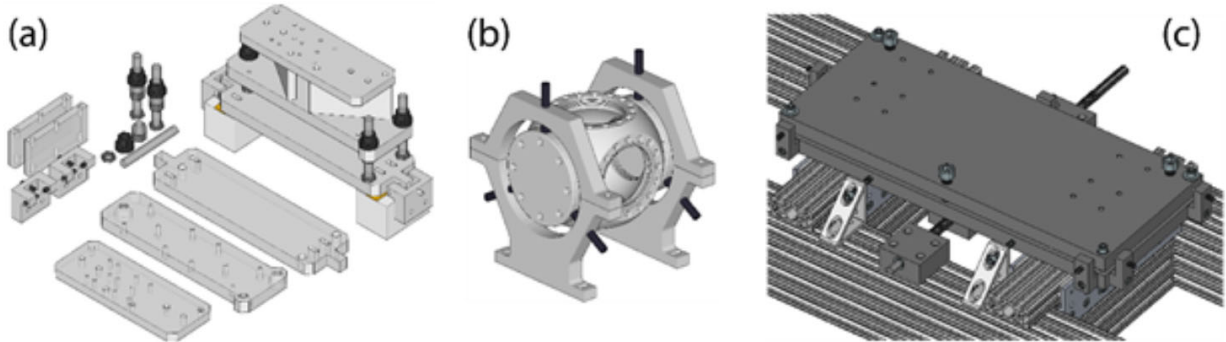

Figure 69: (a) 6-DOF universal support stand for the OSC magnets: (b) concentric chamber/flange supports holding a Kimball Physics spherical cube; (c) 6-DOF support stand for the OSC undulators.

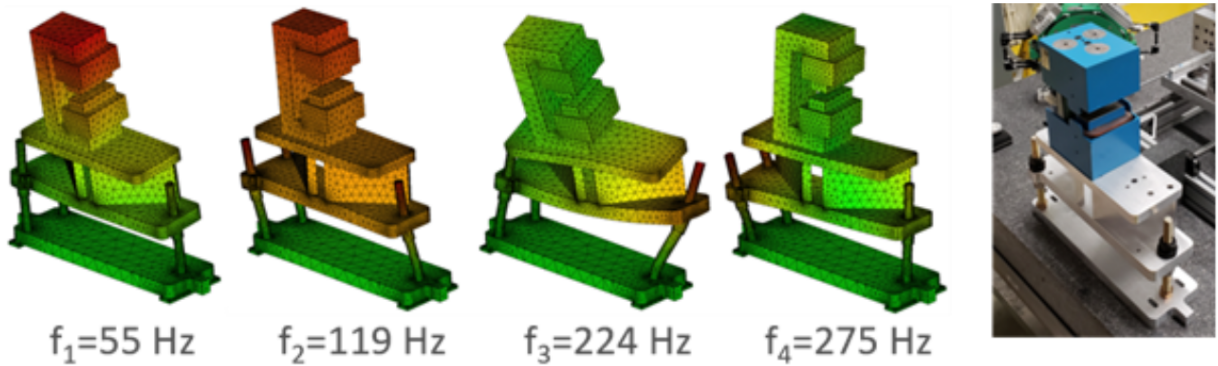

Figure 70: Universal, 6-DOF magnet support stand and FEM eigenmode analysis of the support stand under the load of an OSC dipole. Coloring represents the absolute displacement.

### 12.4. Mechanical Supports

**Universal 6-DOF supports**

The universal support stand shown in Figure 69 (a) was designed to provide 6-dimensional (6-DOF) adjustment for the OSC dipoles, quadrupoles and sextupoles. The design emphasizes compactness, high eigenfrequencies, low-cost and ease of assembly. Precision tip-tilt control and vertical translation is accomplished using three pairs of open-cap nuts and spherical flange nuts. The cap nuts rest against conical insets in the central plate, which provides improved registration of the assembly relative to a configuration with dual flange nuts. Longitudinal and horizontal translation and yaw is provided by two pusher blocks located on the sides of the stand's base. FEM simulations of a support stand with an OSC chicane dipole were performed to examine its eigenmodes and ensure that the system was sufficiently rigid. The four lowest eigenmodes and a fully assembled dipole and support stand are shown in Figure 70. At ~55 Hz, the lowest-frequency mode should be sufficiently far from any significant sources of mechanical excitation.

### Low-profile 6-DOF stand

A lower profile 6DOF stand was developed to support the vertical correctors and the coupling quad in the OSC insertion. This stand is a single plate, supported by 8020 extrusion, that uses the same pusher blocks as the universal stand described above, but tip-tilt and vertical translation are provided by three captured ball-tipped screws in the plate that register against steel plates on the 8020 support frame.

### Undulator 6-DOF stage

The undulator supports are pictured in Figure 69 (c). Tip tilt and vertical translation is provided via three ball-tipped screws captured in the top plate. Yaw and fine translation in the longitudinal and horizontal directions is provided by pusher blocks with captured ball-tipped screws around the perimeter of the support plates. Additionally, compatibility with the UHV bakeout procedure requires a long-range (~10 cm) horizontal-motion capability. This is accomplished using prefabricated 8020 bearing blocks and coupling of a lead screw to the main support plate via a threaded drive block.

### Concentric chamber supports

Simple concentric chamber supports, shown in Figure 61 and Figure 69, are used as a low-profile means of providing additional support to the vacuum envelope. These aluminum supports use non-magnetic hardware and are split in the vertical center to allow for insertion and removal. During bakeout, the top screw is loosened to allow the chambers to expand and slide freely.

## 12.5. Precision-Motion Systems

### SmarAct 70.42 Smarpod

It is desirable to have a precision 6-DOF motion capability for the primary OSC optics in both the amplified and non-amplified experiments. In the amplified case, where multiple focusing elements are situated on a long moment arm, this precision control is especially important. The SmarAct 70.42 Smarpod is a compact, hexapod-like motion solution that provides high-accuracy, closed-loop motion with large translational and angular ranges ($[x,y,z]$=[10,10,5] mm; $[\theta_x,\theta_y,\theta_z]$=[14,16,28] deg). The Smarpod is also UHV compatible to the $10^{-11}$ Torr level, and while a non-magnetic version is available, in the OSC experiments, the Smarpod's location is sufficiently far from the beam that the standard version is adequate. The maximum normal-load capacity of the stage (5 N), is more than sufficient for the lightweight, telescopic optical system that is envisioned for the amplified-OSC experiments.

### SmarAct SR-2013 Rotary Stages

The delay stage, which is described in the instrumentation section, includes two SR-2013 in-vacuum rotary stages. These stages feature continuous, precision angular motion (70-nrad step) with a high rotation speed (~45 deg/s), UHV compatibility ($10^{-11}$ Torr) and an ultra-compact form factor (22.5x20x10.2 mm).  The non-magnetic version is used due to the stages' close proximity to the electron beam (~2 cm).

## 12.6. Optics for Radiation Focusing

This section details the focusing optics that have been developed for the non-amplified OSC experiments.  Specific design details for the amplified-OSC optical systems have not been developed. Preliminary consideration of these systems is discussed in Sections 8 and 10 of this report.

### **Primary lens**

As demonstrated previously, while a single lens does not suppress depth-of-field effects, it can produce nearly the same integrated kick as a three-element telescope and enable the non-amplified OSC experiment at 0.95 μm.  The sensitivity of the maximum OSC cooling force to the focal length and positioning of a single lens was given in Section 8.2.  Due to the location of the coupling quad and the small aperture of the bypass chamber, the lens position was chosen to be 10 cm downstream of the bypass center.  As seen in Figure 48, when paired with a focal length of 85 cm, the reduction in kick value is limited to 6% and is relatively flat over the anticipated variations of position and focal length.

The lens is fabricated from fused silica (CORNING-HPFS-7980) in a spherical, plano-convex form factor with 20-10 surface quality.  It has a 16-mm diameter and a clear aperture of 14 mm when mounted.  Accounting for the smallest upstream aperture, which is the inner surface of the bypass chamber at the transition to the optics cross, the incident undulator-radiation spot has a diameter of approximately 13 mm.  The total optical delay of 0.648 mm, which must be split between the lens and the delay plates, corresponds to 1.397 mm of fused silica at a wavelength of 0.95 μm.  The lens' central thickness was chosen to be 0.817 mm, which keeps its aspect ratio within the limits of standard manufacturing techniques while allowing for two delay plates with exactly 0.25-mm thickness, a standard offering for precision optical flats.

Monte Carlo simulations of the maximum OSC energy exchange were performed in SRW for the lens' expected manufacturing tolerances (central thickness: +/- 0.02 mm; radius of curvature: +/- 0.1%). Variation of the optical delay due to the CT tolerance is small compared to the tuning range of the apparatus and is assumed to be corrected for in the actual experiment.  The results of these simulations, shown in Figure 71, demonstrate negligible deviation (~1%) from the design performance.

Both anti-reflection-coated and uncoated lenses were produced.  The coating is centered at 1050 nm to maximize the transmitted power of the fundamental band (1% loss at 1050 nm and 5% loss at 900 and 1300 nm).  The coating creates additional loses of the second and third-harmonic undulator radiation, which is used for diagnostic purposes; however, there is already orders of magnitude more radiation than what is required for this purpose.

The lens is secured in an aluminium holder with a combination of clamps and a set screw.  Indium wire is used as a buffer at all contact points, and care is taken to avoid significant stresses that could distort the lens. The indium is compatible with the bakeout temperature of these areas (90 °C), and its sublimation, which could contaminate the surface of the optics, should be negligible. The assembled lens holder and motion system is shown in Figure 72. The lens holder couples to the Smarpod via a 10-cm post to augment the translational range of the Smarpod and allow the lens to be moved clear of the alignment laser (optical axis). The lens-holder assembly has a weight of 0.32 N, far below the 5 N, normal-load capacity of the Smarpod.

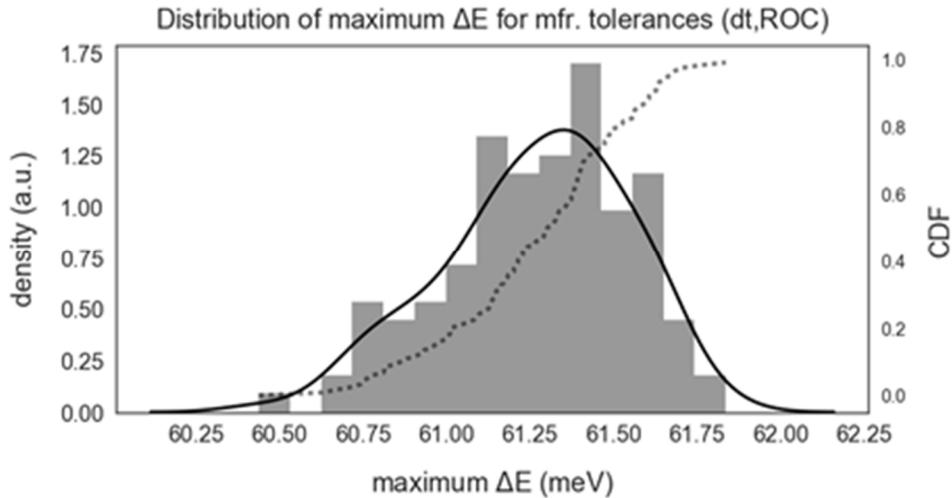

Figure 71: Maximum OSC energy exchange for the 0.95 μm case with rms radius-of-curvature and central-thickness deviations of +/-0.1% and +/- 0.02 mm respectively; a kernel density estimation (solid line) and the cumulative distribution function (dotted line) are shown as well.

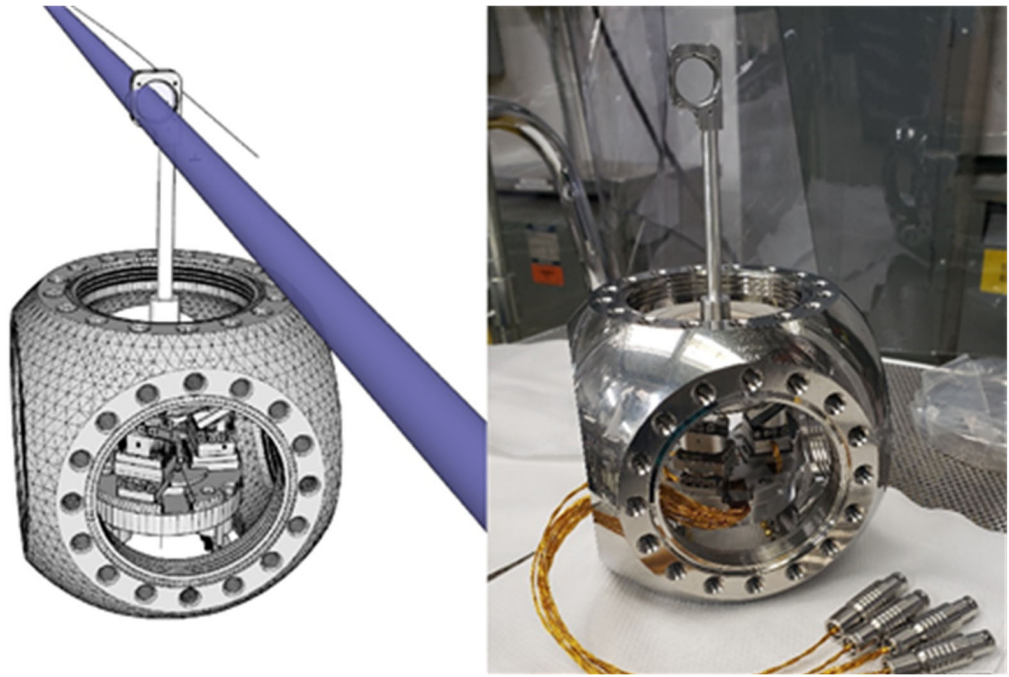

Figure 72: OSC lens holder and Smarpod 6-DOF motion system (left); OSC optical delay stage and pickoff mirror (right); blue shows the outer envelope of the undulator radiation from the pickup.

# 13. Instrumentation of the OSC Experiment and OSC Tuning

## 13.1. Optical Diagnostics and Alignment for the OSC Straight

**Undulator Radiation BPMs and Temporal Alignment**

Synchrotron-radiation (SR) BPMs have been an important tool for lattice correction and beam characterization in IOTA, especially during low-charge operation where the ring's electrostatic BPMs are ineffective. In the OSC straight these BPMs have additional very important role: bringing together the radiation of pickup undulator and the electron beam in the kicker undulator.

A system of undulator-radiation (UR) BPMs for both the pickup and kicker undulators (PU; KU) has been developed to support correction of the OSC lattice. A basic model of the two lightboxes housing the UR BPMs is shown in Figure 73.

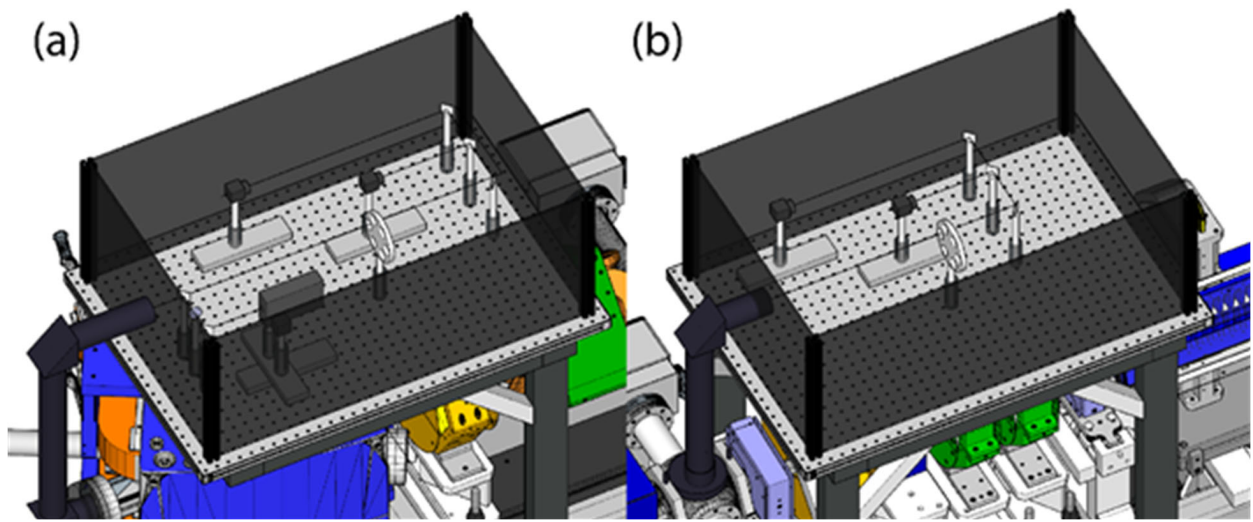

Figure 73: Kicker (a) and pickup (b) lightboxes housing the undulator radiation BPMs and fundamental-band detectors and optics.

For the PU lightbox, shown schematically in Figure 74, an actuated first-surface mirror, located in vacuum on the upstream side of the OSC bypass, directs the PU radiation upwards through a fused-quartz window and is focused with an achromatic lens into the light-tight box above. Two CMOS cameras (BlackFly-PGE-23S6M-C) on motorized longitudinal stages (~15-cm range) are positioned at different distances downstream from a filter wheel and a 50/50 beam splitter. These cameras, Cam1 and Cam2, are located at the image planes for two separate source points a quarter of the way ($L_U/4$) from either end of the undulator. Although radiation from the electron beam's entire trajectory is present at the cameras, each camera's image has a single high-intensity beam spot from the corresponding source point. As with IOTA's SR BPMs, the differential motion of these spots can be used in LOCO-based lattice and orbit correction. The position and angle errors of the beam orbit at the undulator center can be computed using the known transfer matrix of the optical system and the position of the light spots relative an alignment laser, which is described below. Furthermore, the cameras' motion stages enable more detailed mapping of the orbit along the length of the undulator, which may be helpful in optimizing/zeroing of the undulator's field integrals.

The UR BPM concept was first demonstrated using ray-tracing simulations and subsequently confirmed with full simulations in SRW. Similar performance was observed for both the fundamental and across the second and third-harmonic bands using both perfect achromats and standard singlets (N-BK7 glass). Figure 74 includes an example SRW simulation where the beam trajectory has an angular error at the center of the PU of +2 mrad in both $x'$ and $y'$. A hard, bandpass filter was used corresponding to the range where the cameras have significant QE (380-680 nm). For clarity, the two simulated images have been summed and displayed on a single plot; however, only a single high-intensity spot is visible in either camera's image. The images show only the energy density at the sensor and do not include the effects of QE on the intensity map. The nominal position of the alignment laser is shown as a red dot in the center of the image. The PU lightbox also includes additional space to accommodate single-electron and photon-science experiments.

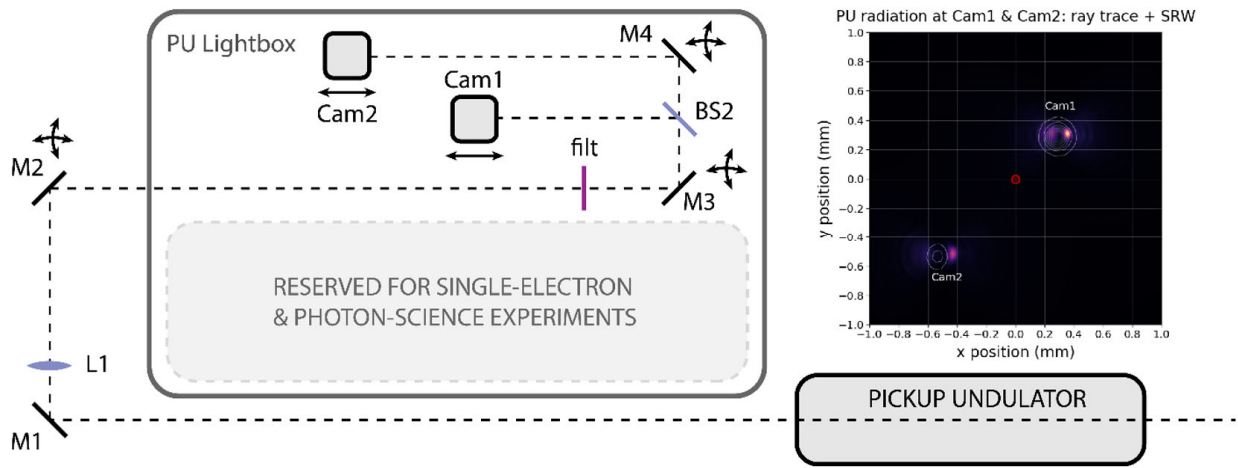

Figure 74: PU lightbox diagram: inset shows a composite SRW simulation of the visible radiation at the nominal positions of Cam1 and Cam2 in the UR BPM. The alignment laser location is shown as a red dot and the overlaid contour plots are the result of simplified ray-tracing calculations. These simulations used a single particle that had an angular error in the KU of +2 mrad in both $x'$ and $y'$. The PU lightbox has a large area available to accommodate planned single-electron and photon-science experiments.

The KU lightbox, shown in Figure 75, has a similar layout for the PU light box, but it includes an additional light path specifically for the detection of radiation in the fundamental band. In this line we can use one of two infrared photodetectors: avalanche photo-diode capable to register radiation of single photon (SPAD: Excelitas SPCM-AQRH-14-TR-BR2) or NIR InGaAs PIN photodiode (Hamamatsu G12180-010A) to register radiation of a bunch which includes many electrons. The avalanche photo-diode will be used in the single-electron OSC experiments (see Section 11); while the PIN photodiode will be used in regular (many-particle) OSC experiments. These diodes are used to observe interference between the radiation from the two undulators, which signals temporal and transverse alignment of the beam and the light in the KU. They are mounted on horizontal and longitudinal stages to enable switching between detectors and fine tuning of the UR spot size at the detector plane. The lens (installed in L2) is required to focus the beam radiation to the small sensitive area of the photodiodes (⌀180 µm for the SPAD; ⌀1 mm for the photodiode). The focal length of the lens was chosen so that during betatron oscillations with the maximum expected amplitude, on the order of 300 µm, the fundamental radiation is still fully collected on the small active element of the SPAD. The smaller inset in Figure 75 shows an SRW simulation of the focused fundamental radiation in the band where the SPAD has significant QE (950-1050 nm). For reference, the full scale of the image is 200 µm in both x and y, approximately the size of the SPAD's active element. Assuming a 50/50 beam splitter, zero losses from mirrors and lenses and an average QE of 12% over this band, a single electron in IOTA should produce on the order of 8000 counts per second from the SPAD from one undulator. For reference, focal lengths and element positions for the PU and KU lightboxes are given in Table 19.

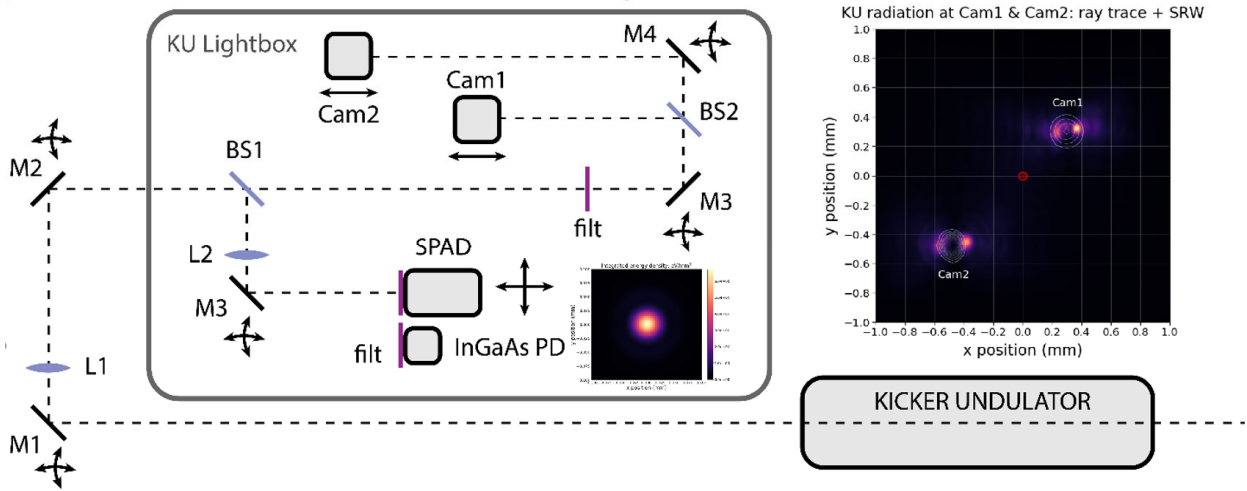

Figure 75: KU lightbox diagram: the two insets show SRW simulations of the focused fundamental radiation at the SPAD/PIN PD detector plane (left) and a composite of the visible radiation at the nominal positions of Cam1 and Cam2 in the UR BPM (right). The alignment laser location is shown as a red dot and the overlaid contour plots are the result of simplified ray-tracing calculations. These simulations used a single particle that had an angular error in the KU of +2 mrad in both $x'$ and $y'$.

**Table 19: Distances and focal lengths for PU and KU lightboxes**

|  | PU lightbox | KU lightbox |
|---|---|---|
| Und. Center to L1 (m) | 1.5 | 2.06 |
| L1 to L2 (m) | N/A | 0.83 |
| L2 to detectors (m) | N/A | 0.173 |
| L1 to Cam1/Cam2 (m) | 1.324/1.731 | 1.769/2.119 |
| L1 focal length @ 587.6 nm (m) | 0.748 | 0.997 |
| L2 focal length @ 587.6 nm (m) | N/A | 0.199 |

### Transverse Alignment

The optical axis of the OSC apparatus is established using pinholes that are surveyed onto the axis of IOTA's E straight with an accuracy of 50 μm. An alignment laser is aligned on this axis in position and angle and matching optics are used to focus the laser near the center of the pickup undulator, emulating the undulator radiation for the alignment and referencing of all diagnostic systems and optics. Figure 76 shows an example ray trace for the alignment laser from the focal spot in the pickup undulator to the imaging sensors. The four simulated beam spots show the expected size of the alignment laser spot at each of the four cameras (L to R: KU Cam1, KU Cam2, PU Cam1, PU Cam2); the full scale of each image is 3-mm in both $x$ and $y$ and the FWHM of each beam is ~1.5 mm, which is small compared to the size of the CMOS sensor ($\Delta x$ = 11.25 mm, $\Delta y$ = 7.03 mm). The centroid of each beam spot is used as a reference for the undulator light at each camera, and from these relative positions, the trajectory error in the undulators can be computed.

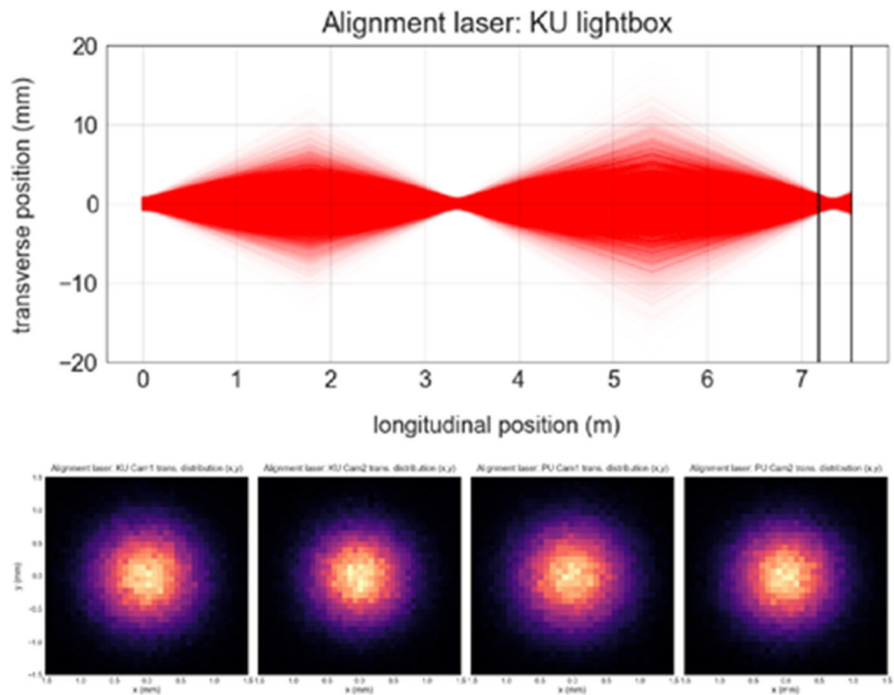

Figure 76: (top) Ray tracing of the alignment laser through the OSC apparatus to the KU BPM camera locations (vertical black lines); (bottom) intensity map of the alignment-laser spot at each UR BPM camera for both KU and PU lightboxes (L to R: KU Cam1, KU Cam2, PU Cam1, PU Cam2); the beam spots are ~1-mm FWHM, much smaller than the cameras' detector element and should provide a clear centroid position.

### Optical-Delay System

Fine tuning of the bypass's optical delay is accomplished by the independent rotation of two precision optical flats. These plates are made of 0.25-mm, uncoated fused silica (CORNING-HPFS-7980). They have a central-thickness tolerance of +/-0.005 mm and a surface quality of 20-10. The plates are oriented such that the horizontally polarized undulator radiation (which only makes the kick in OSC) is p-polarized and incident near the Brewster angle. To determine the optimal orientation, the Fresnel equations for the p-polarized light are solved as a function of the independently variable angles of the two plates. The angles can be mapped onto the delay and horizontal shift of the radiation as shown in Figure 77. By shifting the nominal design point (which considers the CT of the lens) away from the Brewster angle by 7 degrees, the transmittance surface is flattened over a larger range of the rotary motion. At zero horizontal shift, the full delay range without significant transmission loss is ~50 μm, approximately 8% of the total optical delay in the bypass. For zero delay offset, the maximum horizontal shifts that can be produced are on the order of +/- 100 μm, comparable to the focused beam size in the kicker undulator.

The delay plates are secured in low-profile aluminium clamps using indium wire as a mechanical buffer. The clamps are then attached to the SmarAct SR-2013 rotary stages described previously. At maximum rotation speed, the OSC cooling/heating zones can be inverted (*i.e.* the delay changed by $\lambda/2$) in ~8 ms, which is much faster than the OSC cooling time and comparable to the synchrotron period for the planned OSC lattice and experimental conditions.

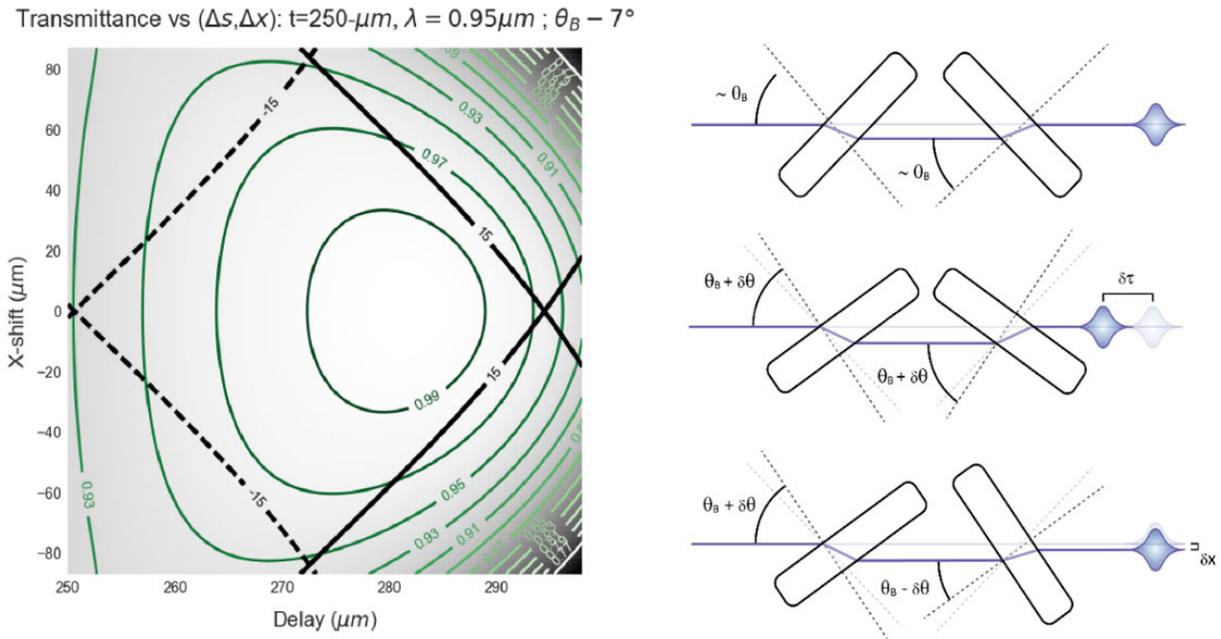

Figure 77: (left) Transmittance surface for the delay stage as a function of delay and transverse shift of the light; the black lines correspond to isopleths for the angle setting of the two plates. (right) The angles of the plates can be changed in concert to produce independent temporal and horizontal shifts of the light.

### 13.2. Upgrade of IOTA SR monitors

To increase the cooling range, the OSC lattice is designed to have a small equilibrium emittance. This necessarily means that the dispersion and lattice functions are small in the main dipoles where the synchrotron radiation (SR) monitors receives the image from. Accordingly, the *x* and *y* beam sizes are small and approach the diffraction limit of the SR BPM optical system. Figure 78 shows the rms beam sizes throughout the ring circumference for sufficiently small beam current where IBS can be neglected (see Section 9). As one can see in the absence of OSC the typical beam size does not exceed 50 μm. With OSC on, both rms beam sizes are reduced to ~10 μm or smaller which is well below the theoretical limit of present SR monitors. Therefore, an improvement of their resolution is a high priority task. We pursue this in two routes. The first one is an improvement of resolution for traditional beam image system, and the second one is based on a usage of π-mode image.

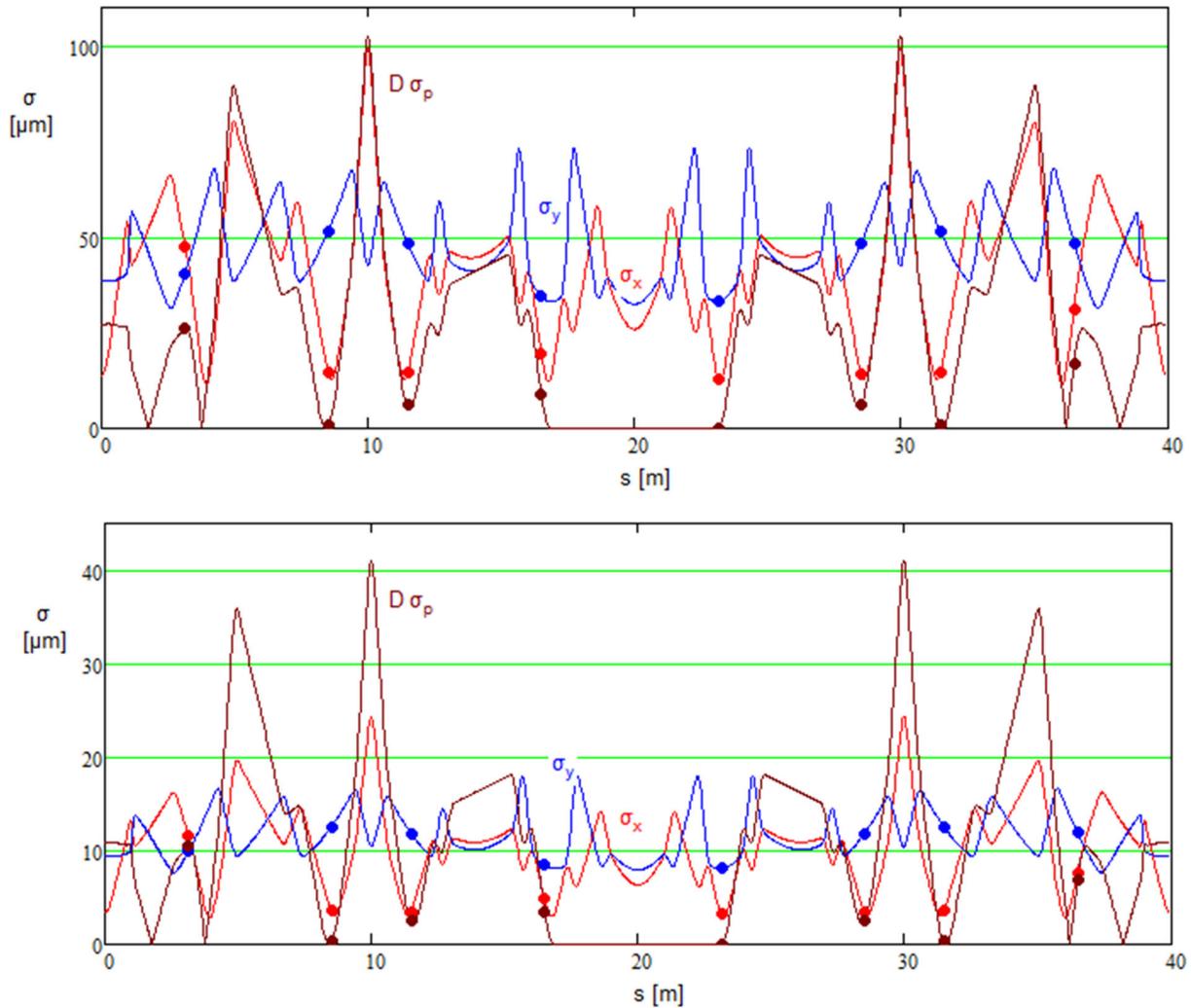

Figure 78: Rms beam sizes along the IOTA circumference with (bottom) and without (top) OSC for sufficiently small beam current when IBS can be neglected: red – $\sigma_x$, blue – $\sigma_y$, brown – beam size due to momentum spread. Dots mark positions of SR monitors: SL4L, SL3L, SL2L, SL1L, SL1R, SL2R, SL3R, SL4R in order of increasing *s*-coordinate.

**σ-mode image**

The resolution of the present SR monitors is determined by the three major constituents: the diffraction, the chromatic aberration of focusing lens and the image widening due to contribution of radiation with vertical polarization – so called the π-mode radiation. Calculations show that the rms resolution of present SR monitor is about 20 μm. That is consistent with the minimum vertical rms beam size of 40 μm observed in IOTA Run II which is mainly determined by coupling between vertical and horizontal motions.

For narrow band radiation with horizontal polarization and the wavelength of $\lambda$ the diffraction results in a Gaussian like distribution in the vertical direction of the focal plane. For point like beam the vertical rms spot size referenced to the point of radiation can be approximated by the following equation:

$$\sigma_v \approx 0.162 \lambda_c \gamma \left(\frac{\lambda}{\lambda_c}\right)^{0.6}, \quad 0.1 \leq \frac{\lambda}{\lambda_c} \leq 50, \tag{82}$$

where $\lambda_c = 2\pi c/\omega_c$ is the critical wavelength of radiation coming from a dipole, and $\gamma$ is the relativistic factor. As one can see for a fixed beam energy a transition to the shorter wavelength improves the resolution as $\lambda^{0.6}$ and the transition to a shorter wavelength is the main remedy which can be used to improve the resolution. Another important consideration is the radiation polarization. Although the vertical polarization contributes to only ~15% of the radiation power for the present SR monitor it increases the width of the distribution by ~13%.

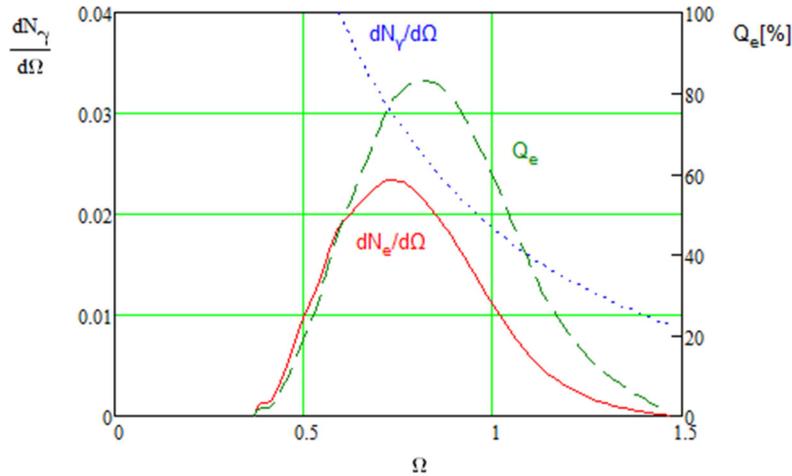

Figure 79: Dependencies of spectral density of photons per turn at the CCD camera entrance (blue), the spectral density of CCD signal per turn (red) and quantum efficiency of CCD (green) for IOTA SR monitor, where $\Omega=\lambda_c/\lambda$, and $\lambda_c$ = 405 nm is the critical wave length of the SR from IOTA dipole. The beam energy is 100 MeV, and the lens angular acceptance is 0.07 rad.

Taking above into account the upgrade of SR monitor includes two major constituents. The first one is a transition to the shortest possible wavelength to reduce the diffraction contribution; and the second one is filtering out the radiation with vertical polarization. Figure 79 presents the spectral density of photons at the entrance of the CCD camera. Accounting the quantum efficiency of the CCD (FL-20BW) one obtains the spectral density of CCD signal. In the figure it is expressed in the number of electrons born in CCD by a single electron in the ring per turn per unit dimensionless frequency $\Omega=\omega/\omega_c$. For IOTA at 100 MeV $\lambda_c$=405 nm. The choice of $\Omega \approx 1$ ($\lambda \approx 405$ nm) results in a reasonably small loss of the sensitivity and allows one to obtain the vertical rms resolution of about 13 μm which is sufficient to observe with confidence the beam size decrease due to the OSC. Figure 80 shows the beam image on the CCD camera computed with SRW for 405 nm wavelength. One can see that a usage of narrow band results in even better resolution for the horizontal plane but the beam trajectory curvature in the dipole results in an interference rings in horizontal plane. The reason of better resolution in horizontal plane is related to smaller diffraction due to larger light width on the lens which diameter is ~50 mm, while the vertical size on the lens is 6 mm (4σ) for $\lambda$=405 nm.

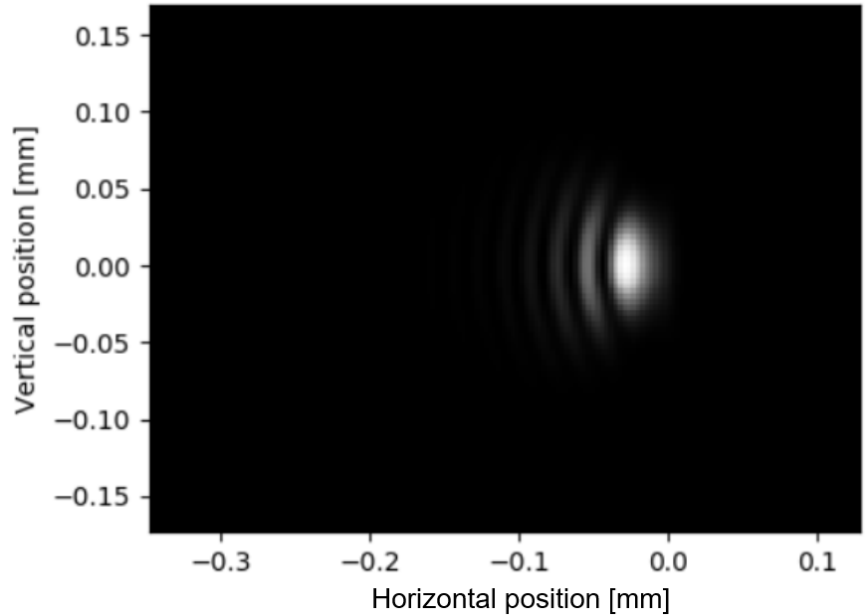

Figure 80: The beam image in the focal plane of the lens (CCD camera) computed with SRW for the beam energy 100 MeV and the wavelength of 405 nm. Magnification is 1.133.

To prevent the image widening due to chromatic aberration the bandwidth of the pass filter should not exceed ~20-30 nm. Taking into account that a single electron could be observed in the course of IOTA Run II the loss of SR monitor sensitivity due to bandwidth limitation is uncritical if a small intensity bunch is observed.

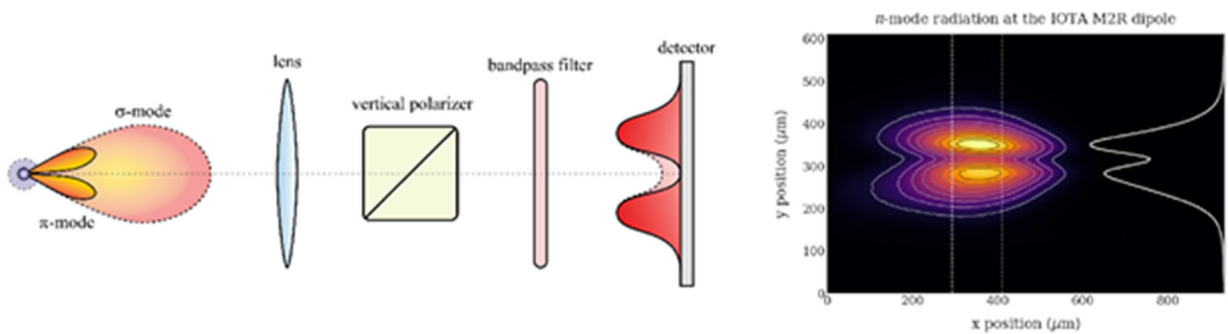

Figure 81: Conceptual diagram of a π-mode diagnostic (left); example π-mode data from IOTA's M2R dipole SR monitor (right). Raw image data is integrated between the dotted white lines to produce the solid-white, vertical cross section on the right.

### π-mode image

Further improvement of the resolution can be obtained with π-mode imaging which is shown schematically in Figure 81. It is a well-known technique for measuring vertical beam sizes below the diffraction limit [39]. For a single particle, the vertically polarized component of the synchrotron radiation from a bending magnet features a null in the bend plane. For a beam of particles, the null is blurred due to the nonzero vertical beam size; therefore, the depth of this on-axis minimum can be used to infer the vertical beam size, even if that size is below the diffraction limit for normal imaging. During the first science run at the IOTA ring, a π-mode diagnostic was prototyped and tested; an example π-mode image from this testing is shown in Figure 81.

To infer the vertical beam size from a π-mode image, the intensity distribution for a single particle must be computed or simulated. For the case of a bending magnet and small observation angles, this "filament-beam-spread function" (FBSF) is given by

$$I(x,y) = \left[ E_0 \operatorname{sinc}\left(\frac{2\pi x_a}{\lambda L_i} x\right) \int_0^\infty \sqrt{1+h^2}\, h K_{1/3}\left(\frac{\lambda_c}{2\lambda}(1+h^2)^{3/2}\right) \sin\left(\frac{2\pi L_0}{\lambda \gamma L_i} yh\right) dh \right]^2 \qquad (83)$$

where $E_0$ is a normalization constant, $L_0$ and $L_i$ are the source-to-lens and lens-to-detector distances, $x_a$ is the (half) aperture at the lens, and $K_{1/3}(x)$ is a modified Bessel function of fractional order [39]. Figure 82 (left) shows the FBSF for using the typical optics in an IOTA SR station. Convolving the FBSF with the electron-beam distribution, and accounting for the magnification of the optics, results in the blurred π-mode images shown in Figure 82 (center), where each curve corresponds to a different vertical beam size. Finally, the ratio of the on-axis to peak intensities for each vertical size is shown in Figure 82 (right). Such a curve can be used to estimate the vertical beam size, and thus vertical emittance, during the OSC experiments.

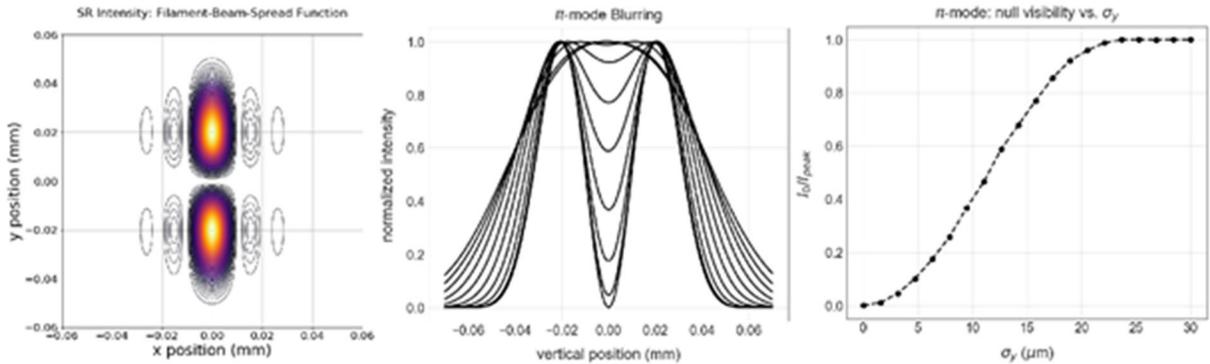

Figure 82: (left) Computed FBSF for typical configuration of an IOTA SR station at 550 nm; (center) simulated, integrated vertical profiles for beams with varying vertical emittance; (right) ratio of valley-to-peak intensity as a function of rms vertical size.

## Appendix A. Ultimate OSC rates and the optimal gain of OA

In this section we assume that the bandwidth of amplifier is sufficiently large, so that the cooling bandwidth is determined by number of undulator periods, and the depth of field in radiation focusing is suppressed, so that the amplitude of the electromagnetic wave stays constant along the kicker undulator. Then, the kick value obtained by a particle is:

$$\Delta p = e\beta c \theta_e E_0 \, \text{Re}\left( \int \exp\left(i\left(\omega_0 t - k_0 z(t) - \psi\right)\right) \sin\left(k_w \beta c t\right) dt \right), \tag{84}$$

where $E_0$ is the amplitude of e.-m. wave generated by a particle in the pickup undulator and then amplified in the OA, $\psi$ is its phase and $\omega_0 = k_0 c$ is its angular frequency, $\theta_e = K/\gamma$ is the amplitude of particle angle variation in the kicker undulator, $K$ is the undulator parameter, $\beta$ and $\gamma$ are the relativistic parameters, $k_w = k_0 / (2\gamma^2)$ is the wave number of undulator, and the integration is performed over the kicker undulator length. We also assume that K << 1 although, as will be seen later, it does not limit the generality of the result. Performing integration, one obtains:

$$\Delta p = \frac{eE_0 \theta_e L_u}{4\pi n_w c} \begin{cases} (2\pi n_w - |\psi|) \sin\psi, & |\psi| \le 2\pi n_w \\ 0, & |\psi| \ge 2\pi n_w \end{cases} \tag{85}$$

where $n_w$ is the number of periods in the undulator, and $L_u$ is the undulator length. For longitudinal cooling the phase $\psi$ is related to the particle momentum as

$$\psi \equiv k_0 s = k_0 S_{pk} \frac{\Delta p}{p}, \tag{86}$$

where $S_{pk}$ is determined by Eq. (4).

To obtain the diffusion we need to average kicks of the nearby particles. Then, the rms momentum spread increase per turn is:

$$\frac{d}{dn} \overline{\Delta p^2} = \frac{dN}{ds} \int_{-L_u/2\gamma^2}^{L_u/2\gamma^2} \left(\Delta p\big|_{\psi=ks}\right)^2 ds \tag{87}$$

where $dN/ds$ is the particle longitudinal density. Substituting Eq. (85) into Eq. (87) and performing integration we obtain:

$$\frac{d}{dn}\sigma_p^2 \equiv \frac{d}{dn}\frac{\overline{\Delta p^2}}{p^2} = \pi n_w \frac{\left(eE_0 \theta_e L_u\right)^2}{6 k_0 p^2 c^2} \frac{dN}{ds}\left(1 - \frac{3}{8\pi^2 n_w^2}\right). \tag{88}$$

Averaging over the Gaussian distribution,

$$\frac{dN}{ds} = \frac{N_e}{\sqrt{2\pi}\sigma_s} \exp\left(-\frac{s^2}{2\sigma_s^2}\right), \tag{89}$$

and accounting that the synchrotron motion decreases the longitudinal emittance growth rate for a bunch by factor of 2 one obtains an increase of bunch momentum spread per turn due to particle interaction through cooling system:

$$\frac{d}{dn}\sigma_p^2 = \frac{\sqrt{\pi} n_w N_e}{24 k_0 \sigma_s} \left(\frac{eE_0 K L_u}{\gamma pc}\right)^2, \tag{90}$$

where we also neglected the second term in the parenthesis in Eq. (88) which is insignificant for large number of periods.

In computation of the optimal gain we will neglect all other diffusion mechanisms. Adding effects of diffusion and the cooling and accounting Eq. (5) we obtain:

$$\frac{d}{dn}\sigma_p^2 = -\left(\frac{eE_0 K L_u}{2\gamma pc}\right) k_0 S_{pk} \sigma_p^2 + \frac{\sqrt{\pi} n_w N_e}{6 k_0 \sigma_s} \left(\frac{eE_0 K L_u}{2\gamma pc}\right)^2. \tag{91}$$

$E_0$ is determined by the gain. With gain increase the second addend in Eq. (91) (representing diffusion) grows faster than the first one (representing cooling). The maximum damping is achieved when the second addend is equal to the half of the first one. That determines $E_0$ at the optimal gain and, subsequently, the emittance cooling rate per turn:

$$\lambda_{s\_opt} = \frac{3\mu_{01}^2 k_0 \sigma_s}{2\sqrt{\pi} n_w N_e n_{\sigma s}^2}, \tag{92}$$

where $n_{\sigma s}$ is determined by Eq. (18), and we expressed $S_{pk}$ through the cooling acceptance of Eq. (17).